\documentclass{aa}

\usepackage{epsfig}
\usepackage {graphicx}
\usepackage{amssymb}
\usepackage{amsmath}
\usepackage{multirow}
\usepackage {txfonts}
\usepackage{natbib}
\usepackage{pifont}
\usepackage{color}
\usepackage{float}
\usepackage{ulem}
\usepackage{tabularx}

\begin{document}
\newcommand{\boldvec}[1]{\vec{\mbox{\boldmath{$#1$}}}}


\title{Constraints from Faraday rotation on the magnetic field structure
in the Galactic halo}
\author{Philippe Terral\inst{1} \and Katia Ferri\`ere\inst{1}}
\institute{$^{1}$ IRAP, Universit\'e de Toulouse, CNRS,
9 avenue du Colonel Roche, BP 44346, F-31028 Toulouse Cedex 4, France\\
Email: katia.ferriere@irap.omp.eu}
\date{Received  ; accepted }
\titlerunning{Magnetic field in the Galactic halo}
\authorrunning{Philippe Terral}

\abstract
{}
{We examine the constraints imposed by Faraday rotation measures 
of extragalactic point sources on the structure of the magnetic field 
in the halo of our Galaxy.
Guided by radio polarization observations of external spiral galaxies,
we look in particular into the possibility that field lines in the Galactic halo 
have an X shape.}
{We employ the analytical models of spiraling, possibly \rm X-shape magnetic fields 
derived in a previous paper to generate synthetic all-sky maps 
of the Galactic Faraday depth, which we fit to an observational reference map 
with the help of Markov Chain Monte Carlo simulations. 
}
{We find that the magnetic field in the Galactic halo is slightly more likely 
to be bisymmetric (azimuthal wavenumber, $m = 1$) than axisymmetric ($m = 0$).
If it is indeed bisymmetric, it must appear as X-shaped in radio polarization maps 
of our Galaxy seen edge-on from outside, 
but if it is actually axisymmetric, it must instead appear as nearly 
parallel to the Galactic plane.}
{}


\maketitle

\section{\label{intro}Introduction}

Interstellar magnetic fields are an important component 
of the interstellar medium (ISM) of galaxies. 
They play a crucial role in a variety of physical processes, 
including cosmic-ray acceleration and propagation, 
gas distribution and dynamics, star formation$\ldots$ 
However, their properties remain poorly understood.
The main difficulty is the lack of direct measurements,
apart from Zeeman-splitting measurements in dense, neutral clouds.
In addition, all existing observational methods provide only partial information, 
be it the field strength, the field direction/orientation, 
or the field component parallel or perpendicular to the line of sight.

In the case of our own Galaxy, magnetic field observations 
are particularly difficult to interpret, 
because the observed emission along any line of sight 
is generally produced by a number of structures, 
whose contributions are often hard to disentangle 
and locate along the line of sight.
In contrast, observations of external galaxies can give us a bird's eye view 
of the large-scale structure of their magnetic fields.
High-resolution radio polarization observations of spiral galaxies 
have shown that face-on galaxies have spiral field lines, 
whereas edge-on galaxies have field lines that are parallel to the galactic plane 
in the disk \citep[e.g.,][]{wielebinski&k_93, dumke&kwk_95} 
and inclined to the galactic plane in the halo, 
with an inclination angle increasing outward in the four quandrants;
these halo fields have been referred to as X-shape magnetic fields
\citep{tullmann&dsu_00, soida_05, krause&wd_06, 
krause_09, heesen&kbd_09, braun&hb_10, soida&kdu_11, haverkorn&h_12}. 

In a recent paper, \cite{ferriere&t_14} (Paper~1) presented
a general method to construct analytical models of divergence-free
magnetic fields that possess field lines of a prescribed shape,
and they used their method to obtain four models of spiraling, possibly X-shape 
magnetic fields in galactic halos as well as two models of spiraling, 
mainly horizontal (i.e., parallel to the galactic plane) 
magnetic fields in galactic disks.
Their X-shape models were meant to be quite versatile
and to have broad applicability; 
in particular, they were designed to span the whole range of field orientation,
from purely horizontal to purely vertical.

Our next purpose is to resort to the galactic magnetic field models derived in Paper~1 
to explore the structure of the magnetic field in the halo of our own Galaxy.
The idea is to adjust the free parameters of the different field models
such as to achieve the best possible fits to the existing observational data
-- which include mainly Faraday rotation measures (RMs)
and synchrotron (total and polarized) intensities --
and to determine how good the different fits are.
Evidently, a good field model must be able to fit both Faraday-rotation 
and synchrotron data simultaneously.
However, we feel it is important to first consider both types of data separately 
and examine independently the specific constraints imposed by each kind 
of observations.
This is arguably the best way to understand either why a given model 
is ultimately acceptable or on what grounds it must be rejected.

In the present paper, we focus on fitting our galactic magnetic field models 
to the existing Faraday-rotation data.
In practice, since we need to probe through the entire Galactic halo, 
we retain exclusively the RMs of extragalactic point sources 
(as opposed to Galactic pulsars).
For each considered field model, we simulate an all-sky map of the Galactic
Faraday depth (FD), which we confront to an observational reference map 
based on the reconstructed Galactic FD map of \cite{oppermann&jge_15}.
The fitting procedure relies on standard $\chi^2$ minimization, 
performed through a Markov Chain Monte Carlo (MCMC) analysis.

Our paper is organized as follows:
In Sect.~\ref{FD_data}, we present the all-sky map of the Galactic FD 
that will serve as our observational reference.
In Sect.~\ref{models}, we review the magnetic field models derived in Paper~1
and employed in this study.
In Sect.~\ref{method}, we describe the fitting procedure.
In Sect.~\ref{results}, we present our results 
and compare them with previous halo-field models. 
In Sect.~\ref{discussion}, we summarize our work 
and provide a few concluding remarks.

Two different coordinate systems are used in the paper.
The all-sky maps are plotted in Galactic coordinates $(\ell,b)$,
with longitude $\ell$ increasing eastward (to the left) 
and latitude $b$ increasing northward (upward).\footnote{
Throughout the paper, the north, south, east, and west directions 
always refer to Galactic coordinates.
}
In contrast, the magnetic field models are described 
in Galactocentric cylindrical coordinates, $(r,\varphi,z)$, 
with azimuthal angle $\varphi$ increasing in the direction of Galactic rotation,
i.e., clockwise about the $z$-axis, 
from $\varphi = 0^\circ$ in the azimuthal plane through the Sun.
As a result, the coordinate system $(r,\varphi,z)$ is left-handed.
For the Galactocentric cylindrical coordinates of the Sun, we adopt 
$(r_\odot = 8.5~{\rm kpc}, \varphi_\odot = 0, z_\odot = 0)$.

\section{\label{FD_data}Observational map of the Galactic Faraday depth}

\subsection{\label{FD_data_maps}Our observational reference map}

\begin{figure}[t]
\resizebox{\hsize}{!}{\includegraphics{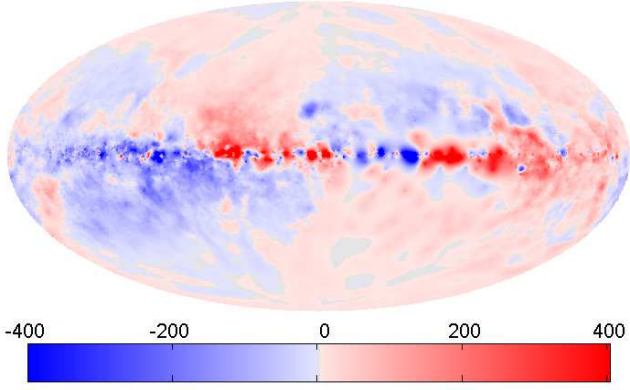}}
\caption{ 
All-sky map, in Aitoff projection, of the observational Galactic Faraday depth, 
${\rm FD}_{\rm obs}$, as reconstructed by \cite{oppermann&jge_15}.
The map is in Galactic coordinates $(\ell,b)$,
centered on the Galactic center $(\ell=0^\circ,b=0^\circ)$,
and with longitude $\ell$ increasing to the left and latitude $b$ increasing upward.
Red [blue] regions have positive [negative] ${\rm FD}_{\rm obs}$, 
corresponding to a magnetic field pointing on average toward [away from] the observer.
The color intensity scales linearly with the absolute value of ${\rm FD}_{\rm obs}$
up to $|{\rm FD}_{\rm obs}| = 400~{\rm rad~m^{-2}}$
and saturates beyond this value.
}
\label{figure_FD}
\end{figure}

\begin{figure}[t]
\centering
\includegraphics[trim=1.9cm 2.2cm 1.4cm 2.2cm, clip, width=0.95\linewidth]{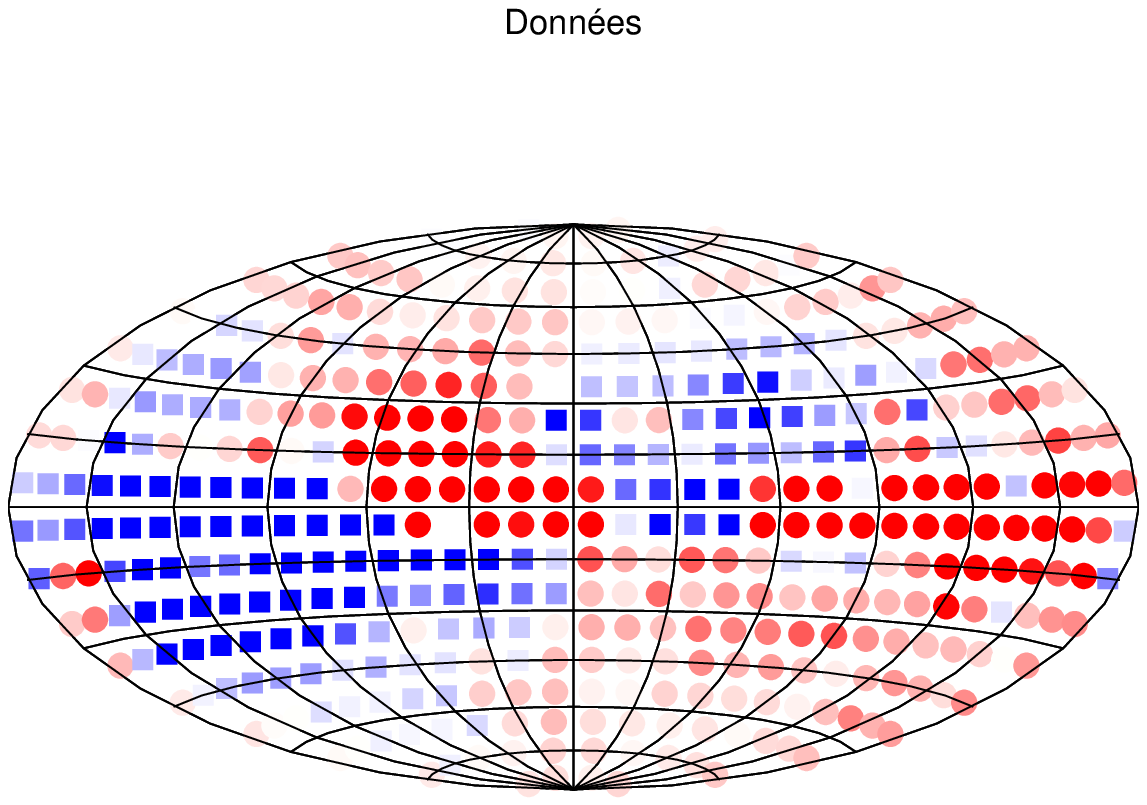}
\caption{ 
All-sky map, in Aitoff projection, showing the disposition 
of the 428 bins covering the celestial sphere,
together with their average observational Galactic Faraday depth,
before subtraction of the contribution from 
\citeauthor{wolleben&flc_10}'s (\citeyear{wolleben&flc_10}) magnetized bubble.
Positive [negative] values are plotted with red circles [blue squares], 
following the same color intensity scale as in Fig.~\ref{figure_FD}.
The coordinate system is also the same as in Fig.~\ref{figure_FD}.
}
\label{figure_FD_bin}
\end{figure}

Faraday rotation is the rotation of the polarization direction of a linearly-polarized
radio wave that passes through a magneto-ionic medium.
The polarization angle $\theta$ rotates by the angle
$\Delta \theta = {\rm RM} \ \lambda^2$,
where $\lambda$ is the observing wavelength and RM is the rotation measure given by
\begin{equation}
\label{eq_RM}
{\rm RM} = (0.81~{\rm rad~m^{-2}}) \ \int_0^L n_{\rm e} \ B_\parallel \ ds \ ,
\end{equation}
with $n_{\rm e}$ the free-electron density in cm$^{-3}$, 
$B_\parallel$ the line-of-sight component of the magnetic field in $\mu{\rm G}$ 
(positive [negative] for a magnetic field pointing toward [away from]
the observer) and $L$ the path length from the observer to the source 
measured in pc.\footnote{
The original expression of RM is an integral from the source to the observer,
which explains the sign convention for $B_\parallel$.
Here, however, it proves more convenient to have the origin
of the line-of-sight coordinate at the observer
and to integrate from the observer ($s=0$) to the source ($s=L$).
This does not affect the sign of RM 
provided one keeps the original sign convention for $B_\parallel$.
}
Clearly, RM is a purely observational quantity, which can be defined only
for a linearly-polarized radio source located behind the Faraday-rotating medium.

More generally, one may  use the concept of Faraday depth
\citep{burn_66,brentjens&d_05},
\begin{equation}
\label{eq_FD}
{\rm FD}(d) = (0.81~{\rm rad~m^{-2}}) \ \int_0^d n_{\rm e} \ B_\parallel \ ds \ ,
\end{equation}
a truly physical quantity, which has the same formal expression as RM
but can be defined at any point of the ISM, independent of any background source.
FD simply corresponds to the line-of-sight depth of the considered point, $d$,
measured in terms of Faraday rotation.
Here, we are only interested in the FD of the Galaxy,
in which case $d$ represents the distance from the observer
to the outer surface of the Galaxy along the considered line of sight.

As an observational reference for our modeling work, we adopt the all-sky map
of the Galactic FD reconstructed by \cite{oppermann&jge_15},
denoting the observational Galactic FD by ${\rm FD}_{\rm obs}$
(see Fig.~\ref{figure_FD}).
To create their map, \citeauthor{oppermann&jge_15} compiled the existing catalogs 
of extragalactic RMs, for a total of 41\,632 data points
(37\,543 of which are from the NVSS catalog of \cite{taylor&ss_09}, 
which covers fairly homogeneously the sky at declination $\delta \ge -40^\circ$),
and they employed a sophisticated signal reconstruction algorithm
that takes spatial correlations into account.

The all-sky map of ${\rm FD}_{\rm obs}$ in Fig.~\ref{figure_FD}, 
like previous all-sky RM maps, shows some coherent structure on large scales.
This large-scale structure in the RM sky has often been assumed to reflect,
at least to some extent, the large-scale organization
of the Galactic magnetic field
\citep[e.g.,][]{simard&k_80, han&mbb_97, taylor&ss_09}.
In reality, however, nearby small-scale perturbations in the magneto-ionic ISM 
can also leave a large-scale imprint in the RM sky
\citep[see][and references therein]{frick&sss_01, mitra&wkj_03, wolleben&flc_10, 
mao&ghz_10, stil&ts_11, sun&lgc15}. 
The most prominent such perturbation identified to date 
is the nearby magnetized bubble uncovered by \cite{wolleben&flc_10} 
through RM synthesis (a.k.a. Faraday tomography) on polarization data
from the Global Magneto-Ionic Medium Survey (GMIMS).
This bubble, estimated to lie at distances around $\sim 150~{\rm pc}$,
is centered at $(\ell \simeq +10^\circ, b \simeq +25^\circ)$
and extends over $\Delta \ell \simeq 70^\circ$ in longitude 
and $\Delta b \simeq 40^\circ$ in latitude,
thereby covering $\simeq 5\%$ of the sky.

Another potential source of strong contamination in the RM sky 
is the North Polar Spur (NPS),
which extends from the Galactic plane at $\ell \approx +20^\circ$
nearly all the way up to the north Galactic pole.
However, \cite{sun&lgc15} showed, through Faraday tomography,
that the actual Faraday thickness ($\Delta$FD) of the NPS 
is most likely close to zero.
They also showed that the $\Delta$FD of the Galactic ISM behind the NPS 
cannot account for the entire Galactic FD
toward the section of the NPS around $b \simeq 30^\circ$.
From this they concluded that if the NPS is local 
(such that the $\Delta$FD of the ISM in front of the NPS is very small), 
the Galactic FD must be dominated by the $\Delta$FD of
\citeauthor{wolleben&flc_10}'s (\citeyear{wolleben&flc_10}) magnetized bubble,
which must then be larger than estimated by \citeauthor{wolleben&flc_10} 
Clearly, the argument can be taken the other way around:
if the $\Delta$FD of the magnetized bubble is as estimated by 
\citeauthor{wolleben&flc_10} (an assumption we will make here), 
the Galactic FD must have a large contribution from the ISM in front of the NPS, 
which implies that the section of the NPS around $b \simeq 30^\circ$ is not local.
This last conclusion is consistent with the results of several recent studies,
which systematically place the lower part of the NPS beyond a few 100~pc 
-- more specifically:
beyond the polarization horizon at 2.3~GHz, $\sim (2-3)~{\rm kpc}$, 
for $b \lesssim 4^\circ$ \citep[from Faraday depolarization;][]{sun&gcp_14}, 
behind the Aquila Rift, at a distance $\gtrsim 1~{\rm kpc}$, 
for $b \lesssim (15^\circ-20^\circ)$ \citep[from X-ray absorption;][]{sofue_15},
and beyond the cloud complex distributed between $300~{\rm pc}$ and $\sim 600~{\rm pc}$ 
and probably much farther away \citep[again from X-ray absorption;][]{lallement&skd_16}.

Since the focus of our paper is on the large-scale magnetic field,
we will subtract from the ${\rm FD}_{\rm obs}$ map of Fig.~\ref{figure_FD}
the estimated contribution from \citeauthor{wolleben&flc_10}'s 
(\citeyear{wolleben&flc_10}) magnetized bubble.\footnote{
The $\Delta$FD values needed to subtract the bubble's contribution to ${\rm FD}_{\rm obs}$
were kindly provided to us by Maik Wolleben.
}
We will then proceed on the assumption that the resulting ${\rm FD}_{\rm obs}$ map
can be relied on to constrain the large-scale magnetic field structure.
As it turns out, the removal of \citeauthor{wolleben&flc_10}'s bubble 
will systematically lead to an improvement in the quality of the fits.

\begin{figure}[t]
\centering
\includegraphics[trim=1.9cm 2.2cm 1.4cm 2.2cm, clip, width=0.95\linewidth]{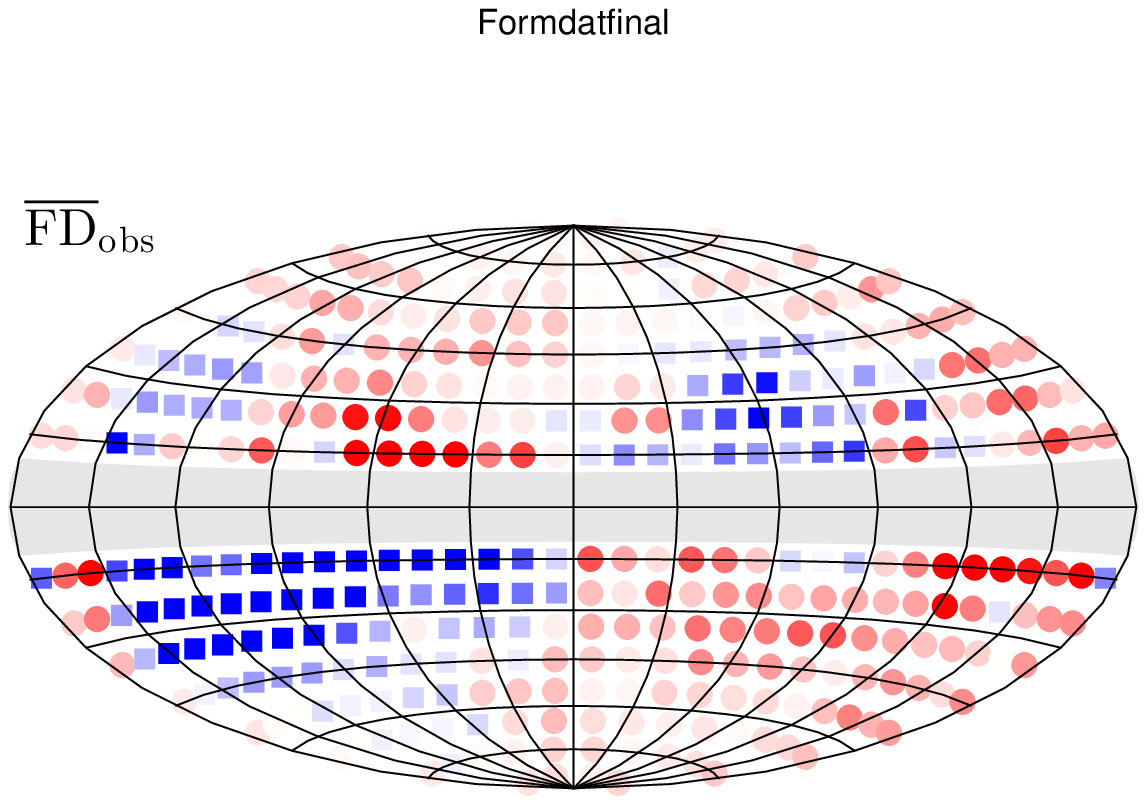}
\includegraphics[trim=1.9cm 2.2cm 1.4cm 2.2cm, clip, width=0.95\linewidth]{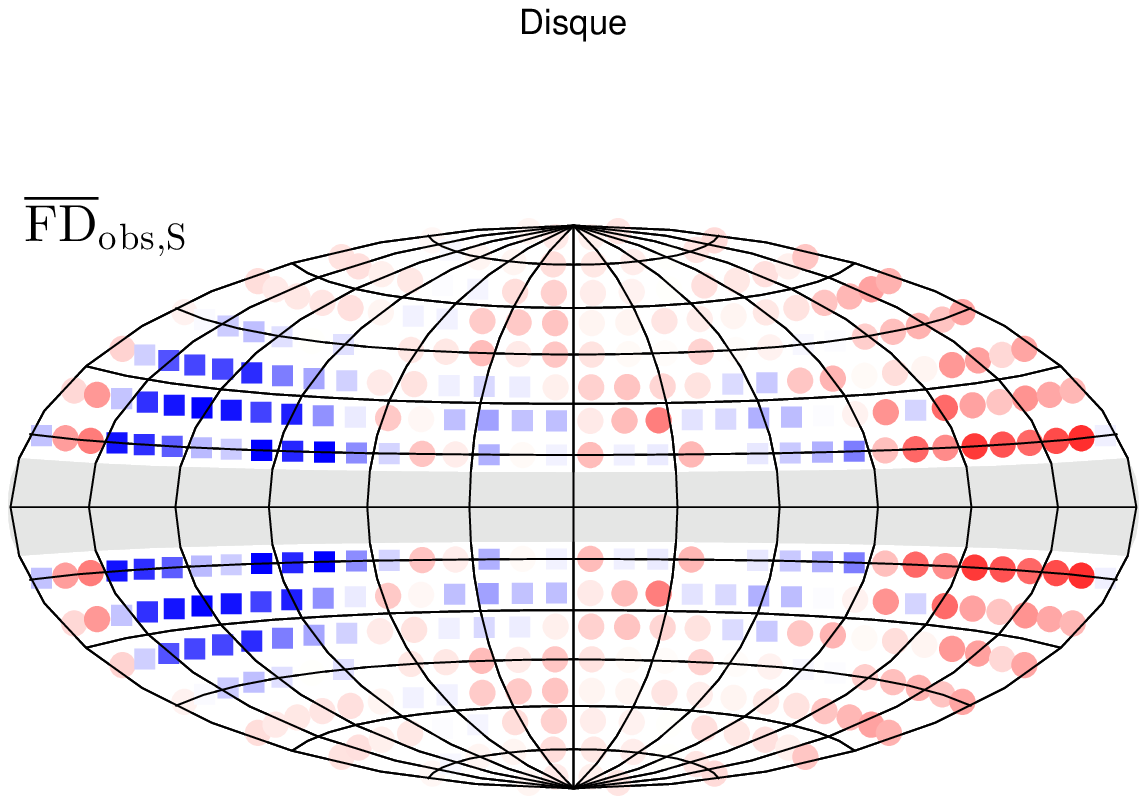}
\includegraphics[trim=1.9cm 2.2cm 1.4cm 2.2cm, clip, width=0.95\linewidth]{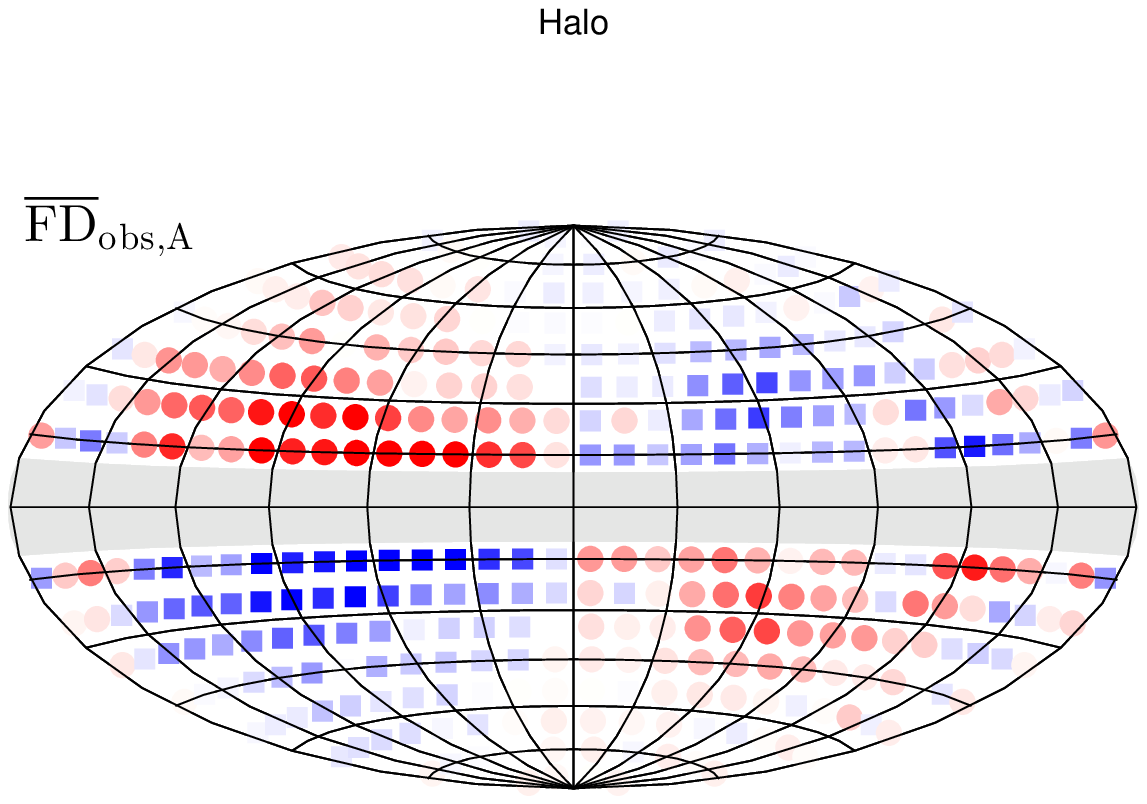}
\caption{All-sky maps, in Aitoff projection, showing the disposition 
of the 356 bins with latitude $|b| \ge 10^\circ$,
which are retained for the fitting procedure,
together with their average observational Galactic Faraday depth,
$\overline{\rm FD}_{\rm obs}$, after subtraction of the contribution from 
\citeauthor{wolleben&flc_10}'s (\citeyear{wolleben&flc_10}) magnetized bubble
(top panel).
Also shown is the decomposition of $\overline{\rm FD}_{\rm obs}$
into a symmetric part, $\overline{\rm FD}_{\rm obs,S}$ (middle panel),
and an antisymmetric part, $\overline{\rm FD}_{\rm obs,A}$ (bottom panel).
The coordinate system and the color code are the same 
as in Figs.~\ref{figure_FD} and \ref{figure_FD_bin}.
The grey band along the Galactic plane masks out the 72 bins with $|b| < 10^\circ$,
which are excluded from the fitting.
}
\label{figure_FD_ref}
\end{figure}

For convenience, we bin the ${\rm FD}_{\rm obs}$ data 
and average them within the different bins, 
both before (Fig.~\ref{figure_FD_bin}) and after (Fig.~\ref{figure_FD_ref}) 
removal of \citeauthor{wolleben&flc_10}'s (\citeyear{wolleben&flc_10}) bubble.
Following a binning procedure similar to that proposed by \cite{pshirkov&tkn_11},
we divide the sky area into 18 longitudinal bands with latitudinal width
$\Delta b = 10^\circ$,
and we divide every longitudinal band into a number of bins chosen 
such that each of the two lowest-latitude bands contains 36 bins 
with longitudinal width $\Delta \ell = 10^\circ$ and all the bins 
have roughly the same area $\simeq (10^\circ)^2$.
This leads to a total of 428 bins. 
For each of these bins, we compute the average ${\rm FD}_{\rm obs}$
before removal of \citeauthor{wolleben&flc_10}'s bubble,
and we plot it in Fig.~\ref{figure_FD_bin}, with a red circle or a blue square 
according to whether it is positive or negative.
In both cases, the color intensity increases with the absolute value 
of the average ${\rm FD}_{\rm obs}$.

We then repeat the binning and averaging steps after removal 
of \citeauthor{wolleben&flc_10}'s bubble,
and for visual purposes, we wash out the obviously anomalous bin 
at $(\ell \simeq 0^\circ - 10^\circ, b = 20^\circ - 30^\circ)$.
The underlying RM anomaly can be identified as the H\,{\sc ii} region 
Sh~2-27 around $\zeta$ Oph \citep{harveysmith&mg_11}.
Our exact treatment of the anomalous bin is of little importance, 
because its weight in the fitting procedure will be drastically reduced 
through an artificial tenfold increase of its estimated uncertainty, $\sigma_i$, 
in the expression of $\chi^2$ (Eq.~\ref{eq_chi}).
In practice, we just blend the anomalous bin into the background 
by replacing its average ${\rm FD}_{\rm obs}$ with that of the super-bin 
enclosing its 8 direct neighbors.
The bin-averaged ${\rm FD}_{\rm obs}$ after removal 
of \citeauthor{wolleben&flc_10}'s bubble and blending of the anomalous bin 
is denoted by $\overline{\rm FD}_{\rm obs}$.

Finally, since we are primarily interested in the magnetic field 
in the Galactic halo, we exclude the 72 bins with $|b| < 10^\circ$, 
which leaves us with a total of 356 bins.
The $\overline{\rm FD}_{\rm obs}$ of each of these bins is plotted 
in the top panel of Fig.~\ref{figure_FD_ref},
again with a red circle if $\overline{\rm FD}_{\rm obs} > 0$
and a blue square if $\overline{\rm FD}_{\rm obs} < 0$.
The $\overline{\rm FD}_{\rm obs}$ map thus obtained 
provides the observational reference against which we will test 
our large-scale magnetic field models in Sect.~\ref{method}.

\subsection{\label{FD_data_trends}General trends}

A few trends emerge from the observational map of the average ${\rm FD}_{\rm obs}$
in Fig.~\ref{figure_FD_bin}:
\begin{itemize}
\item[1.] a rough antisymmetry with respect to 
the prime ($\ell = 0^\circ$) meridian,
\item[2.] a rough antisymmetry with respect to the midplane
in the inner Galactic quadrants away from the plane 
($|\ell| < 90^\circ$ and $|b| \gtrsim 10^\circ$),
\item[3.] a rough symmetry with respect to the midplane
in the inner quadrants close to the plane 
($|\ell| < 90^\circ$ and $|b| \lesssim 10^\circ$)
and in the outer quadrants ($|\ell| > 90^\circ$).
\end{itemize}
These three trends appear to persist 
after removal of \citeauthor{wolleben&flc_10}'s (\citeyear{wolleben&flc_10}) bubble 
(see top panel of Fig.~\ref{figure_FD_ref}, for $|b| \ge 10^\circ$).
We will naturally try to reproduce them with our magnetic field models,
but before getting to the models, a few comments are in order.

First, the rough symmetry with respect to the midplane 
at $|b|\lesssim 10^\circ$ and all $\ell$
suggests that the disk magnetic field is symmetric (or quadrupolar),\footnote{
A magnetic field is symmetric/antisymmetric with respect to the midplane 
(or quadrupolar/dipolar) when its horizontal components, $B_r$ and $B_\varphi$, 
are even/odd functions of $z$ and its vertical component, $B_z$, is an odd/even
function of $z$.
\label{foot_parity}
}
as already pointed out by many authors \citep[e.g.,][]{rand&l_94, frick&sss_01}.

Second, the rough antisymmetry [symmetry] with respect to the midplane 
at $|b| \gtrsim 10^\circ$ and $|\ell| < 90^\circ$ [$|\ell| > 90^\circ$]
suggests one of the two following possibilities:
either the halo magnetic field is antisymmetric in the inner Galaxy
and symmetric in the outer Galaxy,
or the halo magnetic field is everywhere antisymmetric,
but only toward the inner Galaxy does its contribution 
to the Galactic FD at $|b| \gtrsim 10^\circ$
exceed the contribution from the disk magnetic field.
The first possibility is probably not very realistic 
(although it cannot be completely ruled out):
while the disk and halo fields could possibly have different vertical parities
\citep[see, e.g.,][]{moss&s_08, moss&sbk_10},
it seems likely that each field has by now evolved toward a single parity.
The second possibility may sound a little counter-intuitive at first,
but one has to remember that the halo field contribution to the Galactic FD 
is weighted by a lower free-electron density than the disk field contribution 
(especially toward the outer Galaxy); moreover, the halo field can very well 
be confined inside a smaller radius than the disk field.
Such is the case in the double-torus picture originally sketched by \cite{han_02}
and later modeled by \cite{prouza&s_03, sun&rwe_08, jansson&f_12a}:
in these three models, the toroidal field of the halo falls off 
exponentially with $r$ at large radii, i.e., much faster than the disk field, 
which falls off approximately as $(1/r)$.

Third, the rough antisymmetry with respect to the prime meridian
(east-west antisymmetry) can a priori be explained 
by an axisymmetric, predominantly azimuthal magnetic field.
However, the situation is a little more subtle.

\noindent
$\bullet$ For the disk, RM studies (mainly of Galactic pulsars,
and also of extragalactic point sources) converge to show
that the magnetic field is indeed predominantly azimuthal, 
though not with the same sign everywhere throughout the disk, 
i.e., the field must reverse direction with Galactic radius
\citep[e.g.,][]{rand&l_94, han&mq_99, han&mlq_06, brown&hgt_07}.
The exact number and the radial locations of these reversals 
are still a matter of debate, which we do not whish to enter.
Here, we are content to note that field reversals naturally arise 
in a bisymmetric (azimuthal modulation $\propto \cos m \, \varphi$, 
with $m = 1$) or higher-order ($m > 1$) configuration,
but can also be produced in the axisymmetric case \citep{ferriere&s_00}.

\noindent
$\bullet$ For the halo, most existing models have settled on 
a purely azimuthal, and hence axisymmetric, magnetic field.
Here, we inquire into the other possibilities.

\noindent
Let us first ask whether a purely poloidal (possibly, but not necessarily, 
X-shape) magnetic field would be a viable alternative.
Clearly, a purely poloidal field that is axisymmetric 
automatically leads to an FD pattern with east-west symmetry, 
inconsistent with the observed FD distribution.
In contrast, a purely poloidal field that is bisymmetric 
can lead to a variety of FD patterns, 
depending on the azimuthal angle of maximum amplitude:
if the maximum amplitude occurs in the azimuthal plane parallel 
to the plane of the sky ($\varphi = \pm 90^\circ$) 
[in the azimuthal plane through the Sun ($\varphi = 0^\circ$)],
the FD pattern has east-west antisymmetry [symmetry],
in agreement [disagreement] with the observed FD distribution.
Hence, a purely poloidal magnetic field can possibly match the FD data,
provided it is bisymmetric (or higher-order) and favorably oriented.

\noindent
Let us now turn to the more general and more realistic situation 
where the magnetic field has both azimuthal and poloidal components, 
as required by dynamo theory.
If this general field is axisymmetric, the observed east-west antisymmetry in FD 
implies that it must be mainly azimuthal.
In contrast, if the general field is bisymmetric, it could in principle cover 
a broad range of orientation, from mostly azimuthal to mostly poloidal.
What happens here is that, in the presence of an azimuthal field component,
the azimuthal modulation, which is presumably carried along field lines, 
crosses azimuthal planes.
The resulting FD pattern becomes more difficult to predict:
it may become more complex and fluctuating than in the axisymmetric case, 
and large-scale longitudinal trends may be partly washed out,
but a priori the predominance of neither the azimuthal nor the poloidal 
field component can be ruled out.

\noindent
To sum up, the halo magnetic field could either be axisymmetric and mainly azimuthal 
or bisymmetric (or higher-order) with no clear constraint 
on its azimuthal-versus-poloidal status.

\section{\label{models}Magnetic field models}

\subsection{\label{models_halo}Magnetic fields in galactic halos}

\begin{figure*}[t]
\begin{tabularx}{\linewidth}{XcX}
\includegraphics[trim=1.5cm 0.5cm 1cm 0cm, clip, width=\linewidth]{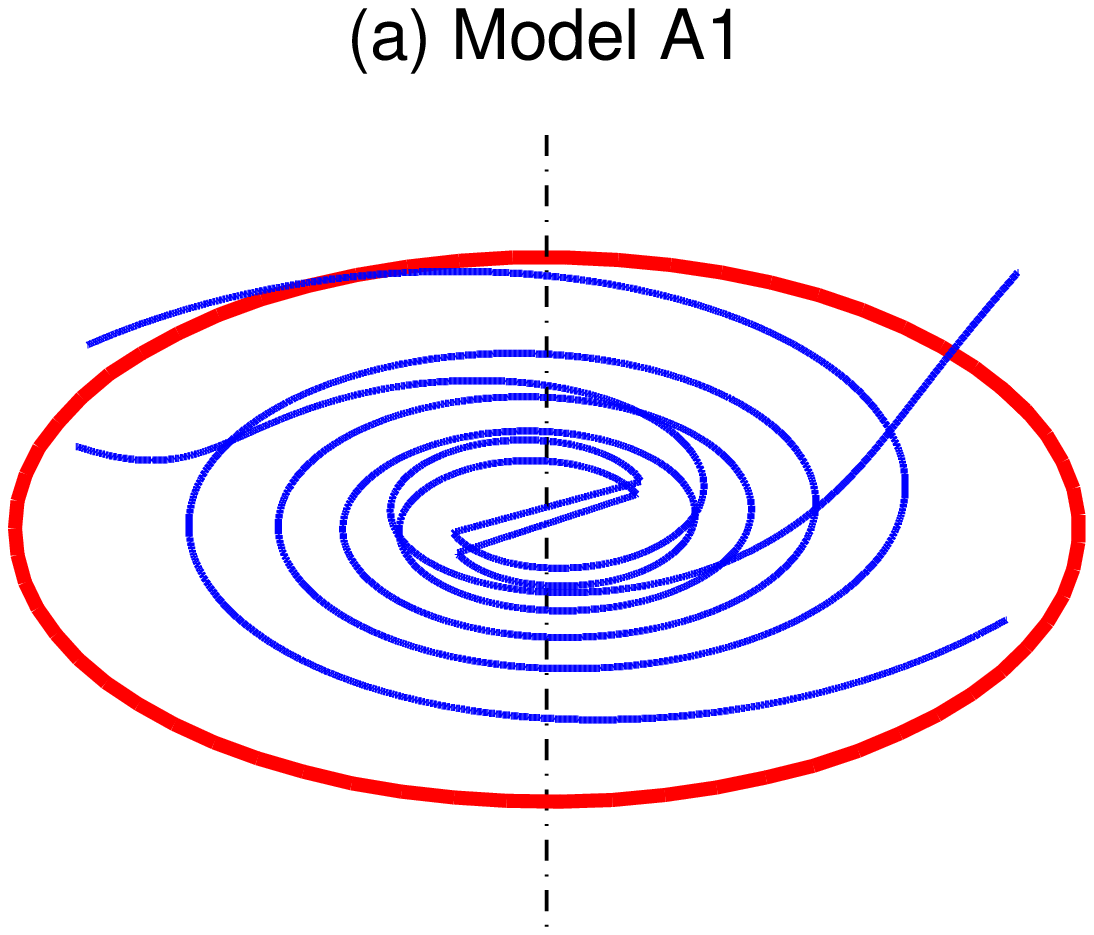}
& &
\includegraphics[trim=1.5cm 0.5cm 1cm 0cm, clip, width=\linewidth]{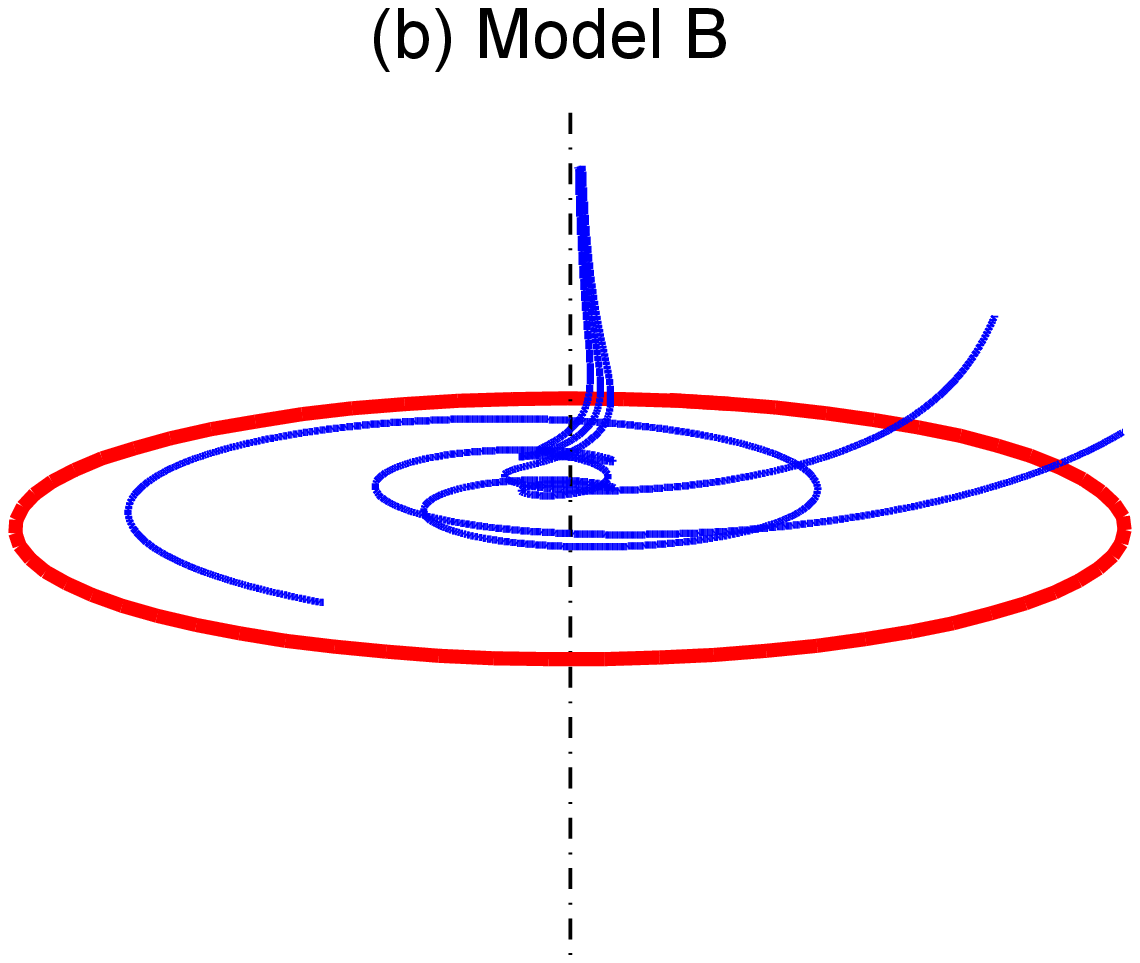}
\bigskip \\
\includegraphics[trim=1.5cm 0.5cm 1cm 0cm, clip, width=\linewidth]{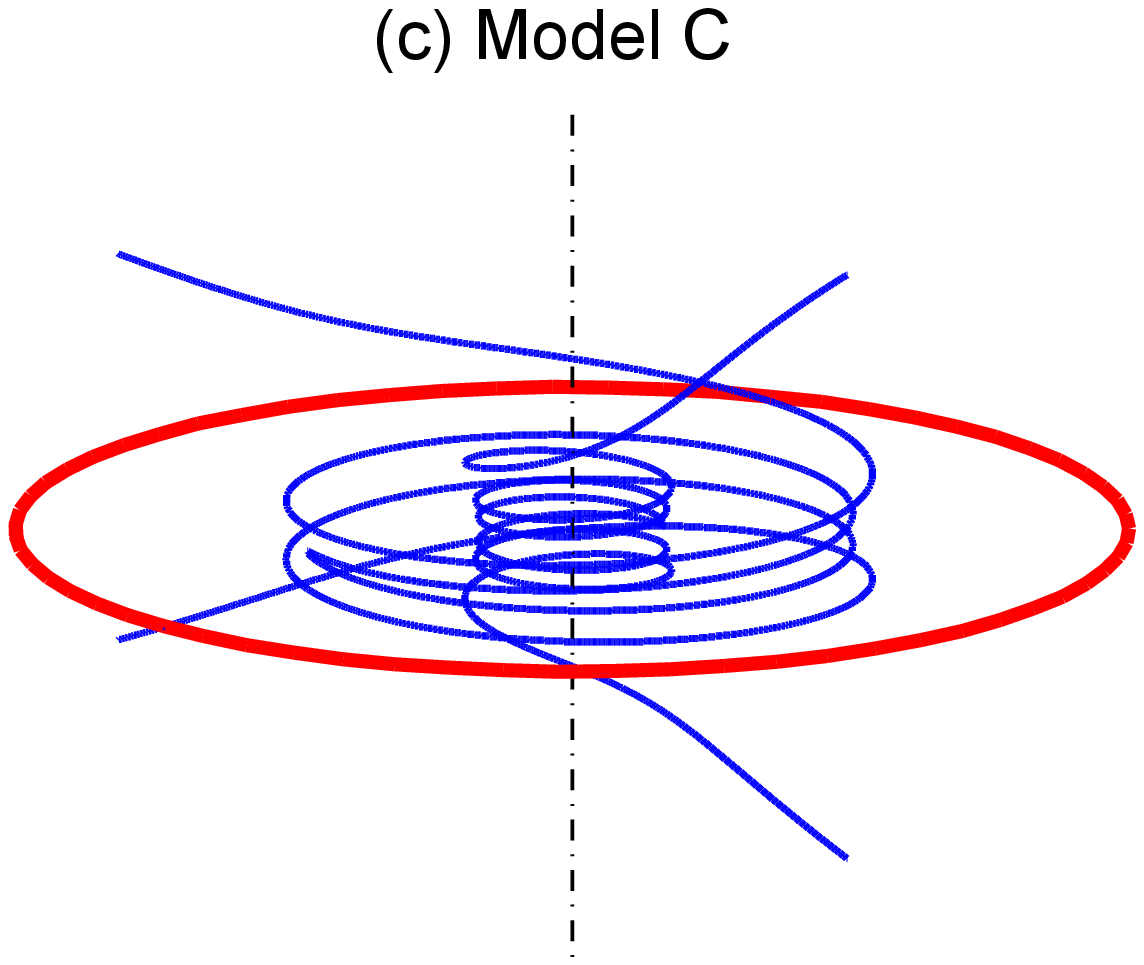}
& &
\includegraphics[trim=1.5cm 0.5cm 1cm 0cm, clip, width=\linewidth]{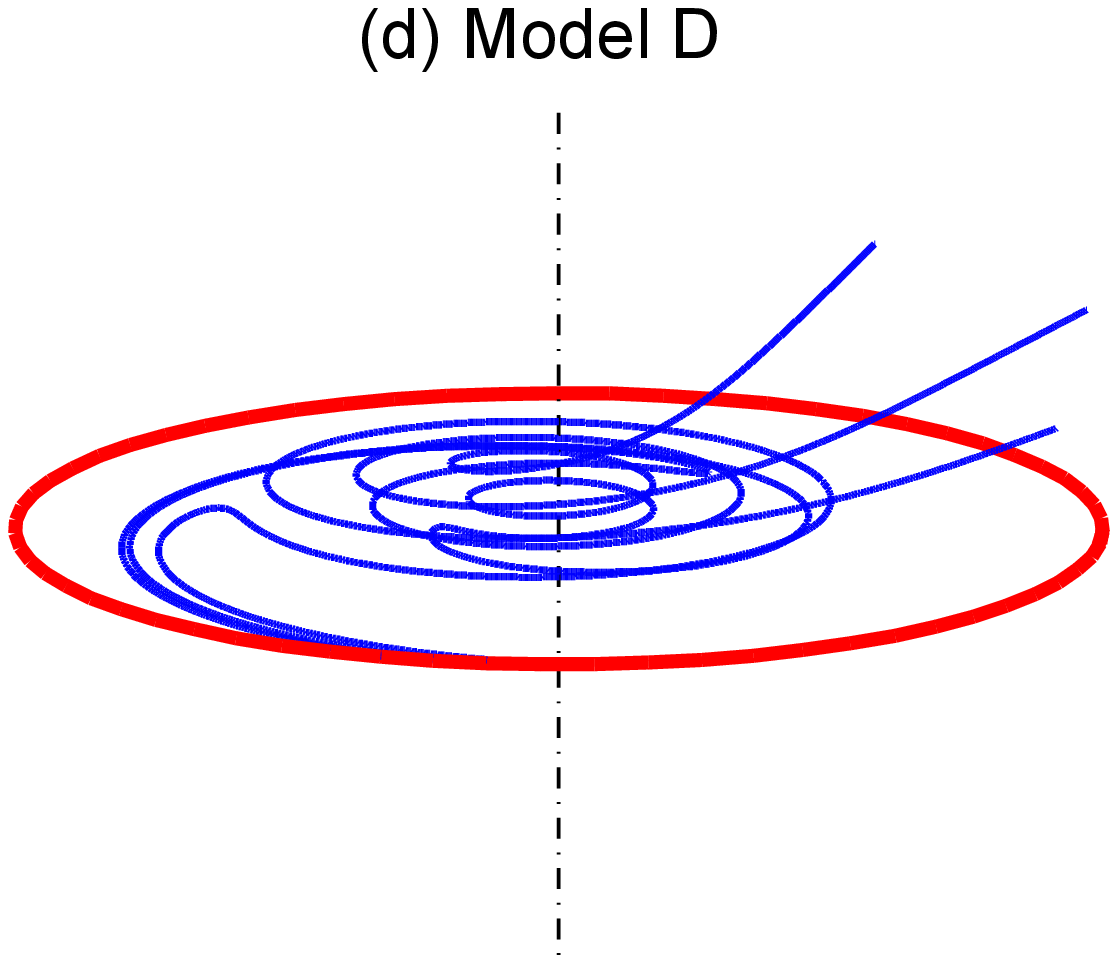}
\end{tabularx}
\caption{Small set of field lines for each of our four models of spiraling, 
possibly X-shape magnetic fields in galactic halos, as seen from an oblique angle.
The shapes of field lines for the three halo-field models A1, B, and D
are also representative of our three disk-field models Ad1, Bd, and Dd.
All the plotted field lines lie on the same winding surface 
(Eq.~(\ref{eq_mfl_winding_surf}) with $\varphi_\infty = 0^\circ$ 
and $g_\varphi (r,z)$ given by Eq.~(\ref{eq_shiftedwindingfc})
with $p_0 = -9^\circ$, $H_p = 1.5~{\rm kpc}$, and $L_p = 45~{\rm kpc}$),
and their footpoints, $(r_1,\varphi_1,z_1)$, on the relevant reference surface 
($r = r_1$ in models~A1 and B, and $z = z_1$ in models~C and D) are given by:
(a) $(3~{\rm kpc},172^\circ,0.5~{\rm kpc})$ 
and $(3~{\rm kpc},357^\circ,2~{\rm kpc})$ 
in model~A1 (with $a = 0.05~{\rm kpc}^{-2}$);
(b) $(3~{\rm kpc},326^\circ,1~{\rm kpc})$, 
$(3~{\rm kpc},357^\circ,2~{\rm kpc})$, 
and $(3~{\rm kpc},198^\circ,3~{\rm kpc})$ 
in model~B (with $n = 2$);
(c) $(2.5~{\rm kpc},336^\circ,0)$ 
and $(7.5~{\rm kpc},318^\circ,0)$ 
in model~C (with $a = 0.01~{\rm kpc}^{-2}$);
and (d) $(2.5~{\rm kpc},168^\circ,1.5~{\rm kpc})$, 
$(5~{\rm kpc},47^\circ,1.5~{\rm kpc})$, 
and $(7.5~{\rm kpc},339^\circ,1.5~{\rm kpc})$ 
in model~D (with $n = 0.5$).
The galactic plane is represented by the red, solid circle of radius $15~{\rm kpc}$,
and the rotation axis by the vertical, black, dot-dashed line.
}
\label{figure_fieldlines}
\end{figure*}

In Paper~1, we derived four different models of spiraling, possibly 
X-shape magnetic fields in galactic halos, which can be applied to the halo 
of our own Galaxy.
Each of these models was initially defined by the shape of its field lines
together with the distribution of the magnetic flux density
on a given reference surface.
Using the Euler formalism \citep[e.g.,][]{northrop_63, stern_66},
we were able to work out the corresponding analytical expression 
of the magnetic field vector, ${\boldvec B} = (B_r,B_\varphi,B_z)$,
as a function of Galactocentric cylindrical coordinates, $(r,\varphi,z)$.
The main characteristics of the four halo-field models are summarized below,
and a few representative field lines are plotted in Fig.~\ref{figure_fieldlines}. 

In models~A and B, we introduced a fixed reference radius, $r_1$,
and we labeled field lines by the height, $z_1$, 
and the azimuthal angle, $\varphi_1$,
at which they cross the centered vertical cylinder of radius $r_1$.
In models~C and D, we introduced a fixed reference height, $z_1$
(more exactly, one reference height, $z_1 = 0$, in model~C
where all field lines cross the galactic midplane,
and two reference heights, $z_1 = \pm |z_1|$, in model~D
where field lines do not cross the midplane),
and we labeled field lines by the radius, $r_1$, 
and the azimuthal angle, $\varphi_1$,
at which they cross the horizontal plane 
(or one of the two horizontal planes) of height $z_1$.
Thus, in all models, the point $(r_1,\varphi_1,z_1)$ of a given field line
can be regarded as its footpoint on the reference surface
(vertical cylinder of radius $r_1$ in models~A and B
and horizontal plane(s) of height $z_1$ in models~C and D).

\subsubsection{\label{models_poloidal}Poloidal field}

The four models are distinguished by the shape of field lines associated
with the poloidal field (hereafter referred to as the poloidal field lines),
and hence by the expressions of the radial and vertical field components.

In {\bf model~A},
the shape of poloidal field lines is described by the quadratic function
\begin{equation}
\label{eq_mfl_A}
z = z_1 \ \frac{1 + a \, r^2}{1 + a \, r_1^2} \ ,
\end{equation}
where $a$ is a strictly positive free parameter governing the opening of field lines
away from the $z$-axis,
$r_1$ is the prescribed reference radius,
and $z_1$ is the vertical label of the considered field line.
Conversely, the vertical label of the field line passing 
through $(r,\varphi,z)$ is given by
\begin{equation}
\label{eq_z1_A}
z_1 = z \ \frac{1 + a \, r_1^2}{1 + a \, r^2} \ \cdot
\end{equation}
It then follows (see Paper~1 for the detailed derivation)
that the poloidal field components can be written as
\begin{eqnarray}
B_r & = & \frac{r_1}{r} \ \frac{z_1}{z} \ B_r(r_1,\varphi_1,z_1)
\label{eq_Br_A} \\
B_z & = & \frac{2 \, a \, r_1 \, z_1}{1 + a \, r^2} \ B_r(r_1,\varphi_1,z_1) \ ,
\label{eq_Bz_A}
\end{eqnarray}
with, for instance, $B_r(r_1,\varphi_1,z_1)$ obeying Eq.~(\ref{eq_B1_AB}).

In {\bf model~B}, the corresponding equations read
\begin{equation}
\label{eq_mfl_B}
z = \frac{1}{n+1} \ z_1 \ 
\left[ \left( \frac{r}{r_1} \right)^{-n} + n \, \frac{r}{r_1} \right] \ ,
\end{equation}
with $n$ a power-law index satisfying the constraint $n \ge 1$,
\begin{equation}
\label{eq_z1_B}
z_1 = (n+1) \ z \ 
\left[ \left( \frac{r}{r_1} \right)^{-n} + n \, \frac{r}{r_1} \right]^{-1} \ ,
\end{equation}
and
\begin{eqnarray}
B_r & = & \frac{r_1}{r} \ \frac{z_1}{z} \ B_r(r_1,\varphi_1,z_1)
\label{eq_Br_B} \\
B_z & = & - \frac{n}{n+1} \ \frac{r_1 \, z_1^2}{r^2 \, z} \
\left[ \left( \frac{r}{r_1} \right)^{-n} - \frac{r}{r_1} \right] \
B_r(r_1,\varphi_1,z_1) \ \cdot
\label{eq_Bz_B}
\end{eqnarray}

In both models~A and B, the radial field component on the vertical cylinder 
of radius $r_1$ is chosen to have a linear-exponential variation with $z_1$
and a sinusoidal variation with $\varphi_1$:
\begin{eqnarray}
B_r(r_1,\varphi_1,z_1) & = &
B_1 \ f_{\rm sym} \ 
\left[ \frac{|z_1|}{H} \ \exp \left( - \frac{|z_1| - H}{H} \right) \right]
\nonumber \\
& & \times \ \cos \Big( m \, \big( \varphi_1 - g_\varphi (r_1,z_1) 
- \varphi_\star \big) \Big) \ ,
\label{eq_B1_AB}
\end{eqnarray}
where $B_1$ is the normalization field strength, 
$f_{\rm sym}$ is a factor setting the vertical parity of the magnetic field
($f_{\rm sym} = 1$ for symmetric fields 
and $f_{\rm sym} = {\rm sign}~z_1$ for antisymmetric fields;
see footnote~\ref{foot_parity}),
$H$ is the exponential scale height,
$m$ is the azimuthal wavenumber,
$g_\varphi$ is the shifted winding function defined by Eq.~(\ref{eq_shiftedwindingfc}),
and $\varphi_\star$ is the orientation angle of the azimuthal pattern.
The term $g_\varphi (r_1,z_1)$, which will be discussed in more detail 
in Sect.~\ref{models_azimuthal}, ensures that the phase of the sinusoidal modulation 
remains constant on the winding surfaces defined by Eq.~(\ref{eq_mfl_winding_surf}) 
(see comment following Eq.~(\ref{eq_mfl_winding_surf})).

Models~C and D are the direct counterparts of models~A and B, respectively,
with the roles of the coordinates $r$ and $z$ inverted in the equations of field lines.

In {\bf model~C}, the reference height is set to $z_1 = 0$,
and the shape of poloidal field lines is described by the quadratic function
\begin{equation}
\label{eq_mfl_C}
r = r_1 \ (1 + a \, z^2) \ ,
\end{equation}
with $a$ a strictly positive free parameter governing the opening of field lines
away from the $r$-axis 
and $r_1$ the radial label of the considered field line.
Conversely, the radial label of the field line passing
through $(r,\varphi,z)$ is given by
\begin{equation}
\label{eq_r1_C}
r_1 = \frac{r}{1 + a \, z^2} \ \cdot
\end{equation}
The poloidal field components can then be written as
\begin{eqnarray}
B_r & = & \frac{2 \, a \, r_1^3 \, z}{r^2} \ B_z(r_1,\varphi_1,z_1)
\label{eq_Br_C} \\
B_z & = & \frac{r_1^2}{r^2} \ B_z(r_1,\varphi_1,z_1) \ ,
\label{eq_Bz_C}
\end{eqnarray}
with, for instance, $B_z(r_1,\varphi_1,z_1)$ obeying Eq.~(\ref{eq_B1_CD}).

In {\bf model~D}, where every field line remains confined to one side 
of the galactic midplane, a reference height, $z_1 = |z_1| \ {\rm sign}~z$,
is prescribed on each side of the midplane, such that the ratio $(z/z_1)$
is always positive.
We then have
\begin{equation}
\label{eq_mfl_D}
r = \frac{1}{n+1} \ r_1 \ 
\left[ \left( \frac{z}{z_1} \right)^{-n} + n \, \frac{z}{z_1} \right] \ ,
\end{equation}
with $n \ge 0.5$,
\begin{equation}
\label{eq_r1_D}
r_1 = (n+1) \ r \ 
\left[ \left( \frac{z}{z_1} \right)^{-n} + n \, \frac{z}{z_1} \right]^{-1} \ ,
\end{equation}
and
\begin{eqnarray}
B_r & = & - \frac{n}{n+1} \ \frac{r_1^3}{r^2 \, z} \
\left[ \left( \frac{z}{z_1} \right)^{-n} - \frac{z}{z_1} \right] \
B_z(r_1,\varphi_1,z_1)
\label{eq_Br_D} \\
B_z & = & \frac{r_1^2}{r^2} \ B_z(r_1,\varphi_1,z_1) \ \cdot
\label{eq_Bz_D}
\end{eqnarray}

In both models~C and D, the vertical field component on the horizontal plane(s)
of height $z_1$ is chosen to have an exponential variation with $r_1$
and a sinusoidal variation with $\varphi_1$:
\begin{equation}
\label{eq_B1_CD}
B_z(r_1,\varphi_1,z_1) =
B_1 \ \bar{f}_{\rm sym} \ \exp \left( - \frac{r_1}{L} \right) \
\cos \Big( m \, \big( \varphi_1 - g_\varphi (r_1,z_1) 
- \varphi_\star \big) \Big) \ ,
\end{equation}
with $B_1$ the normalization field strength,
$\bar{f}_{\rm sym}$ a factor setting the vertical parity of the magnetic field
($\bar{f}_{\rm sym} = 1$ in model~C, which is always antisymmetric,
and in the antisymmetric version of model~D,
and $\bar{f}_{\rm sym} = {\rm sign}~z_1$ in the symmetric version of model~D; 
see footnote~\ref{foot_parity}),
$L$ the exponential scale length,
$m$ the azimuthal wavenumber,
$g_\varphi$ the shifted winding function (Eq.~\ref{eq_shiftedwindingfc}),
and $\varphi_\star$ the orientation angle of the azimuthal pattern.

\subsubsection{\label{models_azimuthal}Azimuthal field}

In all four models, field lines are assumed to spiral up or down
according to the equation
\begin{equation}
\label{eq_mfl_spiral}
\varphi = \varphi_1 + f_\varphi (r,z) \ ,
\end{equation}
where $f_\varphi (r,z)$ is a winding function starting from the field line's
footpoint on the reference surface and, therefore, satisfying 
$f_\varphi (r_1,z_1) = 0$.
If we consider that field lines are somehow anchored in the external
intergalactic medium,\footnote{
We emphasize that this is only one possibility.
Another possibility would be that galactic field lines loop back on themselves,
without connecting to an extragalactic magnetic field.
In Paper~1, we took a different approach,
based on the magnetic pitch angle rather than the winding function itself,
which did nor require discussing the anchoring of field lines.
}
it proves more convenient to shift the starting point of the winding function 
to infinity. 
This can be done by letting
\begin{equation}
\label{eq_windingfc}
f_\varphi (r,z) = g_\varphi (r,z) - g_\varphi (r_1,z_1) \ ,
\end{equation}
where $g_\varphi (r,z)$ is a shifted winding function starting from 
the field line's anchor point at infinity\footnote{
In models~B and D, field lines actually have two anchor points at infinity. 
By construction, both anchor points have the same azimuthal angle, $\varphi_\infty$.
\label{foot_footpoint}
}
and, therefore, satisfying
$g_\varphi (r,z) \to 0$ for $r \to \infty$ and for $|z| \to \infty$.
A reasonably simple choice for the shifted winding function is
\begin{equation}
\label{eq_shiftedwindingfc}
g_\varphi (r,z) = 
\cot p_0 \
\frac{\displaystyle \ln \left[ 1 - \exp \left(-\frac{r}{L_p} \right) \right]}
{\displaystyle 1+ \left( \frac{|z|}{H_p} \right)^2} \ ,
\end{equation}
with $p_0$ the pitch angle (i.e., the angle between the horizontal projection 
of a field line and the local azimuthal direction) at the origin ($(r,z) \to 0$), 
$H_p$ the scale height, and $L_p$ the exponential scale length. 
With this choice, $g_\varphi (r,z)$ defines, in horizontal planes ($z = {\rm c^{st}}$),
spirals that are logarithmic (constant pitch angle) at small $r$
and become increasingly loose (increasing pitch angle) at large $r$.
The spirals also loosen up with increasing $|z|$.

Combining Eqs.~(\ref{eq_mfl_spiral}) and (\ref{eq_windingfc}), 
we can rewrite the equation describing the spiraling of field lines 
in the form
\begin{equation}
\label{eq_mfl_spiral_g}
\varphi = g_\varphi (r,z) + \left[ \varphi_1 - g_\varphi (r_1,z_1) \right] \ ,
\end{equation}
with $\left[ \varphi_1 - g_\varphi (r_1,z_1) \right]$ constant along field lines.
Conversely, the field line passing through $(r,\varphi,z)$ can be traced back 
to the azimuthal label
\begin{equation} 
\label{eq_phi1}
\varphi_1 = \varphi - g_\varphi (r,z) + g_\varphi (r_1,z_1) \ ,
\end{equation}
where $r_1$ and $z_1$ either have prescribed values or are given functions 
of $(r,z)$ (see Eqs.~(\ref{eq_z1_A}), (\ref{eq_z1_B}), (\ref{eq_r1_C}), 
and (\ref{eq_r1_D}) in Sect.~\ref{models_poloidal}).
It then follows from Eqs.~(\ref{eq_B1_AB}) and (\ref{eq_B1_CD}) 
that the sinusoidal modulation of the magnetic field goes as 
$\cos \Big( m \, \big( \varphi - g_\varphi (r,z) - \varphi_\star \big) \Big)$.

If we now consider the anchor point 
at infinity$^{\mbox{\scriptsize \ref{foot_footpoint}}}$
of the field line passing through $(r,\varphi,z)$, 
denote its azimuthal angle by $\varphi_\infty$, 
and recall that $g_\varphi (r,z) \to 0$ for $r, |z| \to \infty$,
we find that the spiraling of field line can also be described by 
\begin{equation}
\label{eq_mfl_winding_surf}
\varphi = g_\varphi (r,z) + \varphi_\infty \ \cdot
\end{equation}
Eq.~(\ref{eq_mfl_winding_surf}) defines a so-called winding surface,
formed by all the field lines with the same azimuthal angle at infinity, 
$\varphi_\infty$, and thus the same phase in the sinusoidal modulation,
$\varphi_\infty - \varphi_\star$. 
The crest of the modulation occurs at phase $\varphi_\infty - \varphi_\star = 0^\circ$,
i.e., at $\varphi_\infty = \varphi_\star$,
which means that the free parameter $\varphi_\star$ can be interpreted as
the azimuthal angle at infinity of the crest surface.

Since the magnetic field is by definition tangent to field lines,
Eq.~(\ref{eq_mfl_spiral_g}) implies that its azimuthal component 
is related to its radial and vertical components through
\begin{equation}
\label{eq_Bphi_g}
B_\varphi = 
\left( r \ \frac{\partial g_\varphi}{\partial r} \right) \ B_r 
+ \left( r \ \frac{\partial g_\varphi}{\partial z} \right) \ B_z \ \cdot
\end{equation}
Roughly speaking, the two terms on the right-hand side of Eq.~(\ref{eq_Bphi_g}) 
represent the azimuthal field generated through shearing 
of radial and vertical fields by radial and vertical gradients 
in the galactic rotation rate, respectively.
With the choice of Eq.~(\ref{eq_shiftedwindingfc}), Eq.~(\ref{eq_Bphi_g}) becomes 
\begin{eqnarray}
\label{eq_Bphi_p}
B_\varphi & = & 
\cot p_0 \
\frac{1}{\displaystyle 1+ \left( \frac{|z|}{H_p} \right)^2} \
\frac{\displaystyle \frac{r}{L_p} \ \exp \left(-\frac{r}{L_p} \right)}
{\displaystyle 1 - \exp \left(-\frac{r}{L_p} \right)} \ B_r 
\nonumber \\
& - & 
2 \ \cot p_0 \ 
\frac{\displaystyle \frac{z}{H_p^2}}
{\displaystyle \left[ 1+ \left( \frac{|z|}{H_p} \right)^2 \right]^2} \ 
r \ \ln \left[ 1 - \exp \left(-\frac{r}{L_p} \right) \right] \ B_z \ \cdot
\end{eqnarray}
It is easily seen that the factor of $B_r$ has the same sign 
as $\cot p_0$ (usually negative) everywhere, 
while the factor of $B_z$ has the same sign as $\cot p_0$ 
above the midplane ($z>0$) and the opposite sign below the midplane ($z<0$).

\subsubsection{\label{models_total}Total field}

To sum up, the poloidal field is described by 
Eqs.~(\ref{eq_Br_A}) -- (\ref{eq_Bz_A}) in model~A
and Eqs.~(\ref{eq_Br_B}) -- (\ref{eq_Bz_B}) in model~B,
with $B_r(r_1,\varphi_1,z_1)$ given by Eq.~(\ref{eq_B1_AB}),
and by Eqs.~(\ref{eq_Br_C}) -- (\ref{eq_Bz_C}) in model~C
and Eqs.~(\ref{eq_Br_D}) -- (\ref{eq_Bz_D}) in model~D,
with $B_z(r_1,\varphi_1,z_1)$ given by Eq.~(\ref{eq_B1_CD}).
The azimuthal field is described by Eq.~(\ref{eq_Bphi_p}) in all models. 
Together, the above equations link the magnetic field at an arbitrary point
$(r,\varphi,z)$ to the normal field component 
at the footpoint $(r_1,\varphi_1,z_1)$ on the reference surface
of the field line passing through $(r,\varphi,z)$.
The footpoint, in turn, is determined by the reference coordinate 
($r_1$ in models~A and B, and $z_1$ in models~C and D,
with $z_1 = 0$ in model~C and $z_1 = \pm |z_1|$ in model~D),
the poloidal label of the field line
($z_1$ given by Eq.~(\ref{eq_z1_A}) in model~A and Eq.~(\ref{eq_z1_B}) in model~B,
and $r_1$ given by Eq.~(\ref{eq_r1_C}) in model~C and Eq.~(\ref{eq_r1_D}) in model~D),
and its azimuthal label ($\varphi_1$ given by Eq.~(\ref{eq_phi1}) in all models).

\subsubsection{\label{models_modelA}Regularization of model~A}

An important remark should be made regarding model~A.
Eq.~(\ref{eq_Br_A}) together with Eq.~(\ref{eq_z1_A}) 
imply that $B_r \to \infty$ for $r \to 0$,
which reflects the fact that all field lines, from all azimuthal planes,
converge to the $z$-axis.
The problem does not arise in model~B,
as shown by Eq.~(\ref{eq_Br_B}) together with Eq.~(\ref{eq_z1_B}), 
because all field lines are deflected vertically before reaching the $z$-axis
(see Fig.~\ref{figure_fieldlines}b).
The singularity in model~A is inescapable in the axisymmetric case 
($m = 0$ in Eq.~(\ref{eq_B1_AB})), where $B_r$ has the same sign 
all around the $z$-axis, so that, at every height, a non-zero magnetic flux 
reaches the $z$-axis.
However, the singularity can be removed in non-axisymmetric ($m \ge 1$) 
configurations, where $B_r$ changes sign sinusoidally around the $z$-axis, 
so that, at every height, no net magnetic flux actually reaches the $z$-axis.
It is then conceivable to detach field lines from the $z$-axis,
spread them apart inside a centered vertical cylinder, whose radius can be chosen 
to be $r_1$ (remember that $r_1$ is still a free parameter at this stage),
and connect up two by two field lines with the same $z_1$,
opposite $\varphi_1$ (with respect to a node of the sinusoidal modulation 
in Eq.~(\ref{eq_B1_AB})), and hence opposite $B_r$,
such as to ensure conservation of the magnetic flux as well as continutity in its sign.

For instance, in the bisymmetric ($m = 1$) case, 
field lines with the same $z_1$ and opposite $\varphi_1$ 
(with respect to $g_\varphi (r_1,z_1) + \varphi_\star \pm \pi/2$) 
can be connected up inside the vertical cylinder of radius $r_1$ 
through straight-line segments parallel to the direction 
$\varphi_1 = g_\varphi (r_1,z_1) + \varphi_\star$.
In this case, magnetic flux conservation (or continuity of $B_r$) 
across the surface of the cylinder gives for the magnetic field 
at an arbitrary point $(r,\varphi,z)$ inside the cylinder:
\begin{eqnarray}
\label{eq_Br_A_in}
{\boldvec B} & = &
B_r \big( r_1, g_\varphi (r_1,z_1) + \varphi_\star, z_1 \big) \
\hat{e}_\star(z_1) 
\nonumber \\
& = &
B_1 \ f_{\rm sym} \ 
\left[ \frac{|z_1|}{H} \ \exp \left( - \frac{|z_1| - H}{H} \right) \right] \
\hat{e}_\star(z_1) \ , \qquad r \le r_1 \ ,
\end{eqnarray}
with $z_1 = z$, $\hat{e}_\star(z_1)$ the unit vector in the direction 
$\varphi_1 = g_\varphi (r_1,z_1) + \varphi_\star$,
and the other symbols having the same meaning as in Eq.~(\ref{eq_B1_AB}).

In the following, the regularized bisymmetric version of model~A
is referred to as {\bf model~A1}, where the number 1 indicates 
the value of the azimuthal wavenumber.

\subsection{\label{models_disk}Magnetic fields in galactic disks}

We now present three different models of spiraling, mainly horizontal magnetic fields 
in galactic disks, which can be applied to the disk of our own Galaxy.
These models are directly inspired from the three models for magnetic fields 
in galactic halos that have nearly horizontal field lines at low $|z|$, 
namely, models~A1, B, and D in Sect.~\ref{models_halo},
and they are named models~Ad1, Bd, and Dd, respectively.
Models~Bd and Dd are directly taken up from Paper~1, with a minor amendment 
in model~Dd (see below), while model~Ad1 is a new addition.

{\bf Model~Ad1} has nearly the same descriptive equations as model~A1.
Outside the reference radius, $r_1$, the shape of field lines is described by 
Eq.~(\ref{eq_mfl_A}), with the parameter $a$ assuming a much smaller value 
than for halo fields, and by Eq.~(\ref{eq_mfl_spiral_g});
the vertical and azimuthal labels of the field line passing through $(r,\varphi,z)$
are given by Eqs.~(\ref{eq_z1_A}) and (\ref{eq_phi1});
and the three magnetic field components obey Eqs.~(\ref{eq_Br_A}), (\ref{eq_Bz_A}),
and (\ref{eq_Bphi_p}), with
\begin{equation}
\label{eq_B1_ABd}
B_r(r_1,\varphi_1,z_1) =
B_1 \ f_{\rm sym} \ 
\exp \left( - \frac{|z_1|}{H} \right) \ 
\cos \Big( m \, \big( \varphi_1 - g_\varphi (r_1,z_1) 
- \varphi_\star \big) \Big) \ \cdot
\end{equation}
All the symbols entering Eq.~(\ref{eq_B1_ABd}) have the same meaning 
as in Eq.~(\ref{eq_B1_AB});
the difference between both equations resides in the $z_1$-dependence of $B_r$,
which leads to a peak at $|z_1| = H$ (appropriate for halo fields)
in Eq.~(\ref{eq_B1_AB}) and a peak at $z_1 = 0$ (appropriate for disk fields)
in Eq.~(\ref{eq_B1_ABd}).
Inside $r_1$, field lines are straight, horizontal ($z = z_1$),
and parallel to the unit vector $\hat{e}_\star(z_1)$ in the direction 
$\varphi_1 = g_\varphi (r_1,z_1) + \varphi_\star$,
and the magnetic field vector is given by Eq.~(\ref{eq_Br_A_in})
adjusted to Eq.~(\ref{eq_B1_ABd}):
\begin{eqnarray}
\label{eq_Br_Ad_in}
{\boldvec B} & = &
B_r \big( r_1, g_\varphi (r_1,z_1) + \varphi_\star, z_1 \big) \
\hat{e}_\star(z_1) 
\nonumber \\
& = &
B_1 \ f_{\rm sym} \ 
\exp \left( - \frac{|z_1|}{H} \right) \
\hat{e}_\star(z_1) \ , \qquad r \le r_1 \ \cdot
\end{eqnarray}

{\bf Model~Bd} is a slight variant of model~B.
Given a reference radius, $r_1$, the shape of field lines is described by
\begin{equation}
\label{eq_mfl_Bd}
z = \frac{1}{2n+1} \ z_1 \ 
\left[ \left( \frac{r}{r_1} \right)^{-n} + 2n \, \sqrt{\frac{r}{r_1}} \right] \ ,
\end{equation}
with $n \ge 2$, and by Eq.~(\ref{eq_mfl_spiral_g});
the vertical and azimuthal labels of the field line passing through $(r,\varphi,z)$
are given by 
\begin{equation}
\label{eq_z1_Bd}
z_1 = (2n+1) \ z \ 
\left[ \left( \frac{r}{r_1} \right)^{-n} + 2n \, \sqrt{\frac{r}{r_1}} \right]^{-1} 
\end{equation}
and by Eq.~(\ref{eq_phi1}), respectively;
and the three magnetic field components obey
\begin{eqnarray}
B_r & = & \frac{r_1}{r} \ \frac{z_1}{z} \ B_r(r_1,\varphi_1,z_1)
\label{eq_Br_Bd} \\
B_z & = & - \frac{n}{2n+1} \ \frac{r_1 \, z_1^2}{r^2 \, z} \
\left[ \left( \frac{r}{r_1} \right)^{-n} - \sqrt{\frac{r}{r_1}} \right] \
B_r(r_1,\varphi_1,z_1) \ ,
\label{eq_Bz_Bd}
\end{eqnarray}
and Eq.~(\ref{eq_Bphi_p}), together with Eq.~(\ref{eq_B1_ABd}).

{\bf Model~Dd} is nearly the same as model~D.
Given a reference height on each side of the midplane, $z_1 = |z_1| \ {\rm sign}~z$,
the shape of field lines is described by Eq.~(\ref{eq_mfl_D}), 
with $n = 0.5$, and by Eq.~(\ref{eq_mfl_spiral_g});
the radial and azimuthal labels of the field line passing through $(r,\varphi,z)$
are given by Eqs.~(\ref{eq_r1_D}) and (\ref{eq_phi1});
and the three magnetic field components obey Eqs.~(\ref{eq_Br_D}), (\ref{eq_Bz_D}),
and (\ref{eq_Bphi_p}), with
\begin{equation}
\label{eq_B1_Dd}
B_z(r_1,\varphi_1,z_1) =
B_1 \ \bar{f}_{\rm sym} \ F(r_1) \
\cos \Big( m \, \big( \varphi_1 - g_\varphi (r_1,z_1) - \varphi_\star \big) \Big)
\end{equation}
and
\begin{equation}
\nonumber
F(r_1) = 
\left\lbrace
\begin{array}{ll}
1 \ , & \qquad r_1 \le L \\
\noalign{\smallskip}
\displaystyle \exp \left( - \frac{r_1-L}{L} \right) \ , & \qquad r_1 \ge L
\end{array}
\right. \ \cdot
\end{equation}
Eq.~(\ref{eq_B1_Dd}) is the counterpart of Eq.~(\ref{eq_B1_CD}), 
with the radial exponential cut off for $r_1 \le L$.
The effect of this cutoff is to reduce the crowding of field lines 
along the midplane.

\section{\label{method}Method}

\subsection{\label{method_simul}Simulated Galactic Faraday depth}

As explained earlier, our purpose is to study the structure 
of the magnetic field in the Galactic halo.
We are not directly interested in the exact attributes of the disk magnetic field, 
but we may not ignore the disk field altogether:
since all sightlines originating from the Sun pass through the disk
(if only for a short distance before entering the halo),
we need to have a good description of the disk field 
in the interstellar vicinity of the Sun.

Therefore, all our Galactic magnetic field models must include a disk field 
and a halo field. 
Moreover, in view of the general trends discussed in Sect.~\ref{FD_data_trends},
we expect the disk field to be roughly symmetric and the halo field
roughly antisymmetric with respect to the midplane.
Alternatively, we may consider that the Galactic magnetic field is composed
of a symmetric field, which dominates in the disk,
and an antisymmetric field, which dominates in the halo.
In other words, we may write the complete Galactic magnetic field as
\begin{equation}
\label{eq_B_tot}
{\boldvec B} = {\boldvec B}_{\rm S} + {\boldvec B}_{\rm A} \ ,
\end{equation}
where ${\boldvec B}_{\rm S}$ is the symmetric field, described by
one of the disk-field models (model Ad1, Bd, or Dd; see Sect.~\ref{models_disk}),
and ${\boldvec B}_{\rm A}$ is the antisymmetric field, described by
one of the halo-field models (model A1, B, C, or D; see Sect.~\ref{models_halo}).

Our knowledge of the large-scale magnetic field at the Sun,
${\boldvec B}_\odot$, imposes two constraints on our magnetic field models.
Since the Sun is assumed to lie in the midplane ($z_\odot = 0$), 
where ${\boldvec B}_{{\rm A}}$ vanishes,
these constraints apply only to ${\boldvec B}_{\rm S}$.
From the RMs of nearby pulsars, 
\cite{rand&k_89} derived a local field strength $B_\odot \simeq 1.6~\mu{\rm G}$,
while \cite{rand&l_94, han&q_94} obtained $B_\odot \simeq 1.4~\mu{\rm G}$.
All three studies found ${\boldvec B}_\odot$ to be running clockwise 
about the rotation axis ($(B_\varphi)_\odot > 0$).
\cite{han&q_94} also derived a local pitch angle $p_\odot \simeq -8.2^\circ$
(implying $(B_r)_\odot < 0$), close to the value $p_\odot \simeq -7.2^\circ$ 
inferred from the polarization of nearby stars \citep{heiles_96}. 
In the following, we restrict the range of $B_\odot$ to $[1,2]~\mu{\rm G}$ 
and the range of $p_\odot$ to $[-15^\circ, -4^\circ]$.
The implications of these restrictions are discussed 
at the end of Appendix~\ref{appendix_results_intervals}.

To calculate the Galactic FD (Eq.~\ref{eq_FD}),
we also need a model for the free-electron density, $n_{\rm e}$.
Here, we adopt the NE2001 model of \cite{cordes&l_02},
with revised values for the parameters of the thick disk component. 
\cite{gaensler&mcm_08} found that the scale height of the thick disk
in the NE2001 model was underestimated by almost a factor of 2;
they derived a new scale height of 1.83~kpc (up from 0.95~kpc)
and a new midplane density of 0.014~cm$^{-3}$ (down from 0.035~cm$^{-3}$).
This revision prompted several authors to adopt the NE2001 model
and simply replace the parameters of the thick disk with the new values
derived by \citeauthor{gaensler&mcm_08}, while keeping all the other components 
unchanged \citep{sun&r_10, pshirkov&tkn_11, jansson&f_12a}. 
However, \cite{schnitzeler_12} showed that this simple modification 
entails internal inconsistencies,
as the thick disk of \citeauthor{gaensler&mcm_08} overlaps with 
the local ISM and local spiral arm components of the NE2001 model.
By correctly accounting for all the components of the NE2001 model,
\cite{schnitzeler_12} updated the thick disk with a scale height of 1.31~kpc
and a midplane density of 0.016~cm$^{-3}$.
These are the values that we adopt for our FD calculation.
For future reference, the scale length of the thick disk is $\simeq 11~{\rm kpc}$.

For any considered magnetic field model, we can now
use Eq.~(\ref{eq_FD}) to calculate the modeled Galactic FD,
${\rm FD}_{\rm mod}$, as a function of position in the sky.
To produce a modeled FD map that is directly comparable to the observational map
of $\overline{\rm FD}_{\rm obs}$ displayed in the top panel 
of Fig.~\ref{figure_FD_ref}, we divide the sky area into the same 428 bins 
as in the $\overline{\rm FD}_{\rm obs}$ map,
and we only retain the same $N_{\rm bin} = 356$ bins with $|b| \ge 10^\circ$.
For each of these bins, we compute ${\rm FD}_{\rm mod}$ 
along 20 well-separated sightlines (which is sufficient, 
in view of the smoothness of our large-scale field models,
to reach the required accuracy), and we average the computed ${\rm FD}_{\rm mod}$
to obtain $\overline{\rm FD}_{\rm mod}$.
The resulting modeled map of $\overline{\rm FD}_{\rm mod}$ can then be compared
to the observational map of $\overline{\rm FD}_{\rm obs}$.

To disentangle the impact of the different model parameters
and to guide (and ultimately speed up) the global fitting process, 
it helps to first consider ${\boldvec B}_{\rm S}$ and ${\boldvec B}_{\rm A}$ 
separately, denoting the associated $\overline{\rm FD}_{\rm mod}$
by $\overline{\rm FD}_{\rm mod,S}$ and $\overline{\rm FD}_{\rm mod,A}$,
respectively.
If the slight vertical asymmetry in the free-electron density distribution
(which essentially arises from the local ISM component in the NE2001 model)
is ignored, $\overline{\rm FD}_{\rm mod,S}$ and $\overline{\rm FD}_{\rm mod,A}$
represent the symmetric and antisymmetric parts of $\overline{\rm FD}_{\rm mod}$.
Accordingly, we will first adjust
$\overline{\rm FD}_{\rm mod,S}$ to $\overline{\rm FD}_{\rm obs,S}$
and $\overline{\rm FD}_{\rm mod,A}$ to $\overline{\rm FD}_{\rm obs,A}$,
where $\overline{\rm FD}_{\rm obs,S}$ and $\overline{\rm FD}_{\rm obs,A}$
are the symmetric and antisymmetric parts of $\overline{\rm FD}_{\rm obs}$
(displayed in the middle and bottom panels of Fig.~\ref{figure_FD_ref}).
We will then combine the separate best-fit ${\boldvec B}_{\rm S}$ 
and ${\boldvec B}_{\rm A}$ 
into a starting-guess ${\boldvec B}$ for the actual fit 
of $\overline{\rm FD}_{\rm mod}$ to $\overline{\rm FD}_{\rm obs}$.

\subsection{\label{method_fitting}Fitting procedure}

For each magnetic field model, 
our fitting procedure relies on the least-square method, 
which consists of minimizing the $\chi^2$ parameter defined as
\begin{equation}
\label{eq_chi}
\chi^2 = \sum_{i=1}^{N_{\rm bin}} 
\frac{\left( \overline{\rm FD}_{{\rm obs},i}-\overline{\rm FD}_{{\rm mod},i}
\right)^2}{\sigma^2_i} \ ,
\end{equation}
where $N_{\rm bin} = 356$ is the total number of retained bins,
subscript $i$ is the bin identifier,
$\overline{\rm FD}_{{\rm obs},i}$ and $\overline{\rm FD}_{{\rm mod},i}$
are the average values of the observational and modeled Galactic FDs, 
respectively, within bin~$i$, and $\sigma_i$ is the uncertainty 
in $\overline{\rm FD}_{{\rm obs},i}$ 
(estimated in Appendix~\ref{appendix_uncertainty}). 
We also define the reduced $\chi^2$ parameter,
\begin{equation}
\label{eq_chi_red}
\chi^2_{\rm red} = \frac{\chi^2}{N_{\rm bin}-N_{\rm par}} \ ,
\end{equation}
where $N_{\rm par}$ is the number of free parameters entering 
the considered magnetic field model. 
The advantage of using $\chi^2_{\rm red}$ instead of $\chi^2$ is that
$\chi^2_{\rm red}$ gives a more direct idea of how good a fit is.
In principle, a fit with $\chi^2_{\rm red} \approx 1$ 
[$\chi^2_{\rm red} \gg 1$] is good [bad].
Here, however, because of the large uncertainties in $\sigma_i$
(see Appendix~\ref{appendix_uncertainty}),
we do not ascribe too much reality to the absolute $\chi^2_{\rm red}$,
but we rely on the relative $\chi^2_{\rm red}$ to rank the different
fits by relative likelihood.
In the following, the minimum values of $\chi^2$ and $\chi^2_{\rm red}$
are denoted by $\chi^2_{\rm min}$ and $\chi^2_{\rm red,min}$, respectively.

Because of the relatively large number of free parameters, 
a systematic grid search over the $N_{\rm par}$-dimensional parameter space
would be prohibitive in terms of computing time.
Therefore, we opt for the much more efficient 
MCMC sampling method.
We employ two different versions of this method, 
corresponding to two different implementations 
of the Metropolis-Hastings algorithm
\citep{metropolis&rrt_53, hastings_70}:
in the first version, the proposal distribution is 
centered on an educated guess for the best-fit solution,
whereas in the second version (random walk Metropolis-Hastings algorithm), 
the proposal distribution is centered on the current point of the Markov chain
\citep[see, e.g.,][and references therein]{robert_15}.
After verifying that both versions lead to the same results 
(within the Monte Carlo uncertainties) and remarking 
that the second version is systematically faster, 
we use only the second version in most of our numerous trial runs.
However, in our last series of runs (leading to the final results 
listed in Table~\ref{table_bestfit}),
we use again both versions as a reliability check.

For every $\chi^2$ minimization, we run several, gradually more focused simulations,
whose sequence is optimized by trial and error
to converge as fast as possible to the correct target distribution.
The key is to find a good compromise between 
sufficiently broad coverage of the parameter space
(to make sure the absolute $\chi^2$ minimum is within reach)
and sufficiently narrow bracketing of the minimum-$\chi^2$ solution
(to recover the target distribution in a manageable time).
We begin with a simulation having broad proposal ($q$) and target ($\pi$) distributions.
For the former, we generally adopt 
a $N_{\rm par}$-dimensional Gaussian with adjustable widths,
which we take initially large enough to encompass all plausible parameter values.
For the latter, we adopt $\pi \propto \exp (- \chi^2 / 2T)$,
with $T$ tuned in parallel with the Gaussian widths to achieve the optimal 
acceptance rate for the next proposed point in the Markov chain
\citep[see, e.g.,][]{roberts&gg_97, jaffe&lbl_10}.
Once a likelihood peak clearly emerges, we start a new simulation 
with narrower $q$ (smaller Gaussian widths) and $\pi$ (smaller $T$) 
and with the starting point of the Markov chain close to the emerging peak.
We repeat the procedure a few times until $T = 1$, as ultimately required.
Note that we prefer to keep a handle on 
how the proposal and target distributions are gradually narrowed down, 
rather than implement one of the existing adaptive algorithms 
that can be found in the literature.

We monitor convergence to the target distribution 
by visually inspecting the evolution of the histogram, the running mean, 
and the minimum-$\chi^2$ (i.e., best-fit) value of each parameter
as well as the histogram of $\chi^2$ itself.
We stop sampling when a stationary state appears to have been reached,
and after checking that the best-fit value of each parameter has achieved
an accuracy better than 10\% of the half-width of its $1\sigma$ confidence 
interval.\footnote{
In practice, this second condition is generally amply satisfied.
}
The latter is taken to be the projection onto the parameter's axis
of the $N_{\rm par}$-dimensional region containing the $68\%$ points
of the Markov chain (or second half thereof) with the lowest $\chi^2$.
For our final results (listed in Table~\ref{table_bestfit}),
we simulate multiple Markov chains in parallel, with overdispersed starting points,
we discard the first halves of the chains to skip the burn-in phase,
and relying on the Gelman-Rubin convergence diagnostic \citep{gelman&r_92},
we consider that our chains have converged when, for each parameter, 
the variance of the chain means has dropped below 10\% 
of the mean of the chain variances, or nearly equivalently, 
when the potential scale reduction factor, $\sqrt{\hat{R}}$, 
has dropped below 1.05.

As mentioned at the end of Sect.~\ref{method_simul},
we apply our fitting procedure first to the symmetric and antisymmetric
components of the Galactic FD separately, 
then to the total (symmetric + antisymmetric) Galactic FD.
The uncertainties in $\overline{\rm FD}_{{\rm obs,S},i}$ 
and $\overline{\rm FD}_{{\rm obs,A},i}$,
$\sigma_{{\rm S},i}$ and $\sigma_{{\rm A},i}$,
are related to the uncertainties in $\overline{\rm FD}_{{\rm obs},i}$
and $\overline{\rm FD}_{{\rm obs},{-i}}$, $\sigma_i$ and $\sigma_{-i}$,
through $\sigma^2_{{\rm S},i} = \sigma^2_{{\rm A},i} 
= \frac{1}{4} \ (\sigma^2_i + \sigma^2_{-i})$,
where subscript $-i$ is used to denote the bin that is the symmetric of bin $i$ 
with respect to the midplane.

\section{\label{results}Results}

\subsection{\label{results_remarks}Preliminary remarks}

\begin{table*}[t]
\caption{List of all the free parameters\,$^{a}$ 
entering our magnetic field models,$^{b}$
in the axisymmetric ($m = 0$) case.
}
\label{table_param}
\begin{minipage}{\textwidth}
\centering
\renewcommand{\footnoterule}{}   
\renewcommand{\thefootnote}{\alph{footnote}}
{\renewcommand{\arraystretch}{1.5}   
\begin{tabular}{lp{6cm}lp{6cm}}
\hline 
\hline 
Parameter & Definition & Models & 
Impact on $\overline{\rm FD}_{\rm mod}$ map 
\\ 
\hline 
$B_1$ & 
\parbox[t]{4cm}{Normalization field strength \\
(Eqs.~(\ref{eq_B1_AB}) and (\ref{eq_B1_CD}))} & 
All & 
Overall amplitude 
\\
$H$ & 
\parbox[t]{4cm}{Scale height of $B_r$ at $r_1$ \\
(Eq.~\ref{eq_B1_AB})} & 
A, B & 
Latitudinal extent 
\\ 
$L$ & 
\parbox[t]{4cm}{Scale length of $B_z$ at $z_1$ \\
(Eq.~\ref{eq_B1_CD})} & 
C, D & 
Longitudinal extent 
\\ 
\hline 
$p_0$ & 
\parbox[t]{4cm}{Pitch angle at origin \\
(Eq.~\ref{eq_shiftedwindingfc})} & 
All & 
\parbox[t]{5cm}{Longitudes of peaks \& sign reversals \\ 
over the whole sky} 
\\
$H_p$ & 
\parbox[t]{5cm}{Scale height of winding function \\
(Eq.~\ref{eq_shiftedwindingfc})} &
All & 
\parbox[t]{5cm}{Longitudes of peaks \& sign reversals, \\
mainly at high $|b|$} 
\\
$L_p$ & 
\parbox[t]{5cm}{Scale length of winding function \\
(Eq.~\ref{eq_shiftedwindingfc})} &
All & 
\parbox[t]{5cm}{Longitudes of peaks \& sign reversals, \\ 
mainly at large $|\ell|$} 
\\
\hline 
$r_1$ & 
Reference radius & 
A, B & 
\\ 
$|z_1|$ & 
Positive reference height & 
C, D & 
\\ 
$a$ & 
\parbox[t]{6cm}{Opening parameter for poloidal field lines \\
(Eqs.~(\ref{eq_mfl_A}) and (\ref{eq_mfl_C}))} & 
A, C & 
\parbox[t]{6cm}{Amplitude \& location of dominant patches; \\
location of sign reversal line} 
\\ 
$n$ & 
\parbox[t]{6cm}{Power-law index for poloidal field lines \\
(Eqs.~(\ref{eq_mfl_B}) and (\ref{eq_mfl_D}))} & 
B, D & 
\\
\hline 
\footnotetext[1]{For the reasons explained in Appendix~\ref{appendix_param}, 
three of the initially free parameters ($r_1$, $|z_1|$, $n$) 
are actually assigned a fixed value before the fitting process.}
\footnotetext[2]{Only the halo-field models are listed here, 
but all the comments applying to one of the halo-field models A, B, or D 
also apply to the corresponding disk-field model Ad, Bd, or Dd 
(the halo-field model C has no disk counterpart).
Moreover, model~A was retained for completeness, 
even though its axisymmetric version diverges at $r \to 0$.}
\end{tabular} 
}
\end{minipage}
\end{table*}

Before considering any particular magnetic field model, 
let us make a few general remarks that will guide us 
in our exploration of the different parameter spaces
and help us optimize the Metropolis-Hastings MCMC algorithm 
used for our fitting procedure (see Sect.~\ref{method_fitting})
-- in particular, by making an informed initial choice 
for the proposal distribution, $q$.
These remarks will also enable us to check a posteriori 
that the best-fit solutions do indeed conform to our physical expectation.

First, the all-sky maps of the symmetric and antisymmetric components
of the bin-averaged observational Galactic Faraday depth, 
$\overline{\rm FD}_{\rm obs,S}$ and $\overline{\rm FD}_{\rm obs,A}$,
displayed in the middle and bottom panels of Fig.~\ref{figure_FD_ref},
show that $\overline{\rm FD}_{\rm obs,A}$ is broadly distributed,
both in latitude and in longitude,
while the distribution of $\overline{\rm FD}_{\rm obs,S}$
is thinner in latitude\footnote{
The thin distribution of $\overline{\rm FD}_{\rm obs,S}$ is in fact 
largely hidden under the grey band along the Galactic plane 
that masks out the region $|b| < 10^\circ$.
}
and more heavily weighted toward the outer Galaxy.
This suggests that the symmetric field, ${\boldvec B}_{\rm S}$,
has a flatter distribution, which extends out to larger radii,
while the antisymmetric field, ${\boldvec B}_{\rm A}$,
has a more spherical distribution, which extends up to larger heights.
Clearly, these different distributions support our choosing 
disk-field models (Ad1, Bd, Dd) to describe ${\boldvec B}_{\rm S}$ 
and halo-field models (A1, B, C, D) to describe ${\boldvec B}_{\rm A}$.
Furthermore, out of our four halo-field models, 
we may already anticipate that model~C, where the field strength
has the most spherical distribution (see Fig.~\ref{figure_fieldlines}c),
is the best candidate to represent the antisymmetric field.
This expectation is in line with the notion that models A, B, and D,
where low-$|z|$ field lines are nearly horizontal
(see Figs.~\ref{figure_fieldlines}a, \ref{figure_fieldlines}b,
\ref{figure_fieldlines}d), could easily be explained by a disk dynamo
(together with a galactic wind advecting the disk-dynamo field into the halo),
which is known to favor symmetric fields (see Paper~1).\footnote{
Although models A, B, and D do not seem to be well suited to describe 
the {\it antisymmetric} halo field of our own Galaxy, they remain good candidates 
to describe the {\it symmetric} halo fields that could be observed in external galaxies.
}

Second, the multiple sign reversals with longitude observed in the map of 
$\overline{\rm FD}_{\rm obs,S}$ (middle panel of Fig.~\ref{figure_FD_ref})
suggest that ${\boldvec B}_{\rm S}$ is non-axisymmetric.
In reality, an axisymmetric magnetic field could very well lead to 
longitudinal FD reversals (in addition to the two reversals, 
toward $\ell \approx 0^\circ$ and $\ell \approx 180^\circ$, 
naturally associated with an axisymmetric, nearly azimuthal field)
if it reverses direction with radius.
As a general rule, if the axisymmetric field reverses $n$ times with radius,
FD can reverse up to $2\,(n+1)$ times with longitude.
A case in point is provided by the concentric-ring model of \cite{rand&k_89},
where the field is purely azimuthal and reverses direction 
from one ring to the next.
What about our three disk-field models?
As explained in Sect.~\ref{models_disk}, model~Ad1 is inherently bisymmetric.
Model Dd has no radial field reversals 
(see Eqs~(\ref{eq_Br_D})-(\ref{eq_Bz_D}) for the poloidal field
and Eq.~(\ref{eq_Bphi_p}) for the azimuthal field),
so it cannot lead to more than two longitudinal FD reversals.
Model Bd has one reversal of $B_z$ at radius $r_1$ (see Eq.~(\ref{eq_Bz_Bd})), 
which must be accompanied by one reversal of $B_\varphi$ at a slightly smaller radius 
(see Eq.~(\ref{eq_Bphi_p}));
the resulting radial reversal of the total field could account for 
a maximum of four longitudinal FD reversals,
which is still short of the observed number of reversals.
Ultimately, it appears that none of our axisymmetric disk-field models 
is likely to reproduce the general sign pattern 
of the $\overline{\rm FD}_{\rm obs,S}$ map.
To strengthen this conclusion, we note that our axisymmetric disk-field models
would also be hard-pressed to reproduce the preponderance of the outer Galaxy 
in the $\overline{\rm FD}_{\rm obs,S}$ map.

Altogether, the remaining good candidates are 
model~C for the antisymmetric field, ${\boldvec B}_{\rm A}$,
and model~Ad1 plus the non-axisymmetric versions of models Bd and Dd 
for the symmetric field, ${\boldvec B}_{\rm S}$.
If we retain only axisymmetric ($m = 0$) and bisymmetric ($m = 1$) models,
and label them with the value of the azimuthal wavenumber, $m$,
appended to the letters of the models,
our full list reduces to 
models~A1, B0, B1, C0, C1, D0, and D1 for ${\boldvec B}_{\rm A}$
and models~Ad1, Bd0, Bd1, Dd0, and Dd1 for ${\boldvec B}_{\rm S}$,
amongst which the good candidates are 
models~C0 and C1 for ${\boldvec B}_{\rm A}$
and models~Ad1, Bd1, and Dd1 for ${\boldvec B}_{\rm S}$.

The impact of the model parameters on the $\overline{\rm FD}_{\rm mod}$ map 
is discussed in detail in Appendix~\ref{appendix_param},
and a summary of the discussion of the axisymmetric ($m = 0$) case 
is provided in Table~\ref{table_param}.

\subsection{\label{results_bestfit}Best fits}

\begin{table*}[t]
\caption{List of the 6 best models of the total 
(antisymmetric halo + symmetric disk) magnetic field.
}
\label{table_bestfit}
\begin{minipage}{\textwidth}
\centering
\hspace*{-1cm}
\renewcommand{\footnoterule}{}   
\renewcommand{\thefootnote}{\alph{footnote}}
{\renewcommand{\arraystretch}{1.5}   
\begin{tabular}{c|ccc|cccc|c|ccc|c}
\hline 
\hline 
Model\,\footnotemark[1] & 
\multicolumn{3}{c|}{Fixed parameters\,\footnotemark[2]} &
\multicolumn{8}{c|}{Best-fit free parameters\,\footnotemark[2]} &
$\chi^2_{\rm red,min}$ 
\\
& $\!r_1~[{\rm kpc}]\!$ & $\!|z_1|~[{\rm kpc}]\!$ & $n$ & 
$B_1~[\mu{\rm G}]$ & $H~[{\rm kpc}]$ & $L~[{\rm kpc}]$\,\footnotemark[3] & 
$\varphi_\star~[\rm deg]$ & $a~[{\rm kpc}^{-2}]$ &
$p_0~[\rm deg]$ & $H_p~[{\rm kpc}]$\,\footnotemark[4] & 
$L_p~[{\rm kpc}]$\,\footnotemark[5] &
\\
\hline 
C0 &
& 0 & & 
$0.36_{-0.16}^{+0.28}$ & & $3.0_{-0.7}^{+1.3}$ & & $1.17_{-0.16}^{+0.16}$ &
& & &
\\
\noalign{\vspace{-5pt}}
& & & & & & & & & 
$-7.9_{-2.9}^{+2.2}$ & $>5$ & $>18$ &
$2.37$
\\
\noalign{\vspace{-5pt}}
Ad1 & 
3 & & & 
$19_{-9}^{+17}$ & $0.055_{-0.016}^{+0.019}$ & & 
$-54_{-180}^{+180}$ &$0.90_{-0.13}^{+0.19}$ &
& & &
\\
\hline 
C0 &
& 0 & & 
$0.29_{-0.15}^{+0.21}$ & & $3.4_{-0.7}^{+1.8}$ & & $0.88_{-0.14}^{+0.14}$ &
& & &
\\
\noalign{\vspace{-5pt}}
& & & & & & & & & 
$-7.2_{-2.8}^{+2.1}$ & $>9$ & $>16$ &
$2.42$
\\
\noalign{\vspace{-5pt}}
Bd1 & 
3 & & 2 & 
$2.0_{-0.7}^{+1.7}$ & $0.32_{-0.07}^{+0.10}$ & & $153_{-180}^{+180}$ & &
& & &
\\
\hline 
C0 &
& 0 & & 
$0.18_{-0.07}^{+0.15}$ & & $4.8_{-2.2}^{+3.2}$ & & $0.61_{-0.38}^{+0.40}$ &
& & &
\\
\noalign{\vspace{-5pt}}
& & & & & & & & & 
$-7.4_{-1.3}^{+0.9}$ & $4.2_{-0.8}^{+1.0}$ & $>22$ &
$2.26$
\\
\noalign{\vspace{-5pt}}
Dd1 & 
& 1.5 & 0.5 & 
$0.065_{-0.013}^{+0.031}$ & & $9.8_{-3.8}^{+\infty}$ & $14_{-180}^{+180}$ & &
& & &
\\
\hline 
C1 &
& 0 & & 
$9.0_{-3.8}^{+2.6}$ & & $2.1_{-0.2}^{+0.4}$ & 
$198_{-51}^{+50}$ & $0.33_{-0.14}^{+0.14}$ &
& & &
\\
\noalign{\vspace{-5pt}}
& & & & & & & & & 
$-9.1_{-0.6}^{+1.1}$ & $1.2_{-0.1}^{+0.1}$ & $>38$ &
$2.02$
\\
\noalign{\vspace{-5pt}}
Ad1 & 
3 & & & 
$32_{-4}^{+3}$ & $0.054_{-0.013}^{+0.040}$ & & 
$-31_{-52}^{+46}$ & $0.031_{-0.029}^{+0.026}$ &
& & &
\\
\hline 
C1 &
& 0 & & 
$8.2_{-5.3}^{+9.3}$ & & $2.2_{-0.5}^{+0.7}$ & 
$197_{-68}^{+88}$ & $0.38_{-0.20}^{+0.24}$ &
& & &
\\
\noalign{\vspace{-5pt}}
& & & & & & & & & 
$-9.0_{-1.2}^{+1.2}$ & $1.2_{-0.1}^{+0.2}$ & $>38$ &
$2.02$
\\
\noalign{\vspace{-5pt}}
Bd1 & 
3 & & 2 & 
$24_{-11}^{+22}$ & $0.090_{-0.033}^{+0.043}$ & & $-34_{-65}^{+84}$ & &
& & &
\\
\hline 
C1 &
& 0 & & 
$9.5_{-5.7}^{+5.5}$ & & $2.1_{-0.3}^{+0.6}$ & 
$179_{-71}^{+56}$ & $0.45_{-0.20}^{+0.30}$ &
& & &
\\
\noalign{\vspace{-5pt}}
& & & & & & & & & 
$-8.4_{-0.9}^{+1.3}$ & $1.2_{-0.1}^{+0.2}$ & $>30$ &
$2.08$
\\
\noalign{\vspace{-5pt}}
Dd1 & 
& 1.5 & 0.5 & 
$0.40_{-0.23}^{+0.24}$ & & $2.9_{-0.8}^{+1.6}$ & $120_{-75}^{+53}$ & &
& & &
\\
\hline 
\footnotetext[1]{
For each total-field model, the first line refers to the antisymmetric halo field 
and the second line to the symmetric disk field.
Models~C0 and C1 are the axisymmetric ($m = 0$) and bisymmetric ($m = 1$)
versions of model~C, described in Sect.~\ref{models_halo},
while models~Ad1, Bd1, and Dd1 are the bisymmetric versions 
of models~Ad, Bd, and Dd, described in Sect.~\ref{models_disk}.
}
\footnotetext[2]{
The meaning of the different parameters is reminded in Table~\ref{table_param}.
}
\footnotetext[3]{
When the $1\sigma$ upper confidence limit of $L$ is $\gg L_{\rm e}$,
it cannot be determined with any accuracy, so it is replaced by $\infty$.
}
\footnotetext[4]{
When the best-fit value of $H_p$ is $\gg H_{\rm e}$,
it cannot be determined with any accuracy, 
so only an approximate $1\sigma$ lower limit is provided.
}
\footnotetext[5]{
In all models, the best-fit value of $L_p$ is $\gg L_{\rm e}$,
so it cannot be determined with any accuracy, 
and only an approximate $1\sigma$ lower limit is provided.
}
\end{tabular} 
}
\end{minipage}
\end{table*}

We consider 35 models of the total magnetic field,
corresponding to all the possible combinations 
of one of the 7 antisymmetric halo-field models, A1, B0, B1, C0, C1, D0, and D1, 
with one of the 5 symmetric disk-field models, Ad1, Bd0, Bd1, Dd0, and Dd1.
We apply the fitting procedure described in Sect.~\ref{method_fitting}
to each of our 35 total-field models.
As expected, we find that 6 total-field models have significantly lower 
$\chi_{\rm red,min}^2$ than the other models, and that they correspond 
exactly to the 6 combinations of good candidates identified 
in Sect.~\ref{results_remarks}, namely, the combinations 
of C0 or C1 with Ad1, Bd1, or Dd1. 
The $\chi^2_{\rm red,min}$ values, the best-fit parameter values, 
and the $1\sigma$ confidence intervals (in the sense defined 
in Sect.~\ref{method_fitting}) of these 6 total-field models 
are listed in Table~\ref{table_bestfit}.

A first glance at the last column of Table~\ref{table_bestfit} suggests 
that our $\chi_{\rm red,min}^2$ values might be too large to qualify for good fits. 
However, one has to remember that the exact values of $\chi_{\rm red}^2$
should not be taken too seriously; 
most importantly, they should not be considered individually,
but only relative to each other as a means of ranking the different models.

On the basis of this ranking, it emerges that 
(1) the three total-field models with a bisymmetric halo field 
(described by model~C1; last three combinations in Table~\ref{table_bestfit}) 
perform systematically better than the three total-field models 
with an axisymmetric halo field (described by model~C0; 
first three combinations in Table~\ref{table_bestfit}),
and (2) for each halo-field model, the three disk-field models (Ad1, Bd1, Dd1)
are nearly equally good.
The closeness of the three best fits for each halo-field model 
lends credence to their being the true absolute best fits, 
corresponding to the true absolute $\chi^2$ minima.
The differences between the three best fits are smaller 
when the halo field is bisymmetric than when it is axisymmetric,
and in either case, models~Ad1 and Bd1 tend to lead to the closer best fits.
This follows from the similarity in the shape of their field lines 
outside the centered vertical cylinder of radius $r_1 = 3~{\rm kpc}$
(see Figs.~\ref{figure_fieldlines}a and \ref{figure_fieldlines}b),
it being clear that the interior of the cylinder gives but a small 
overall contribution to the $\overline{\rm FD}_{\rm mod}$ maps.

The preference for a bisymmetric halo field can be understood 
as follows:\footnote{
The argument given here also applies to the disk field,
explaining why bisymmetric disk-field models lead to significantly lower 
$\chi_{\rm red,min}^2$ than their axisymmetric counterparts 
(see first paragraph of Sect.~\ref{results_bestfit}).
}
A bisymmetric field leads to field reversals along most lines of sight.
The exact locations of the field reversals, and hence the net result 
of integrating $B_\parallel$ along the different lines of sight 
(see Eq.~(\ref{eq_FD})),
and ultimately the appearance of the $\overline{\rm FD}_{\rm mod}$ maps,
depend sensitively on the precise orientation angle
of the bisymmetric pattern, $\varphi_\star$.
In other words, $\varphi_\star$ offers a very special degree of freedom, 
which makes it possible to produce a great variety 
of $\overline{\rm FD}_{\rm mod}$ maps.
By fine-tuning the value of $\varphi_\star$, it is generally possible 
to bring the $\overline{\rm FD}_{\rm mod}$ maps into (relative) good agreement 
with the observational $\overline{\rm FD}_{\rm obs}$ map.

Let us now discuss the best-fit values of the free parameters,
starting with the parameters governing the field strength distribution
($B_1$, $H$, $L$, $\varphi_{\star}$) 
and the curvature of parabolic field lines in models~Ad1, C0, and C1 ($\!\sqrt{a}$),
and continuing with the parameters governing the spiral shape of field lines
($p_0$, $H_p$, $L_p$).
The $1\sigma$ confidence intervals and the correlations between parameters 
are discussed in Appendix~\ref{appendix_results}.

When the halo field is bisymmetric, it has a much larger normalization field
strength, $B_1$ (by a factor $\approx 25-50$) than when it is axisymmetric, 
and the associated disk field also has a larger $B_1$ (by a factor $\approx 2-15$).
The larger $(B_1)_{\rm halo}$ and $(B_1)_{\rm disk}$ obtained 
with a bisymmetric halo field are needed to make up for the azimuthal modulation 
of the halo field and for the associated field reversals,
which lead to cancelation in the line-of-sight integration of $B_\parallel$.
The strong increase in $B_1$ is accompanied by a weaker decrease 
(by a factor $\approx 1-4$) in the scale height at $r_1$, $H$, 
and in the scale length at $z_1$, $L$.
At the same time, the curvature parameter of parabolic field lines, $\!\sqrt{a}$,
decreases slightly (by a factor $\approx 1.2-2$) for the halo field
and more significantly (by a factor $\approx 5$) for the disk field in model~Ad1.
The origin of these trends will become clearer in Sect.~\ref{results_faraday}.

More quantitatively, the halo field always has a short scale length at $z_1$ 
($L \approx (2-5)~{\rm kpc}$).
The disk field has a longer scale length at $z_1$ 
($L \approx (3-10)~{\rm kpc}$ in model~Dd1),
or a short scale height at $r_1$ ($H \approx 50~{\rm pc}$ in model~Ad1 
and $H \approx (100-300)~{\rm pc}$ in model~Bd1)
combined with an outward flaring of field lines 
($\!\sqrt{a} \approx 1/(1-5)~{\rm kpc}$ in model~Ad1 and $n=2$ in model~Bd1).
The scale lengths in models~C0, C1, and Dd1 are realistic, 
whereas the scale heights in models~Ad1 and Bd1 are probably too short,
even when their outward increase due to the flaring of field lines 
is taken into account -- for instance, the scale height at the Sun is only
$\approx (140-360)~{\rm pc}$ in model~Ad1
and $\approx (120-440)~{\rm pc}$ in model~Bd1.

With the above values of $L$, $H$, $\!\sqrt{a}$, and $n$,
the ratio $(B_1)_{\rm halo} / (B_1)_{\rm disk}$ can be adjusted 
in such a way that the halo [disk] field imposes its vertical antisymmetry [symmetry]
to the Galactic FD toward the inner [outer] halo,
as required by the north-south symmetry properties 
of the $\overline{\rm FD}_{\rm obs}$ sky (see Sect.~\ref{FD_data_trends}).
The resulting $B_1$ are generally realistic, 
although when the halo field is bisymmetric, the total field outside $r = 3~{\rm kpc}$ 
can become quite strong at the crests of the azimuthal modulation
(up to $\approx 200~\mu{\rm G}$ and $\approx 150~\mu{\rm G}$ 
at $(r = 3~{\rm kpc}, z=0)$ in the combinations C1-Ad1 and C1-Bd1, 
respectively).\footnote{
We are not so much concerned with the even stronger fields that can arise inside 
$r = 3~{\rm kpc}$ -- in particular with model~B1, where all field lines 
unavoidably end up concentrating along the rotation axis 
(see Fig.~\ref{figure_fieldlines}b).
}
On the other hand, the large $(B_1)_{\rm disk}$ ($\approx 30~\mu{\rm G}$) 
in C1-Ad1 and C1-Bd1 are not inconsistent with our imposed range 
of $[1,2]~\mu{\rm G}$ for the local field strength, $B_\odot$
(see Sect.~\ref{method_simul}): indeed, it suffices for the Sun 
to fall close to a node of the sinusoidal variation of the disk field.
Neither is the very small $(B_1)_{\rm disk}$ ($\approx 0.06~\mu{\rm G}$) 
in C0-Dd1 problematic, as $(B_1)_{\rm disk}$ is only the peak vertical component 
of the disk field at $(r_1 \le L, z_1 = \pm 1.5~{\rm kpc})$ 
(see Eq.~(\ref{eq_B1_Dd})), 
which is much weaker than the total field near the Galactic plane.

In each total-field model, the parameters governing the spiral shape of field lines
have common values for the disk and halo fields.
The pitch angle at the origin, $p_0$, is always negative and small 
($\approx (-9^\circ) - (-7^\circ)$), corresponding to a trailing magnetic spiral 
with nearly azimuthal field at the origin (see Eq.~(\ref{eq_Bphi_p})).
$|p_0|$ is slightly smaller when the halo field is axisymmetric, 
and accordingly the field is slightly more azimuthal.
The scale length of the winding function, $L_p$, is always very large 
(best-fit value $\gtrsim 40~{\rm kpc}$), implying that the field 
remains nearly azimuthal out to large radii. 
In contrast, the scale height of the winding function, $H_p$, 
is very sensitive to the azimuthal structure of the halo field:
when the halo field is axisymmetric, $H_p$ is greater than the free-electron 
scale height, $H_{\rm e} \simeq 1.3~{\rm kpc}$ (see Sect.~\ref{method_simul}),
so that the field remains nearly azimuthal across the free-electron layer
(where the Galactic FD arises; see Eq.~(\ref{eq_FD})),
while in the case of a bisymmetric halo field, $H_p$ is comparable to $H_{\rm e}$,
so that the field acquires a significant poloidal component
within the free-electron layer.
The reason for this dichotomy as well as for the slight difference in $p_0$
is easily understood:
in the axisymmetric case, the field must remain nearly azimuthal
in order to reproduce the rough east-west antisymmetry 
in the $\overline{\rm FD}_{\rm obs}$ sky,
whereas in the bisymmetric case, the field can a priori remain 
nearly azimuthal or become more poloidal
(see last paragraph of Sect.~\ref{FD_data_trends}).

\subsection{\label{results_faraday}Faraday-depth maps}

\begin{figure*}[t]
\begin{tabularx}{\linewidth}{XX}
\includegraphics[trim=1.9cm 2.2cm 1.4cm 2.2cm, clip, width=0.95\linewidth]{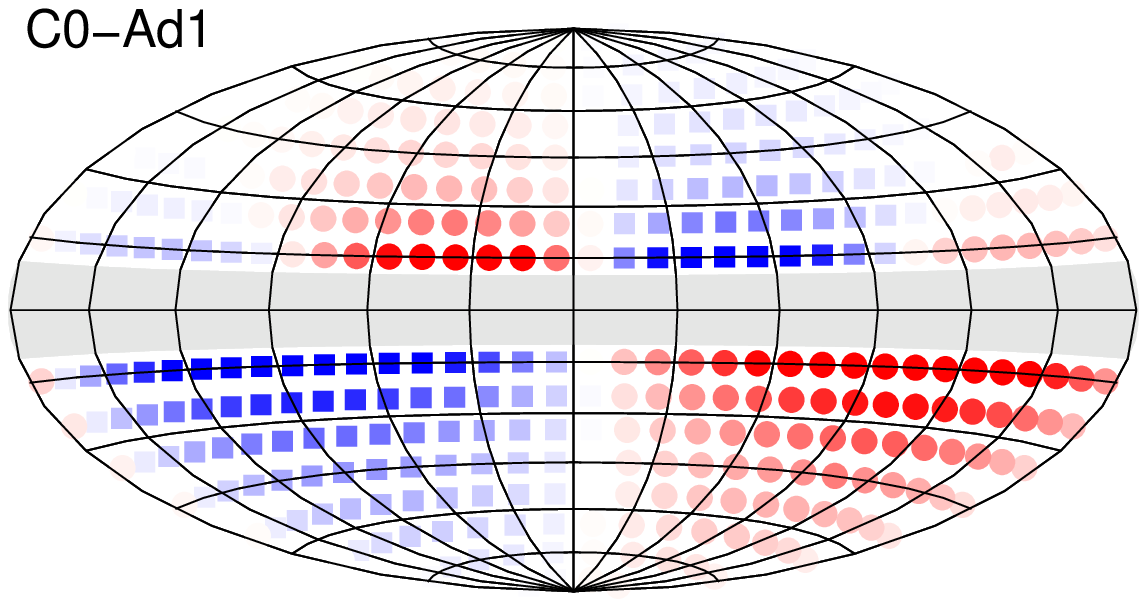} 
& 
\includegraphics[trim=1.9cm 2.2cm 1.4cm 2.2cm, clip, width=0.95\linewidth]{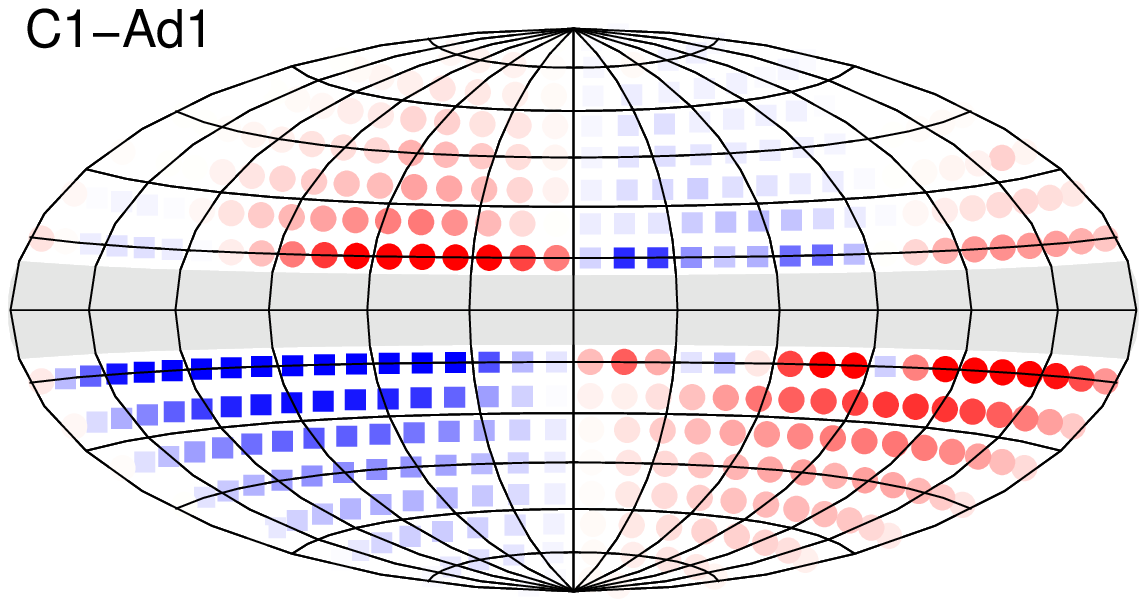} 
\\
\includegraphics[trim=1.9cm 2.2cm 1.4cm 2.2cm, clip, width=0.95\linewidth]{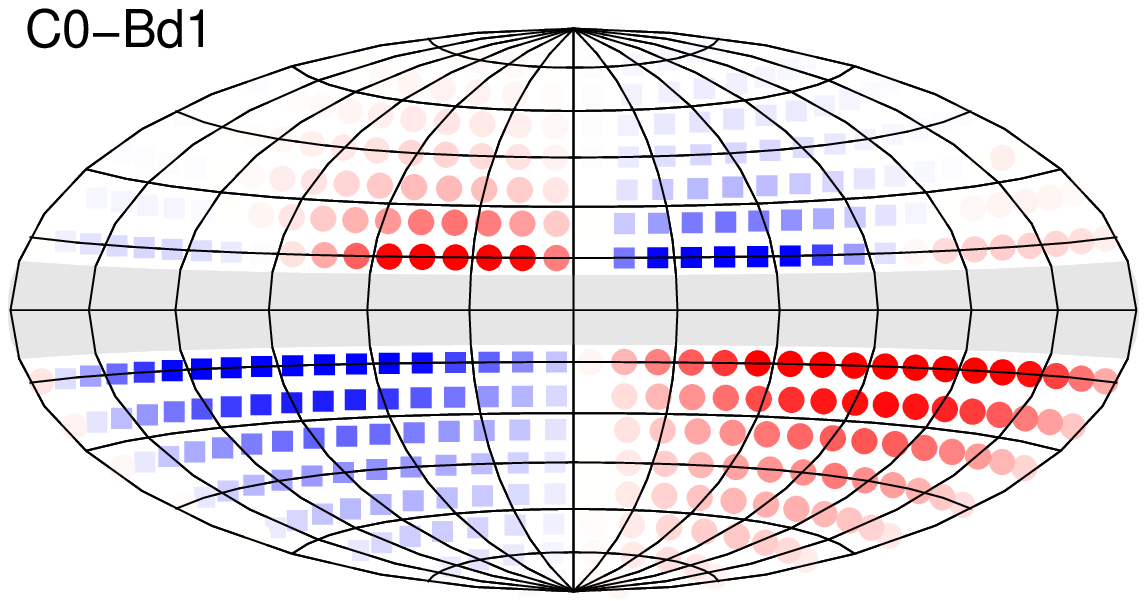} 
& 
\includegraphics[trim=1.9cm 2.2cm 1.4cm 2.2cm, clip, width=0.95\linewidth]{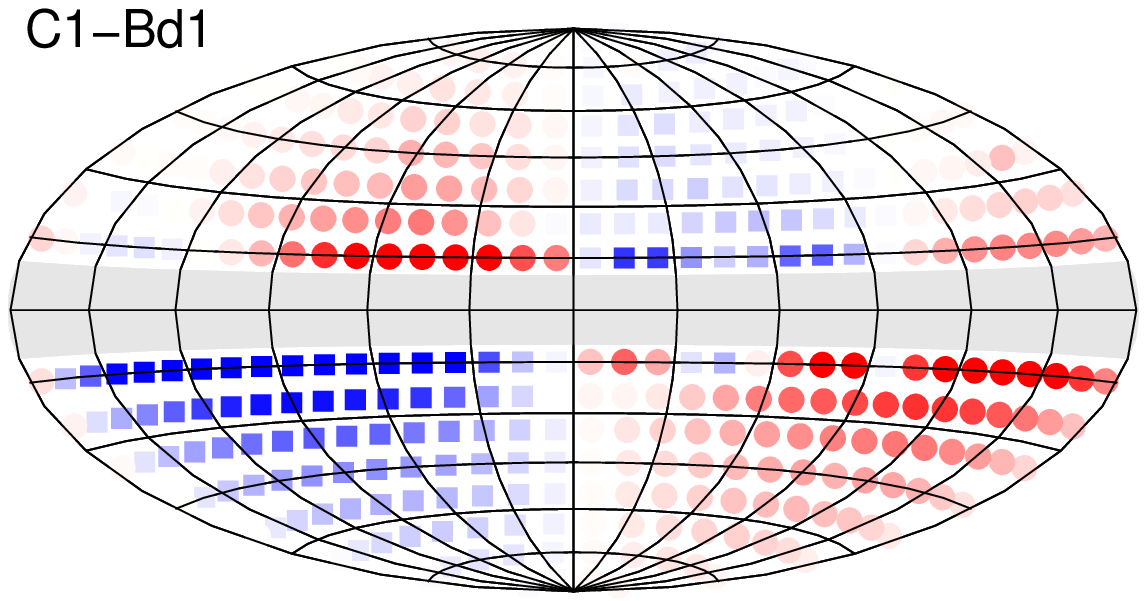} 
\\
\includegraphics[trim=1.9cm 2.2cm 1.4cm 2.2cm, clip, width=0.95\linewidth]{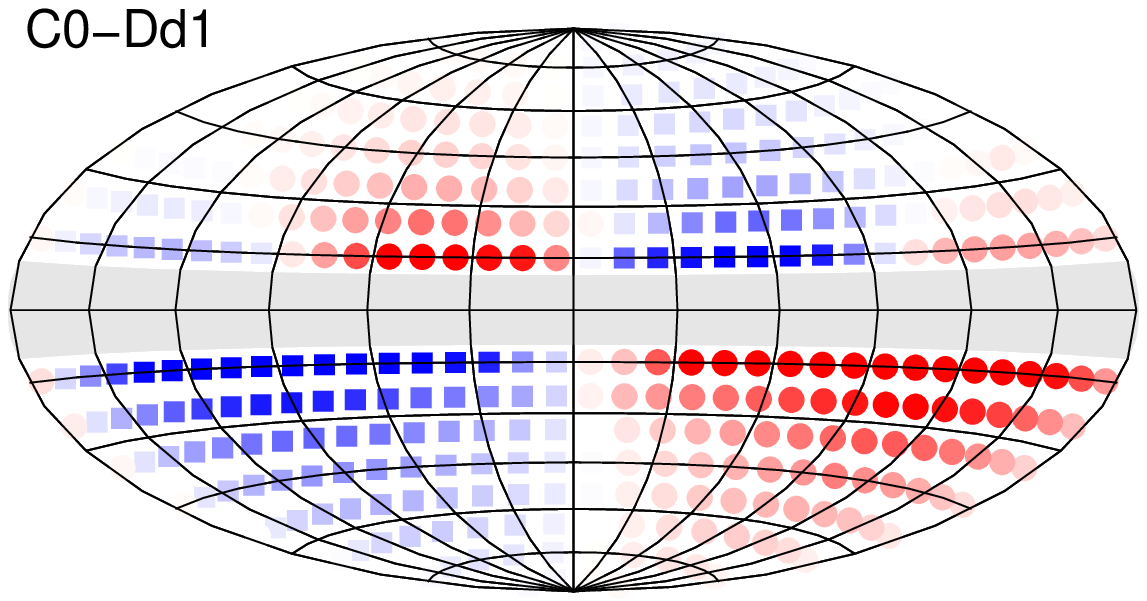} 
& 
\includegraphics[trim=1.9cm 2.2cm 1.4cm 2.2cm, clip, width=0.95\linewidth]{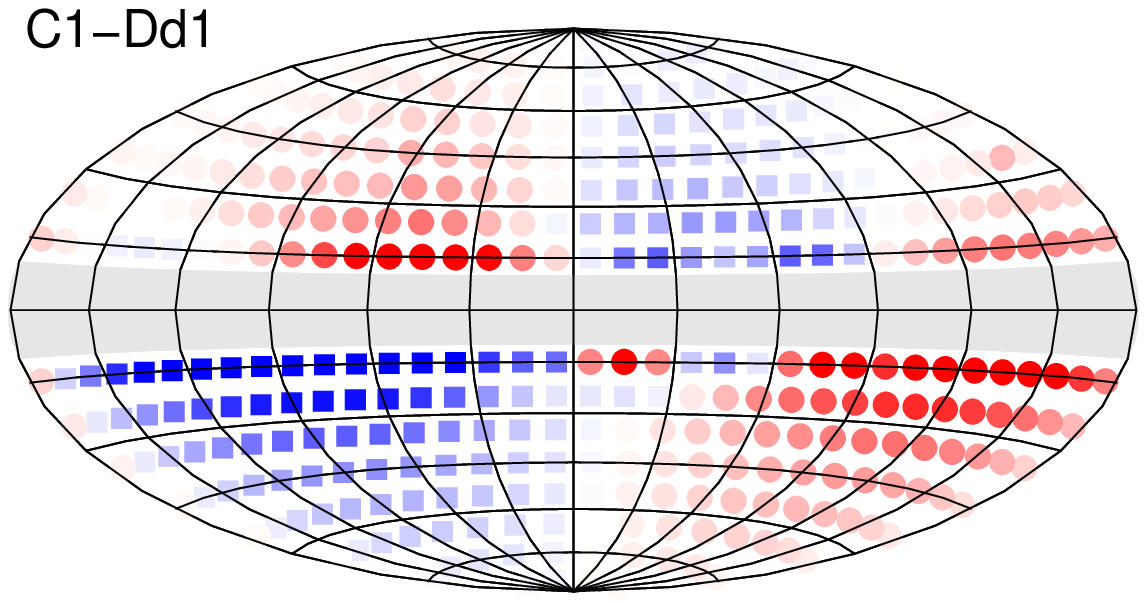} 
\end{tabularx}
\caption{
All-sky maps of the bin-averaged modeled Galactic Faraday depth, 
$\overline{\rm FD}_{\rm mod}$, for the six best-fit magnetic field models 
listed in Table~\ref{table_bestfit}.
The left and right columns are for a halo field described by 
model~C0 (axisymmetric) and C1 (bisymmetric), respectively,
while the top, middle, and bottom rows are for a disk field 
described by model~Ad1, Bd1, and Dd1, respectively.
The coordinate system and the color code of each map 
are the same as in Fig.~\ref{figure_FD_ref}.
}
\label{figure_FD_mod}
\end{figure*}

\begin{figure*}
\begin{tabularx}{\linewidth}{XX}
\includegraphics[trim=2cm 0.5cm 1.5cm 0.5cm, clip, width=\linewidth]{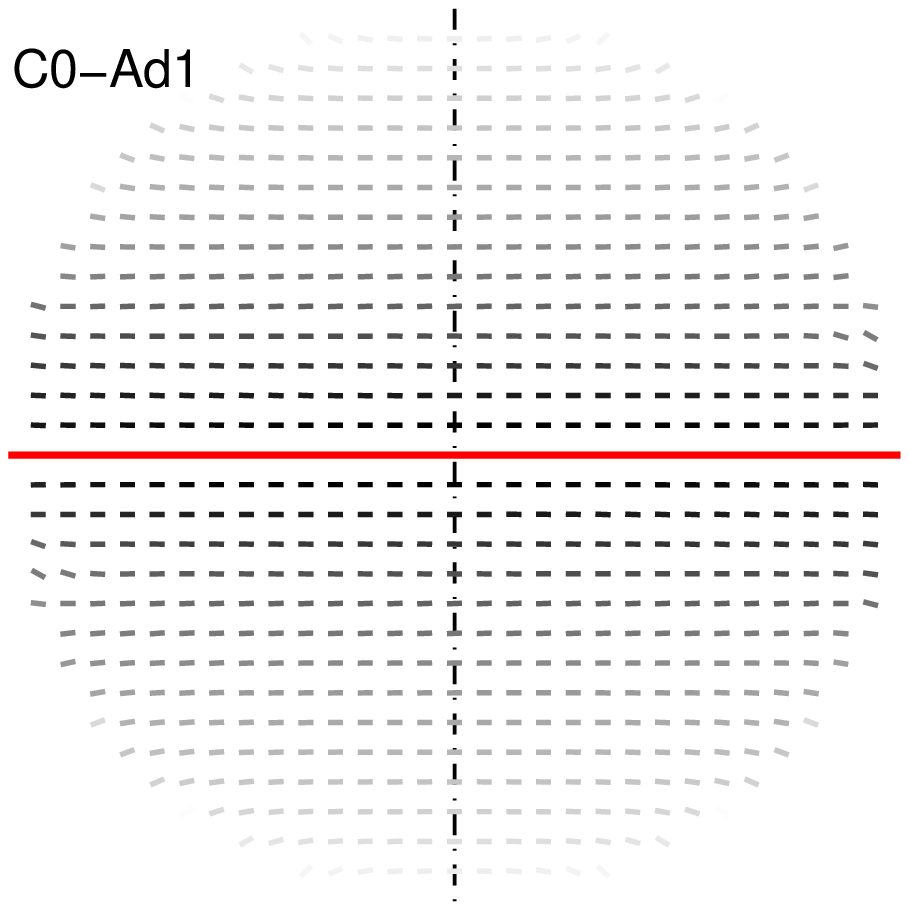} 
& 
\includegraphics[trim=2cm 0.5cm 1.5cm 0.5cm, clip, width=\linewidth]{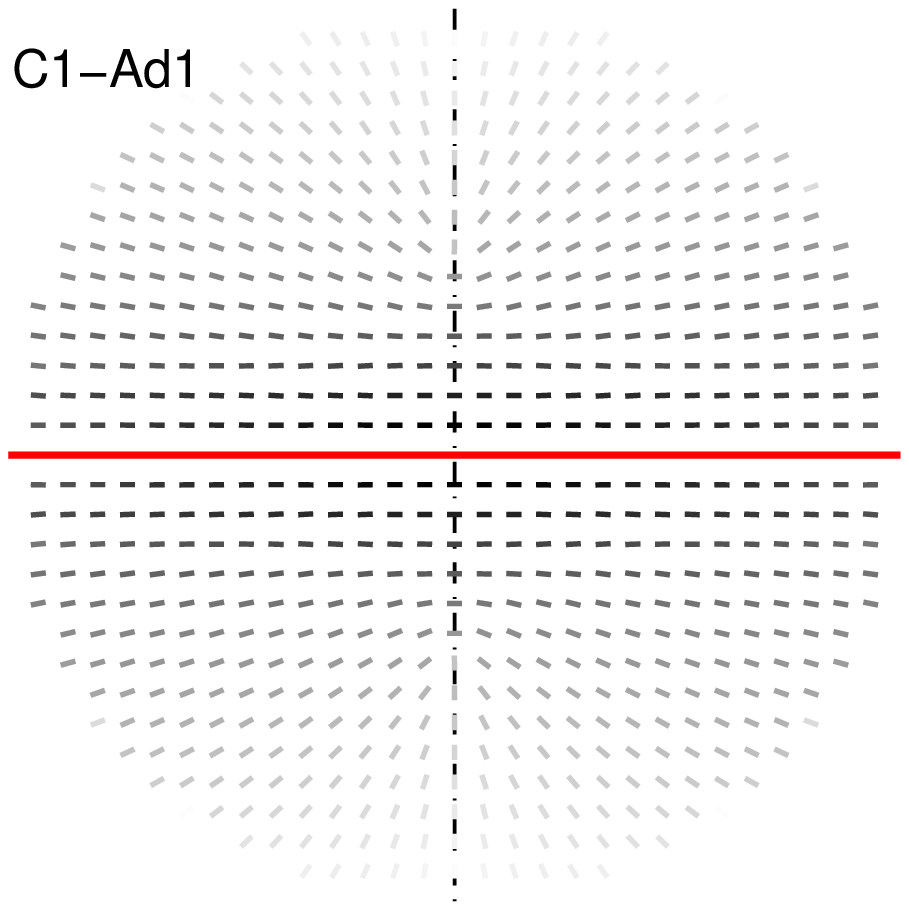} 
\\
\includegraphics[trim=2cm 0.5cm 1.5cm 0.5cm, clip, width=\linewidth]{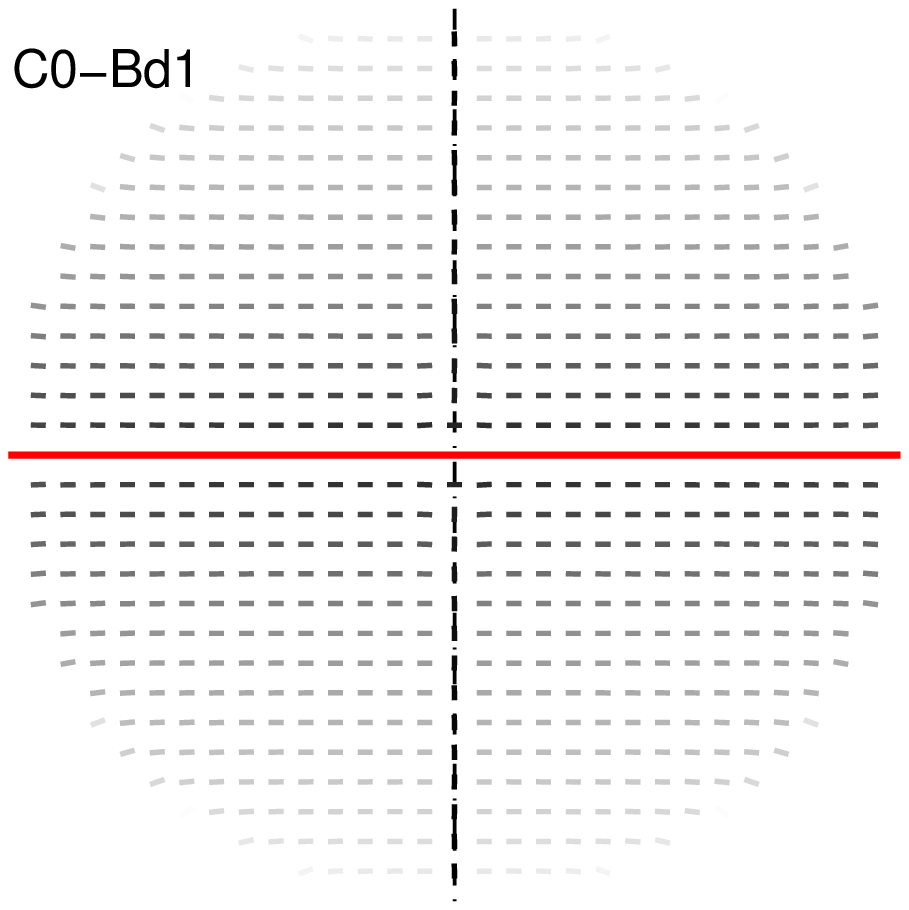} 
& 
\includegraphics[trim=2cm 0.5cm 1.5cm 0.5cm, clip, width=\linewidth]{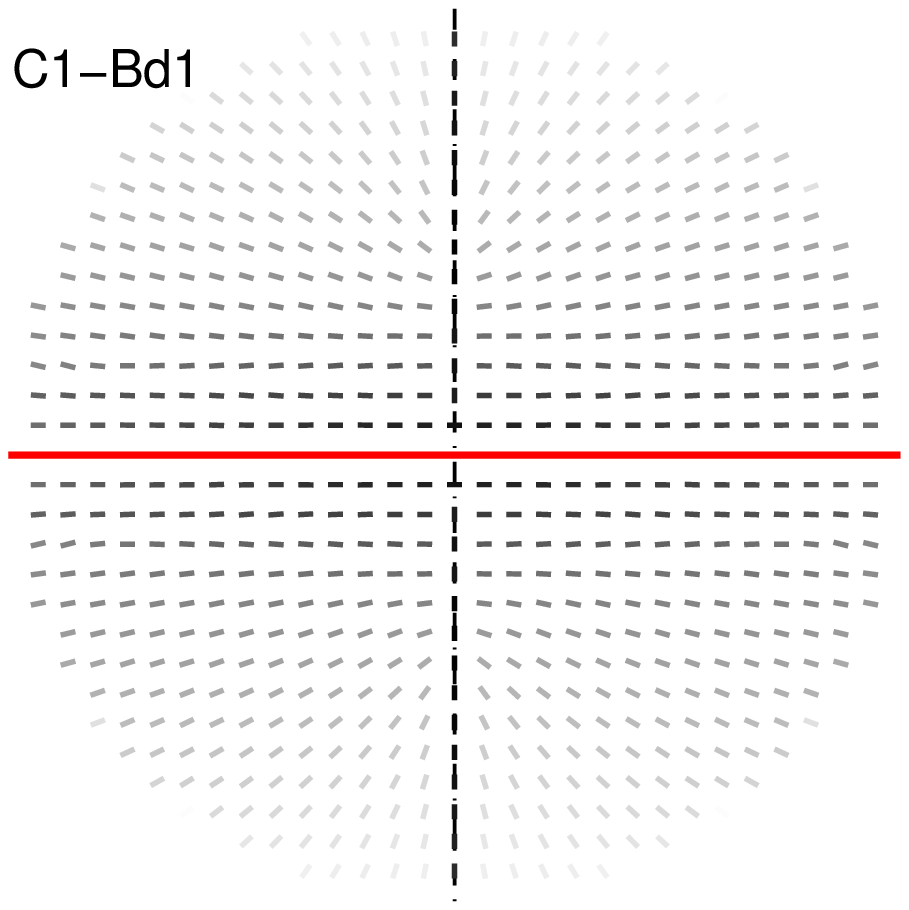} 
\\
\includegraphics[trim=2cm 0.5cm 1.5cm 0.5cm, clip, width=\linewidth]{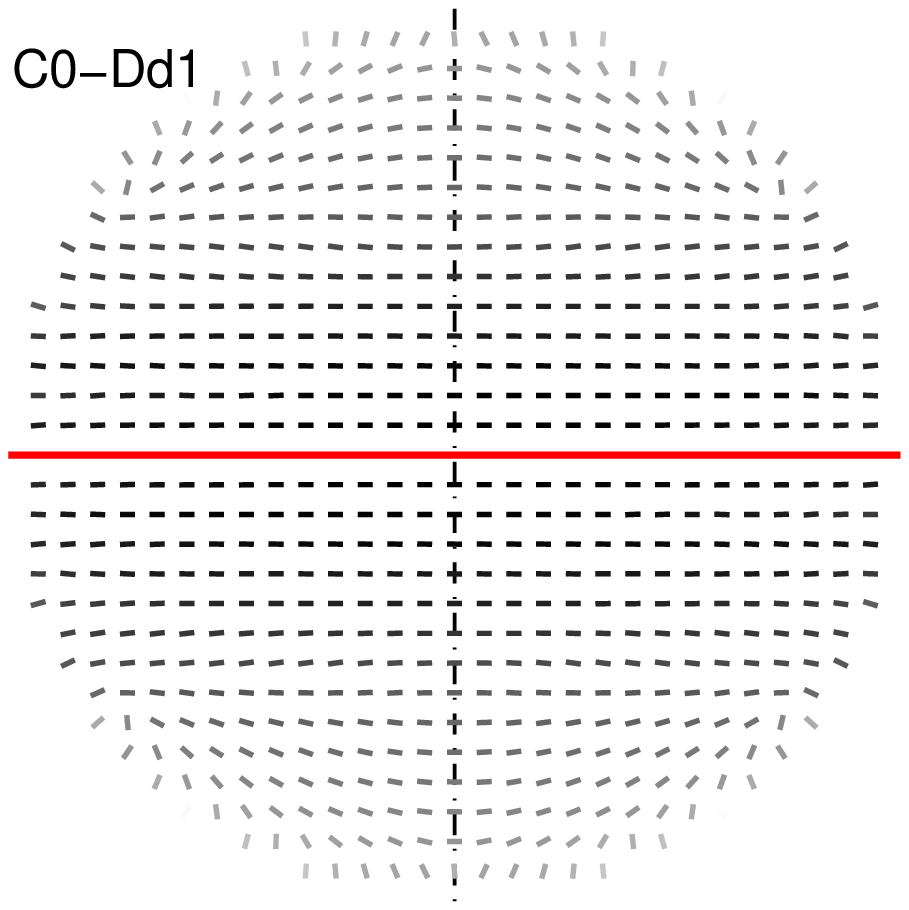} 
& 
\includegraphics[trim=2cm 0.5cm 1.5cm 0.5cm, clip, width=\linewidth]{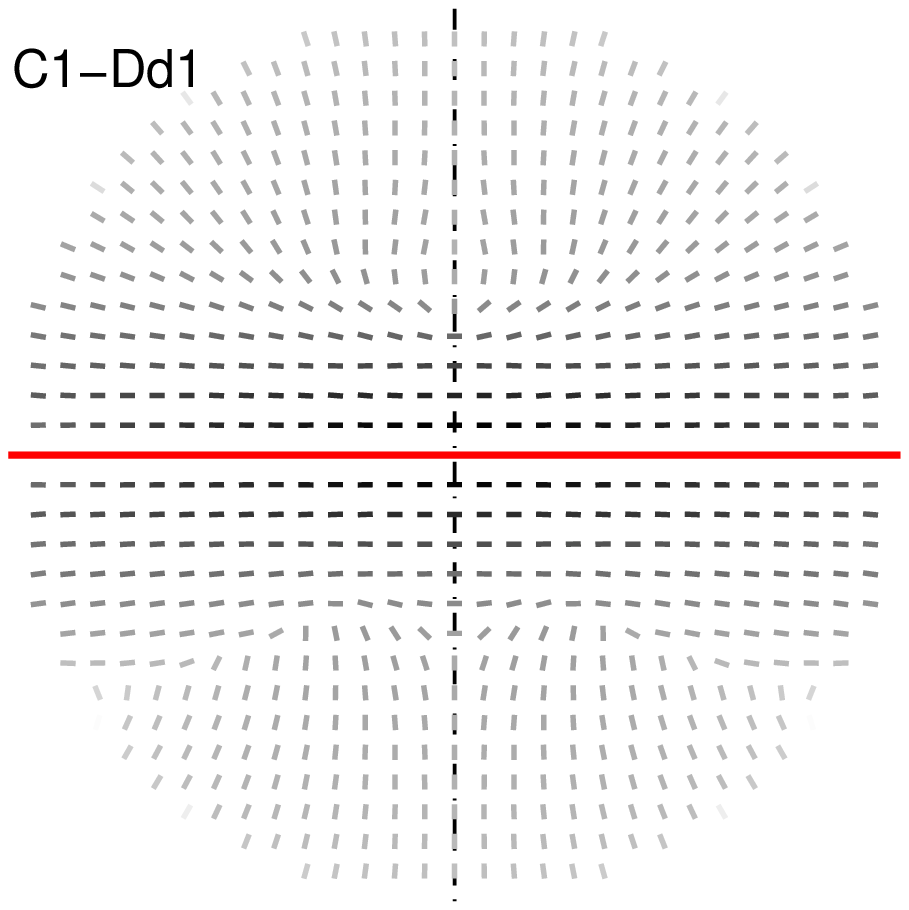} 
\end{tabularx}
\caption{
Synthetic maps of the magnetic orientation bars, inferred from the synchrotron 
polarized emission, of our Galaxy seen edge-on from the position 
$(r \to \infty, \varphi = 90^\circ, z = 0)$,
for the six best-fit magnetic field models listed in Table~\ref{table_bestfit}.
The left and right columns are for a halo field described by 
model~C0 (axisymmetric) and C1 (bisymmetric), respectively,
while the top, middle, and bottom rows are for a disk field 
described by model~Ad1, Bd1, and Dd1, respectively.
Each map covers a circular area of radius 15~kpc,
the trace of the Galactic plane is indicated by the horizontal, red, solid line,
and the rotation axis by the vertical, black, dot-dashed line.
All magnetic orientation bars have the same length, 
and their shade of grey scales logarithmically with the polarized intensity.
}
\label{figure_polarization}
\end{figure*}

Figure~\ref{figure_FD_mod} presents the all-sky $\overline{\rm FD}_{\rm mod}$ maps 
obtained with the best-fit parameter values
of the six magnetic field models listed in Table~\ref{table_bestfit}.
These modeled maps are to be compared with the observational map 
of $\overline{\rm FD}_{\rm obs}$ in the top panel of Fig.~\ref{figure_FD_ref}.

As expected, the $\overline{\rm FD}_{\rm mod}$ maps obtained with 
the axisymmetric halo-field model~C0 (left column in Fig.~\ref{figure_FD_mod}) 
are simpler and show less spatial structure than those obtained 
with the bisymmetric halo-field model~C1 (right column),
and for each halo-field model, the three disk-field models, 
Ad1 (top row), Bd1 (middle row), and Dd1 (bottom row), 
make very little difference, with variations noticeable only at low latitudes. 
Otherwise, all $\overline{\rm FD}_{\rm mod}$ maps reproduce 
the rough east-west antisymmetry as well as the rough north-south 
antisymmetry [symmetry] toward the inner [outer] Galaxy.
The east-west pattern is actually slightly shifted westward (by $\lesssim 10^\circ$), 
with the longitudes of sign reversals occurring not exactly 
at $\ell = 0^\circ, 180^\circ$, but at slightly smaller longitudes.
This westward shift, which is also apparent in the observational 
$\overline{\rm FD}_{\rm obs}$ map, is due to the slightly negative pitch angle.
Furthermore, when the halo field is bisymmetric, the westward shift is compounded 
by an east-west asymmetry arising from the azimuthal modulation 
of the halo field; this asymmetry is most striking between 
the (very dim) north-east and (less dim) north-west outer quadrants.

The $\overline{\rm FD}_{\rm mod}$ values in the north outer quadrants 
are systematically too small, particularly when the halo field is axisymmetric 
(left column in Fig.~\ref{figure_FD_mod}) and in the north-east outer quadrant 
when the halo field is bisymmetric (right column).
The problem arises because the contributions from the symmetric disk field 
and from the antisymmetric halo field partly cancel out in the north outer quadrants, 
while they add up constructively in the south outer quadrants.
Since the disk-field contribution must be dominant toward the outer Galaxy 
(see Sect.~\ref{FD_data_trends}), the problem can be alleviated 
either by enhancing the disk-field contribution 
or by reducing the halo-field contribution toward the outer Galaxy, 
while maintaining the dominance of the halo field toward the inner Galaxy.
Such tuning is more easily achieved with bisymmetric fields,
which possess the degree of freedom to orient their azimuthal pattern 
in favor of either the outer Galaxy (for the disk field) 
or the inner Galaxy (for the halo field).
This is the main reason why bisymmetric field models 
lead to lower $\chi_{\rm red,min}^2$ than their axisymmetric counterparts
(most strikingly for disk-field models, whose axisymmetric versions 
were dropped from the discussion earlier on)
and why the $\overline{\rm FD}_{\rm mod}$ maps obtained with model~C1 
are in slightly better visual agreement with the observational 
$\overline{\rm FD}_{\rm obs}$ map
than the $\overline{\rm FD}_{\rm mod}$ maps obtained with model~C0.

We are now in a position to better understand two of the more subtle results
presented in Sect.~\ref{results_bestfit}.
First, the possibility to orient the azimuthal pattern of the halo field 
in model~C1 in favor of the inner Galaxy explains why, 
when going from model~C0 to model~C1, $(B_1)_{\rm halo}$ can increase more 
than $(B_1)_{\rm disk}$ 
without the halo field imposing its north-south antisymmetry to the outer Galaxy.
Second, the lack of orientability for the halo field in model~C0 
implies that combinations with C0 (first three in Table~\ref{table_bestfit})
must rely more heavily on the disk field to favor the outer Galaxy; 
this is why they are found to have larger best-fit values for the parameters 
$H_{\rm disk}$ (in model~Bd1), $L_{\rm disk}$ (in model~Dd1), 
and $a_{\rm disk}$ (in model~Ad1) than combinations with C1
(last three in Table~\ref{table_bestfit}).

\subsection{\label{results_synchrotron}Synchrotron polarisation maps}

Having shown that the six best-fit magnetic field models 
listed in Table~\ref{table_bestfit} reproduce the observational 
$\overline{\rm FD}_{\rm obs}$ map reasonably well, 
we now use them to generate synthetic synchrotron polarization maps 
of our Galaxy as would be seen by an external edge-on observer.
The procedure is similar to that described in Paper~1: 
For each of our six best-fit magnetic field models, 
we compute the synchrotron emissivity,
${\cal E} \propto n_{\rm rel} \ B_\perp^{(\gamma + 1)/2}$, throughout the Galaxy,
where ${\boldvec B}_\perp$ is the magnetic field component perpendicular
to the line of sight,
$n_{\rm rel}$ is the density of relativistic electrons (assumed to be $\propto B^2$), 
$\gamma$ is the power-law index of the relativistic-electron energy spectrum 
(set to $\gamma = 3$), 
and the value of the proportionality factor is irrelevant.
We also compute the associated Stokes parameters $U$ and $Q$,
knowing that synchrotron emission is (partially) linearly polarized
perpendicular to ${\boldvec B}_\perp$ and assuming that the intrinsic degree
of linear polarization is uniform throughout the Galaxy.
We then integrate $U$ and $Q$ along a large number of sightlines 
covering the Galaxy to produce grids of polarized intensity
and polarization direction.
Finally, we rotate every polarization direction by $90^\circ$ 
to obtain the so-called magnetic orientation bar, i.e., the headless vector 
giving the line-of-sight--average orientation of ${\boldvec B}_\perp$,
and we plot it in a greyscale that varies logarithmically 
with the polarized intensity.
The resulting synchrotron polarization maps, which do not include 
any Faraday rotation, Faraday depolarization or beam depolarization effects,
are displayed in Fig.~\ref{figure_polarization}
for an external observer located at 
$(r_{\rm obs} \to \infty, \varphi_{\rm obs} = 90^\circ, z_{\rm obs} = 0)$.
Their variation with $\varphi_{\rm obs}$ is briefly discussed 
at the end of the section.

We first note that the polarization maps obtained with the bisymmetric 
halo-field model~C1 (right column in Fig.~\ref{figure_polarization})
are perfectly symmetric with respect to the rotation axis, 
while those obtained with the axisymmetric halo-field model~C0 (left column) 
are only approximately symmetric. 
Indeed, in the former case, the total field is the superposition 
of a bisymmetric halo field + a bisymmetric disk field, 
so the total field vector is antisymmetric with respect to the rotation axis 
and the magnetic orientation bars form a symmetric pattern.
In the latter case, the total field is the superposition
of an axisymmetric halo field + a bisymmetric disk field,
so the total field vector is neither perfectly symmetric 
nor perfectly antisymmetric with respect to the axis 
and the magnetic orientation bars do not form a perfectly symmetric pattern;
the reason why the polarization maps look approximately symmetric is because
(1) most regions in space are dominated by either the axisymmetric halo field 
or the bisymmetric disk field, 
and (2) asymmetries in the polarized intensity are toned down 
by the logarithmic greyscale.

When the halo field is axisymmetric (left column in Fig.~\ref{figure_polarization}),
the magnetic orientation bars remain nearly horizontal almost everywhere
(or at least up to $|z| \approx 8~{\rm kpc}$ in model~C0-Dd1; bottom-left panel)
and no X shape emerges. 
In contrast, when the halo field is bisymmetric (right column), 
the magnetic orientation bars are nearly horizontal only in a thick layer
along the Galactic plane ($|z| \lesssim (4-5)~{\rm kpc}$)
and they turn into a V shape (i.e., the upper or lower part of an X) 
on either side of this layer.
This split result is an immediate consequence of the dichotomy in the values 
of the scale height of the winding function, $H_p$, noted and explained 
in Sect.~\ref{results_bestfit}: 
with an axisymmetric halo field, $H_p$ is large ($\gtrsim 4~{\rm kpc}$)
and the field remains nearly azimuthal up to high $|z|$, 
leading to nearly horizontal magnetic bars;
with a bisymmetric halo field, $H_p$ is much smaller ($\simeq 1.2~{\rm kpc}$)
and the field turns poloidal at much lower $|z|$,
leading to an X-shape pattern.

Here, too, the impact of the disk-field model is much weaker 
than that of the halo-field model.
The most obvious, yet rather minor, difference between the three disk-field models 
is that the region of nearly horizontal magnetic bars is somewhat broader 
with models~Ad1 (top row in Fig.~\ref{figure_polarization}) 
and Bd1 (middle row) than with model~Dd1 (bottom row).
Another difference arises in model~Bd1:
the magnetic bars along the rotation axis are vertical, 
and fine-resolution maps show that they remain nearly vertical 
out to projected distances of a few 100~pc from the axis.
This is because model~Bd1 has a strong concentration of nearly vertical 
field lines around the axis (see Fig.~\ref{figure_fieldlines}b), 
which dominates the line-of-sight integration of the synchrotron emissivity.

The above conclusions are quite general, but the details of the polarization maps 
depend on the viewing angle, $\varphi_{\rm obs}$,
especially when the halo field is bisymmetric.
First, while the layer of nearly horizontal magnetic bars 
maintains a roughly constant thickness ($\simeq (4-5)~{\rm kpc}$), 
its grey-shade distribution, which reflects the polarized-intensity distribution, 
varies with $\varphi_{\rm obs}$.
This is easily understood if one remembers that the polarized intensity
depends sensitively on ${\boldvec B}_\perp$ 
near the crests of the azimuthal modulation.
Second, in the X-shape region, the opening angle of the upper and lower V 
that compose the X changes with $\varphi_{\rm obs}$, 
as expected for a projected bisymmetric structure;
in addition, there may be a range of $\varphi_{\rm obs}$ for which 
magnetic bars near the axis slope down toward the midplane 
instead of pointing outwards,
such that the upper and lower V actually turn into W.
The latter arises when the dominant portions of the crest regions 
occur in the near-east and far-west parts of the Galaxy,
where the Galactic differential rotation tilts the projected orientation 
of an initially X-shape poloidal field toward, and even past, the vertical.

In summary, amongst our six best-fit magnetic field models,
the three models with a bisymmetric halo field (model~C1; 
right column in Fig.~\ref{figure_polarization})
produce an X-shape pattern in synchrotron polarisation maps,
while those with an axisymmetric halo field (model~C0; left column) do not.

\subsection{\label{results_comparison}Comparison with previous work}

Other authors have recently set out to determine, or at least constrain, 
the magnetic field structure in the Galactic halo. 
We now briefly discuss their models, in comparison to ours.

\cite{sun&rwe_08} modeled the large-scale magnetic field of our Galaxy
as the superposition of a disk field and a halo field, which they constrained 
with an all-sky map of extragalactic RMs.
Their disk field was assumed to be purely horizontal,
following a logarithmic spiral (constant pitch angle), 
symmetric in $z$, and either axisymmetric with radial reversals or bisymmetric,
while their halo field was assumed to be purely azimuthal,
in the form of a torus on either side of the Galactic plane, 
antisymmetric in $z$, and axisymmetric.
Clearly, the purely azimuthal, axisymmetric halo field is automatically 
divergence-free, but the purely horizontal disk field is not -- 
for instance, in the axisymmetric case, 
the divergence-free condition implies $B_r \propto 1/r$,
which rules out radial reversals.
\cite{sun&rwe_08} found reasonably good fits to the RM data 
with axisymmetric disk fields having a pitch angle of $-12^\circ$ 
(somewhat larger in absolute value than in our best-fit models),
whereas they found no good fits with bisymmetric disk fields.
This result is at odds with our conclusion that bisymmetric disk fields 
provide better fits.
The discrepancy can be mostly traced back to \citeauthor{sun&rwe_08}'s 
neglect of the divergence-free condition,
which allows their axisymmetric disk field to undergo radial reversals
and, therefore, makes it easier to reproduce the RM data
(see our discussion in the third paragraph of Sect.~\ref{results_remarks}).

Another problem with \citeauthor{sun&rwe_08}'s (\citeyear{sun&rwe_08}) study 
is its reliance on the NE2001 model for the free-electron density, 
which underestimates the free-electron scale height, $H_{\rm e}$ 
(see Sect.~\ref{method_simul}).
An unrealistically strong halo field ($\approx 10~\mu{\rm G}$) is then necessary 
to account for the high-latitude extragalactic RMs, 
which, in turn, requires truncating the relativistic-electron distribution 
at $|z| \approx 1~{\rm kpc}$ to avoid excessive synchrotron emission from the halo.
The problem was fixed by \cite{sun&r_10}, with the adoption of 
\citeauthor{gaensler&mcm_08}'s (\citeyear{gaensler&mcm_08}) 
upward revision of $H_{\rm e}$;
this led to a more realistic halo field ($\approx 2~\mu{\rm G}$)
and obviated the need to artificially truncate the relativistic-electron distribution.
The revised model of \cite{sun&r_10} was then used by \cite{sun&r_12}
to simulate the synchrotron emission of spiral galaxies similar to our Galaxy
and study their polarization properties for various observing angles.

Later, \cite{sun&lgc15} noted that \citeauthor{sun&r_10}'s (\citeyear{sun&r_10}) 
model underpredicted the Galactic FD at latitude $b \gtrsim 50^\circ$.
They argued that the halo field needed to have a vertical component,
and they showed that the inclusion of a dipole field 
with $B_z = -0.2~\mu{\rm G}$ at the Sun
was enough to bring the predicted Galactic FD at $b \gtrsim 50^\circ$ 
up to the observed level.
However, they did not discuss how the Galactic FD at lower latitudes 
was affected, nor did they specify whether the global fit to the RM data
(measured through $\chi^2_{\rm red})$ was actually improved. 

The vertical component of the halo field in the solar neighborhood
has been the subject of a few other studies.
Based on the structure of their NVSS RM map at high $|b|$,
\cite{taylor&ss_09} estimated
$B_z \simeq -0.14~\mu{\rm G}$ above the midplane 
and $B_z \simeq 0.30~\mu{\rm G}$ below the midplane.
\cite{mao&ghz_10}, for their part, relied on a set of more than 1000 
WRST/ATCA extragalactic RMs at $|b| \ge 77^\circ$; 
after discarding outliers and anomalous RM regions, they derived
$B_z \simeq 0.00~\mu{\rm G}$ toward the north Galactic pole
(with a $3\sigma$ upper limit on $|B_z|$ of $0.07~\mu{\rm G}$)
and $B_z \simeq +0.31~\mu{\rm G}$ toward the south Galactic pole.
The south value is very close to that of \cite{taylor&ss_09},
but the north value is incompatible with those of \cite{taylor&ss_09} 
and \cite{sun&lgc15}.
Nevertheless, the results of both \cite{taylor&ss_09} and \cite{mao&ghz_10} 
are consistent with the notion that the large-scale Galactic magnetic field 
is the superposition of a symmetric disk field and an antisymmetric halo field, 
whose contributions to high-$|b|$ RMs cancel out above the midplane 
and add up below it. 
In \cite{taylor&ss_09} the symmetric contribution would dominate
(see footnote~\ref{foot_parity}), 
while in \cite{mao&ghz_10} both contibutions would be comparable.

For comparison, in each of our six best-fit models, the average $B_z$ 
toward either Galactic pole, as inferred from the ratio of Galactic FD 
to free-electron column density, is positive.
The contribution from the halo field is always positive and dominant.
The weaker contribution from the disk field is negative [positive] 
above [below] the midplane in models~Ad1 and Bd1, 
and vice-versa in model~Dd1.
This is a direct consequence of imposing $(B_r)_\odot < 0$
(see Sect.~\ref{method_simul}),
which, in view of the shape of poloidal field lines, implies
$(B_r)_{\rm disk} < 0$ and $(B_z)_{\rm disk} < 0$ [$(B_z)_{\rm disk} > 0$] 
in the northern [southern] halo in models~Ad1 and Bd1,
and the opposite in model~Dd1, where poloidal field lines 
reverse direction at $z = z_1$ (see Eq.~(\ref{eq_Br_D})).
As a result, the average $B_z$ is stronger toward the south Galactic pole
with models~Ad1 and Bd1
(as found by both \cite{taylor&ss_09} and \cite{mao&ghz_10})
and stronger toward the north Galactic pole with model~Dd1 
(in conflict with the measured high-$|b|$ RMs).
Hence, high-$|b|$ RMs give more credence to models~Ad1 and Bd1 than model~Dd1.

Additional constraints on the halo field were obtained by \cite{mao&mgb_12}, 
based on 641 extragalactic RMs at longitude $100^\circ < \ell < 117^\circ$ 
and latitude $|b| < 30^\circ$. 
The vast majority of the RMs are negative, as expected for a predominantly 
azimuthal, clockwise magnetic field.
The RM latitudinal distribution is approximately symmetric up to $|b| \simeq 15^\circ$, 
and it becomes increasingly asymmetric at higher $|b|$, 
with significantly larger $|{\rm RM}|$ in the southern hemisphere.
\cite{mao&mgb_12} noted that the RM distribution at $|b| \lesssim 15^\circ$
is consistent with a symmetric disk field in the Perseus arm.
They also showed that the RM distribution at $|b| \gtrsim 15^\circ$
could be reproduced with a purely azimuthal halo field confined to 
the radial range $[8.8,10.3]~{\rm kpc}$ (assuming $r_\odot = 8.4~{\rm kpc}$),
i.e., between the local and Perseus arms,
and to the vertical ranges $\pm [0.8,2.0]~{\rm kpc}$, 
with $B_\varphi = 2~\mu{\rm G}$ [$7~\mu{\rm G}$] above [below] the midplane.
Here, we see that the RM latitudinal distribution can also be explained 
by the superposition of a symmetric disk field and an antisymmetric halo field, 
such that the disk field contribution largely dominates at $|b| \lesssim 15^\circ$
and the halo field contribution becomes increasingly important at higher $|b|$,
opposing the disk field contribution in the northern hemisphere
and reinforcing it in the southern hemisphere.
In our study, the six best-fit models yield negative Galactic FD 
throughout the $17^\circ \times 60^\circ$ area surveyed by \cite{mao&mgb_12},
except for slightly positive values at $b \gtrsim 15^\circ$ with model~C1, 
and they reproduce the observed trends of RM versus $b$ quite well.

The first, and to our knowledge only, authors to include 
an X-shape component in their model of the large-scale Galactic magnetic field 
are \cite{jansson&f_12a,jansson&f_12b}.
The disk field in their model is symmetric in $z$, purely horizontal, 
and composed of a ring between radii 3~kpc and 5~kpc
plus 8 logarithmic spirals between 5~kpc and 20~kpc,
with a common pitch angle of $-11.5^\circ$ and separate field strengths 
constrained by global magnetic flux conservation.
The halo field has two axisymmetric components:
a purely azimuthal field with opposite signs and different strengths 
on both sides of the midplane
(similar to the halo field of \cite{sun&rwe_08, sun&r_10}, 
but more general because not perfectly antisymmetric in $z$),
and a purely poloidal field with an X shape (similar to the poloidal field 
in our model~C, but with straight field lines and a cusp at $z=0$).
These two components are unrelated, in contradiction with dynamo theory 
which predicts that azimuthal field is generated from poloidal field and vice-versa.
\cite{jansson&f_12a,jansson&f_12b} found that the inclusion of an X-shape component 
improved the global fit to their data set, which contained the WMAP7 Galactic 
synchrotron emission map together with a collection of more than 40\,000 
extragalactic RMs.
The improvement brought about by an X-shape poloidal field is also obvious 
in our study, where the pitch angle is found to have non-zero values 
to a high confidence level.

\section{\label{discussion}Discussion}

In this paper, we took a first important step in our efforts to understand 
the structure of the large-scale magnetic field in the Galactic halo,
with special attention to the possibility of uncovering an X-shape magnetic 
configuration, as observed in external edge-on spiral galaxies.
We applied the analytical magnetic field models developed in Paper~1
to the disk and halo of our Galaxy, on the basis that these models 
can describe a broad range of spiraling, possibly X-shape magnetic fields,
including purely horizontal and purely vertical fields as limiting cases.
We considered 35 models of the total (halo + disk) magnetic field,
each composed of one of our 7 antisymmetric halo-field models,
A1, B0, B1, C0, C1, D0, and D1, plus one of our 5 symmetric disk-field models,
Ad1, Bd0, Bd1, Dd0, and Dd1 (where 0 and 1 denote axisymmetric and bisymmetric
fields, respectively).
For each total-field model, we computed the average Galactic Faraday depth 
in the 356 bins covering the sky area at $|b| \ge 10^\circ$,
and we confronted the resulting modeled map of $\overline{\rm FD}_{\rm mod}$ 
to the observational map of $\overline{\rm FD}_{\rm obs}$ 
displayed in the top panel of Fig.~\ref{figure_FD_ref} 
(based on the reconstructed Galactic FD map of \cite{oppermann&jge_15}, 
from which the contribution from \citeauthor{wolleben&flc_10}'s
(\citeyear{wolleben&flc_10}) magnetized bubble was advantageously removed).
The adjustment of $\overline{\rm FD}_{\rm mod}$ to $\overline{\rm FD}_{\rm obs}$ 
was carried out through standard $\chi^2$ minimization, 
with the help of MCMC simulations. 

We found that the 6 total-field models composed of C0 or C1 (for the halo field)
plus Ad1, Bd1, or Dd1 (for the disk field) (listed in Table~\ref{table_bestfit}) 
had significantly lower $\chi_{\rm red,min}^2$ than any of the other total-field models.
Amongst them, models with C1 have slightly lower $\chi_{\rm red,min}^2$ 
($2.02 - 2.08$) than models with C0 ($2.26 - 2.42$),
and they provide a slightly better visual match to the observational 
$\overline{\rm FD}_{\rm obs}$ map (see Fig.~\ref{figure_FD_mod}).
Neither the value of $\chi_{\rm red,min}^2$ nor the appearance 
of the $\overline{\rm FD}_{\rm mod}$ map is significantly affected 
by the disk-field model.
The only discriminating factor comes from high-$|b|$ RMs,
which tend to favor models~Ad1 and Bd1 over model~Dd1.

Thus, regardless of the bisymmetric disk field, the three models 
with a bisymmetric halo field perform slightly (but systematically) better 
with regard to the Galactic FD than the three models with an axisymmetric 
halo field.
As it turns out, the former also produce an X-shape pattern 
in synchrotron polarisation maps, while the latter lead to nearly horizontal 
magnetic orientation bars throughout most of the maps 
(see Fig.~\ref{figure_polarization}).
As explained in Sects.~\ref{results_bestfit} and \ref{results_synchrotron},
this difference in the polarisation maps stems from the rough east-west 
antisymmetry in the $\overline{\rm FD}_{\rm obs}$ sky,
which has strong implications for the azimuthal-to-poloidal field ratio:
in the axisymmetric case, the field must necessarily be nearly azimuthal 
up to high $|z|$ -- hence the nearly horizontal magnetic orientation bars, 
whereas in the bisymmetric case, the field can turn poloidal 
at relatively low $|z|$ -- hence the X-shape pattern.

In conclusion, the existing RM data, interpreted with the help 
of our (hopefully sufficiently general) magnetic field models, 
suggest that the Galactic halo is slightly more likely 
to have a bisymmetric field than an axisymmetric field.
If the halo field is indeed bisymmetric, it would probably be seen 
as X-shaped by an external edge-on observer, while it would probably 
be seen as nearly horizontal if it is instead axisymmetric.
We emphasize that the preference found here for a bisymmetric, 
X-shape halo field cannot be regarded as definite, 
first because it is based solely on RM data 
and second because the bisymmetric halo field performs only 
slightly better than its axisymmetric counterpart.

The results obtained for the disk field -- in particular, 
the finding that the disk field is more likely to be bisymmetric -- 
are even more subject to caution.
Indeed, since our original interest was in the halo field, 
we excluded all sightlines toward the disk ($|b| < 10^\circ$), 
and the main reason why we needed to model the disk field 
was because even sightlines toward the halo must first pass through the disk.
By essence, these sightlines miss a sizeable fraction of the disk field,
so our analysis can only give partial information on the global properties 
of the disk field, such as its axisymmetric versus bisymmetric status.

A severe limitation of the RM data arises from the relatively rapid fall-off 
of the free-electron density with height, which makes it hard to sample 
the Galactic magnetic field at large distances from the Galactic plane.
Therefore, any method relying exclusively on RM data is not particularly 
well suited to look into the structure of the magnetic field in the Galactic halo.
On the other hand, RM data have the unique advantage of containing 
a sign information, which is required to distinguish between symmetric 
and antisymmetric models or between axisymmetric and bisymmetric models.
Ultimately, RM data provide a unique, albeit limited, set of constraints 
on the properties of the large-scale magnetic field in the Galactic halo.
Obtaining a more complete set of constraints will require fitting 
our magnetic field models to other types of observations, such as
synchrotron total and polarized emission. 

Despite the inherent limitation of the RM data to explore the magnetic field 
structure in the Galactic halo, our investigation 
offers a first important glimpse.
In addition, our search for a good fit to the observational 
$\overline{\rm FD}_{\rm obs}$ map led us to improve and refine 
the analytical magnetic field models of Paper~1. 
Most notably, we derived a more realistic winding function 
(Eq.~\ref{eq_shiftedwindingfc}), applicable to all our field models,
and we regularized the bisymmetric version of model~A by assuming 
a straight, horizontal magnetic field inside a vertical cylinder of radius $r_1$ 
centered on the rotation axis (see Eq.~(\ref{eq_Br_A_in}) for halo fields 
and Eq.~(\ref{eq_Br_Ad_in}) for disk fields).
Finally, the method proposed in this paper is interesting in its own right:
it contains a number of original features, including a detailed error estimation
(see Appendix~\ref{appendix_uncertainty}),
which can be retained for similar investigations.
As a possible application, we suggest using our method
to study the magnetic field structure in targeted regions of the sky.
We also suggest repeating the present analysis
when new RM data in the southern hemisphere become available to fill in 
the gap at $\delta < -40^\circ$ in \citeauthor{taylor&ss_09}'s 
(\citeyear{taylor&ss_09}) NVSS catalog,\footnote{
Niels Oppermann (private communication) did a FD reconstruction 
using a preliminary version of Dominic Schnitzeler's new data from S-PASS 
follow-up observations, and he found no major changes on large scales.
}
and when the nearby objects that significantly perturb the FD sky
have been identified, and their FD quantified, through RM synthesis
\citep[see][]{wolleben&flc_10},
such that a more complete and cleaner all-sky $\overline{\rm FD}_{\rm obs}$ map 
can be produced.

\appendix
\section{\label{appendix_uncertainty}Uncertainty estimation}

\begin{figure*}[t]
\centering
\resizebox{0.8\hsize}{!}{\includegraphics{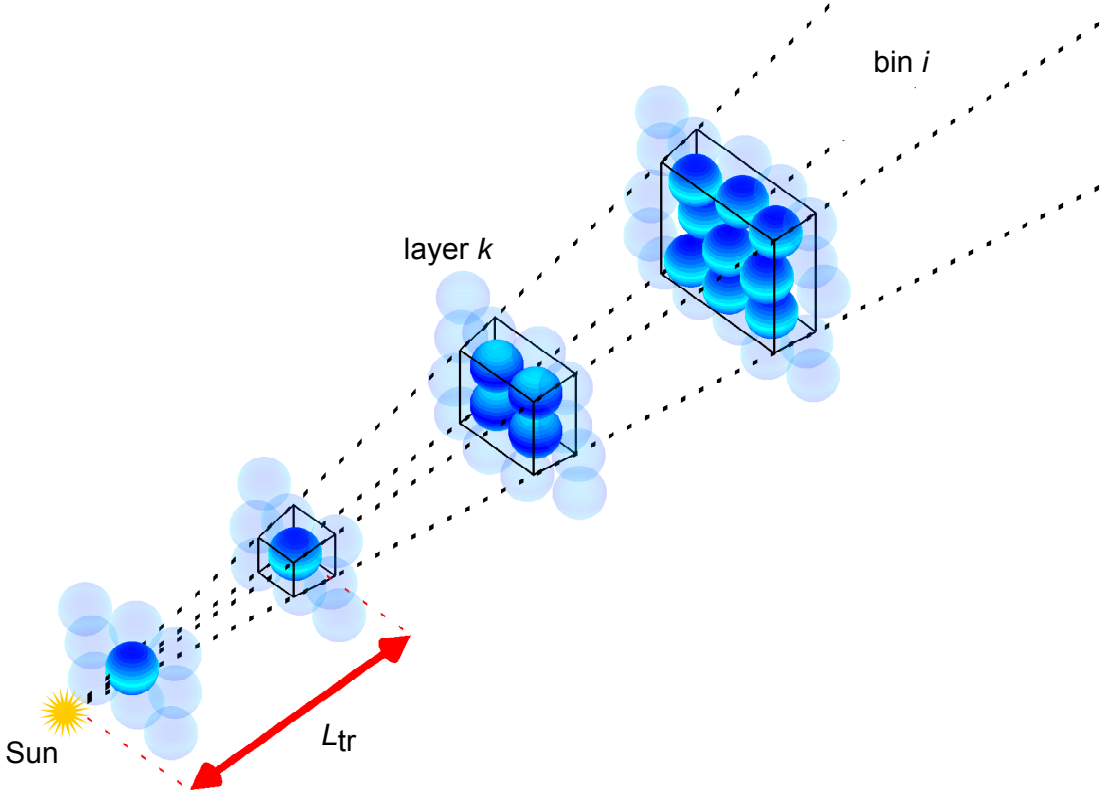}}
\caption{Schematic drawing illustrating the concept of turbulent cells and layers 
used in our estimation of $\sigma_{{\rm turb},i}$, the uncertainty 
in $\overline{\rm FD}_{{\rm obs},i}$ arising from turbulence 
in the magneto-ionic ISM, for a particular bin $i$ 
(see Eq.~(\ref{eq_sigma_turb}) and attendant discussion).
The Galactic volume subtended by bin~$i$ (delimited by the four black dotted lines
originating from the Sun) is divided into transverse layers of thickness $l$ 
(delimited by black solid lines).
These layers contain an increasing number of turbulent cells of size $l$ 
(represented by opaque blue spheres, while transparent blue spheres represent cells 
falling outside the bin).
The distance $L_{\rm tr}$ from the Sun marks the transition between the so-called
nearby layers, which can be completely filled with a single cell, 
and the distant layers, which enclose several cells.
}
\label{figure_turb_cells}
\end{figure*}

\begin{figure*}[t]
\centering
\resizebox{0.8\hsize}{!}{\includegraphics{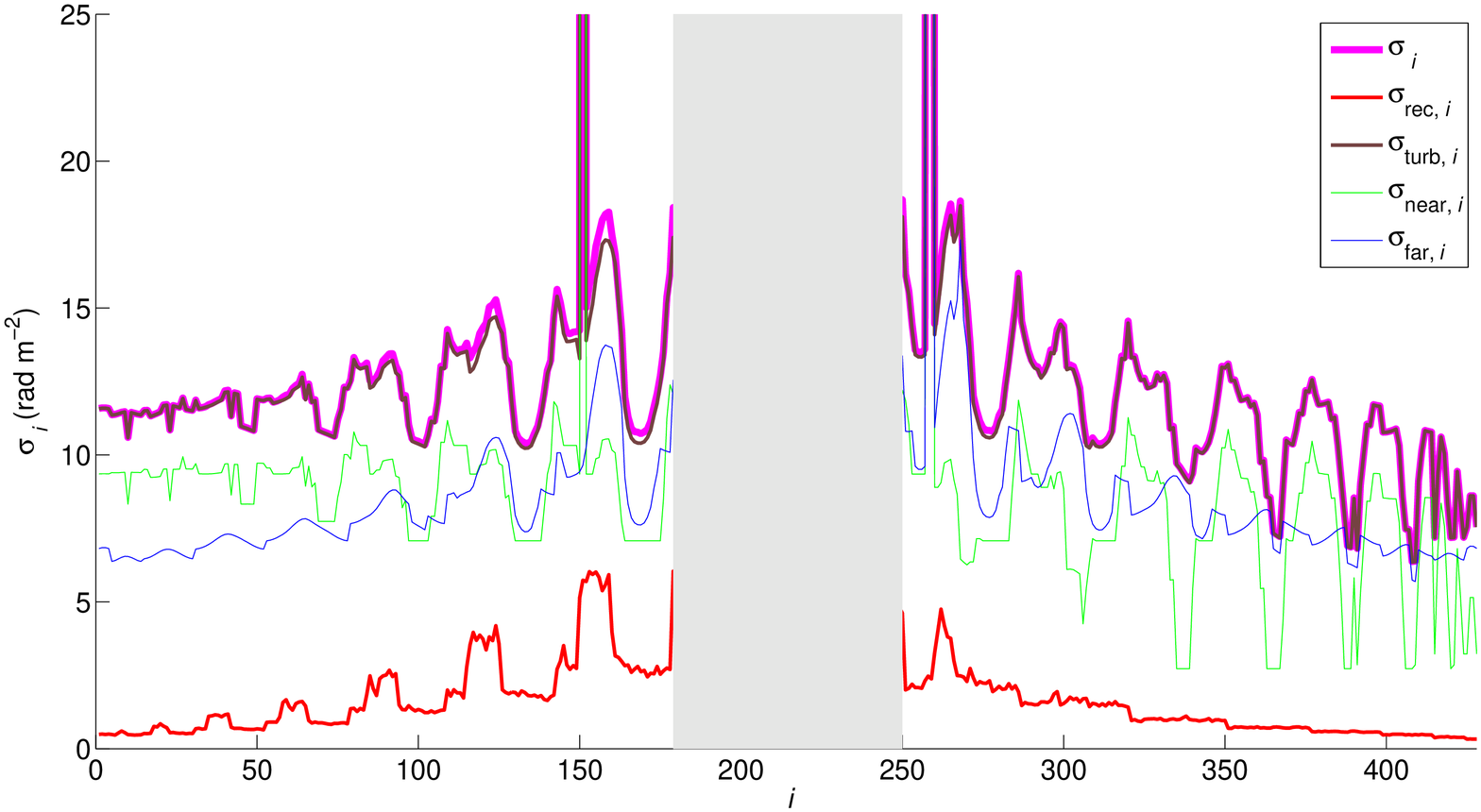}}
\caption{Total uncertainty in $\overline{\rm FD}_{{\rm obs},i}$, $\sigma_i$,
as a function of bin identifier, $i$ (thick magenta line), 
for the 356 bins with latitude $|b| \ge 10^\circ$.
Also plotted are the two contributions to $\sigma_i$ (see Eq.~(\ref{eq_sigma_i})),
which arise from the reconstruction of \cite{oppermann&jge_15} 
($\sigma_{{\rm rec},i}$; red line) and from turbulence in the magneto-ionic ISM 
($\sigma_{{\rm turb},i}$; brown line), respectively.
The latter, in turn, is accompanied by its two contributions
(see Eq.~(\ref{eq_sigma_turb_2terms})), 
which arise from the ${N'}_{\rm \!\!near}$ nearest layers (thin green line) 
and from the $N_{{\rm layer},i} - {N'}_{\rm \!\!near}$ more distant layers 
(thin blue line), respectively.
Bins are ordered by bands of increasing latitude, from $b = -90^\circ$ 
to $90^\circ$, and in each latitude band, by increasing longitude,
from $\ell = -180^\circ$ to $180^\circ$.
The vertical grey band in the middle corresponds to bins with latitude 
$|b| < 10^\circ$, which are excluded from our study.
}
\label{figure_uncertainty}
\end{figure*}

When using the $\overline{\rm FD}_{\rm obs}$ map in the top panel 
of Fig.~\ref{figure_FD_ref} as our observational reference 
to model the large-scale magnetic field in the Galactic halo, 
we have to take two main sources of uncertainty into account:
(1) the uncertainty in the reconstructed Galactic FD map of \cite{oppermann&jge_15},
which includes both the measurement errors in the RM data
and an extragalactic RM contribution,
and (2) the uncertainty due to turbulent fluctuations in the magneto-ionic ISM.
If, for each bin~$i$, we denote the corresponding uncertainties 
in $\overline{\rm FD}_{{\rm obs},i}$ by $\sigma_{{\rm rec},i}$ 
and $\sigma_{{\rm turb},i}$, respectively,
and if we note that the two uncertainties add up quadratically,
we can write for the total uncertainty in $\overline{\rm FD}_{{\rm obs},i}$, 
$\sigma_i$:
\begin{equation}
\sigma^2_i = \sigma^2_{{\rm rec},i} + \sigma^2_{{\rm turb},i} \ \cdot
\label{eq_sigma_i}
\end{equation}
We now successively derive the appropriate expressions 
of $\sigma^2_{{\rm rec},i}$ and $\sigma^2_{{\rm turb},i}$.

The reconstructed Galactic FD map of \cite{oppermann&jge_15}
provides, for each pixel $j$, a value of the observational FD,
${\rm FD}_{{\rm obs},j}$,
and the accompanying uncertainty map provides the associated uncertainty,
$\sigma_{{\rm pixel},j}$.
When the ${\rm FD}_{{\rm obs},j}$ are averaged over bin~$i$,
the individual uncertainties $\sigma_{{\rm pixel},j}$ partially cancel out
(as in a random walk), such that the uncertainty squared in the average value 
$\overline{\rm FD}_{{\rm obs},i}$ is equal to the mean individual uncertainty squared
divided by the number $N_{{\rm data},i}$ of independent data points in bin~$i$.
Note that $N_{{\rm data},i}$ is generally less than the number $N_{{\rm pixel},i}$
of pixels in bin~$i$.
With 41\,632 extragalactic RM data points and 196\,608 pixels distributed
over 428 bins, we find that on average $N_{{\rm data},i} = 97$ 
and $N_{{\rm pixel},i} = 459$.
Altogether, the uncertainty in $\overline{\rm FD}_{{\rm obs},i}$
associated with the reconstruction of \cite{oppermann&jge_15}
is given by
\begin{equation}
\label{eq_sigma_rec}
\sigma^2_{{\rm rec},i} =
\frac{1}{N_{{\rm data},i}} \
\left( \frac{1}{N_{{\rm pixel},i}} \ \sum_{j=1}^{N_{{\rm pixel},i}} 
\sigma^2_{{\rm pixel},j}
\right) \ \cdot
\end{equation}

To estimate the turbulent contribution to $\sigma^2_i$,
we adopt a very crude model of interstellar turbulence, 
which we assume consists of turbulent cells,
all of which have the same size $l$,
cause a magnetic field perturbation $\delta {\boldvec B}$
with constant strength and random direction,
and produce no perturbation in the free-electron density.
The FD associated with an individual turbulent cell can then be written as
\begin{equation}
\label{eq_sigma_cell}
\sigma_{\rm cell} = n_{\rm e} \ \delta B_{\parallel} \ l \ ,
\end{equation}
where $n_{\rm e}$ is the background free-electron density at the cell's location 
(given by \citeauthor{schnitzeler_12}'s (\citeyear{schnitzeler_12}) 
updated version of the NE2001 model, as described in Sect.~\ref{method_simul}) 
and $\delta B_{\parallel}$ is the r.m.s. line-of-sight component of 
$\delta {\boldvec B}$.
For an isotropic turbulent magnetic field, 
$\delta B_{\parallel} = \delta B / \sqrt{3}$.
Here, we adopt $\delta B = 5~\mu{\rm G}$ \citep{rand&k_89, ohno&s_93}
and $l = 100~{\rm pc}$ \citep[][and references therein]{haverkorn&s_13}.

To properly account for the line-of-sight variations of the number and FD 
of turbulent cells contributing to $\sigma^2_{{\rm turb},i}$,
we divide the Galactic volume subtended by bin~$i$ into $N_{{\rm layer},i}$ 
successive layers of thickness $l$.
For future reference, if we denote by $L_i$ the effective path length 
from the observer to the edge of the Galaxy in the direction of bin~$i$, 
we simply have $N_{{\rm layer},i} = L_i / l$.
We identify each layer by its running number $k$ 
(starting from $k=1$ in the closest layer),
and we denote by $N_{{\rm cell},ik}$ the statistical number of turbulent cells 
in layer~$k$ of bin~$i$.
As illustrated in Fig.~\ref{figure_turb_cells}, 
there is an important difference between distant and nearby layers.
A distant layer of bin~$i$ encloses several cells, whose contributions 
to $\overline{\rm FD}_{{\rm obs},i}$ tend to statistically average out 
over the layer.
In contrast, a nearby layer of bin~$i$ can be completely filled 
with a single cell, which then gives a coherent contribution 
(with no averaging-out) across the layer.
Since all the bins have an angular size $\theta_{\rm bin} \simeq 10^\circ$
(see Sect.~\ref{FD_data_maps}), the transition from a single cell to several cells 
per layer occurs at a line-of-sight distance 
$L_{\rm tr} = l / \theta_{\rm bin} \simeq 570~{\rm pc}$, 
i.e., after a number of layers 
$N_{\rm near} = L_{\rm tr} / l = 1 / \theta_{\rm bin} \simeq 5.7$.
Thus, for $k \le N_{\rm near}$, $N_{{\rm cell},ik} = 1$,
while for $k \ge N_{\rm near}$, $N_{{\rm cell},ik} 
= (\theta_{\rm bin} \, k)^2 = (k / N_{\rm near})^2$.

The contribution from layer~$k$ of bin~$i$ to $\sigma^2_{{\rm turb},i}$
is equal to the mean FD squared of the enclosed turbulent cells 
divided by the number $N_{{\rm cell},ik}$ of these cells:
\begin{equation}
\label{eq_sigma_layer}
\sigma^2_{{\rm layer},ik} = 
\frac{1}{N_{{\rm cell},ik}} \ 
\left( \frac{1}{N_{{\rm cell},ik}} \ \sum_{N_{{\rm cell},ik}} \sigma^2_{\rm cell} 
\right) \ \cdot
\end{equation}
Finally, the uncertainty in $\overline{\rm FD}_{{\rm obs},i}$ due to turbulent
fluctuations in the magneto-ionic ISM is the quadratic sum of the contributions 
from the $N_{{\rm layer},i}$ layers of bin~$i$:
\begin{equation}
\label{eq_sigma_turb}
\sigma^2_{{\rm turb},i} = 
\sum_{k=1}^{N_{{\rm layer},i}} \sigma^2_{{\rm layer},ik} \ \cdot
\end{equation}

Let us emphasize that Eq.~(\ref{eq_sigma_turb}) provides only a very rough 
expression for $\sigma_{{\rm turb},i}$, which ignores spatial variations 
in the turbulence parameters (e.g., in the size and field strength 
of turbulent cells) as well as correlations between them 
(e.g., between fluctuations in field strength and in free-electron density).
In that respect, we note that spatial variations in the turbulence parameters 
are partly washed out when summing over all the layers along the line of sight
to obtain $\sigma^2_{{\rm turb},i}$ (Eq.~\ref{eq_sigma_turb}) 
and when summing over all the bins to obtain $\chi^2$ (Eq.~\ref{eq_chi}).
There might also be some cancellation, for instance, between $\delta B$ being larger 
and $l$ being smaller in spiral arms than in interarm regions
\cite[see, e.g.,][respectively]{beck_09, haverkorn&bgm_08}
and between $\delta B$ decreasing (together with $B$) and $l$ increasing
with increasing $|z|$ (Ann Mao, private communication).
Incidentally, the variations of $\delta B$ and $l$ with $|z|$ are probably 
not too critical, as they are weighted down by a decreasing $n_{\rm e}$.

Let us also remark that $\sigma_{{\rm turb},i}$ is sensitive to the poorly 
known values of $l$ and $\delta B$.
To quantify this sensitivity, we write out the full expression 
of $\sigma^2_{{\rm turb},i}$ by inserting Eq~(\ref{eq_sigma_cell}) 
into Eq.~(\ref{eq_sigma_layer}) and Eq.~(\ref{eq_sigma_layer}) 
into Eq.~(\ref{eq_sigma_turb}).
If we denote by $(n_{\rm e}^2)_{ik}$ the mean value of $n_{\rm e}^2$
over layer~$k$ of bin~$i$ and by ${N'}_{\rm \!\!near}$ the closest integer 
below $N_{\rm near}$ (defined above Eq.~(\ref{eq_sigma_layer})),
we readily obtain
\begin{equation}
\label{eq_sigma_turb_2terms}
\sigma^2_{{\rm turb},i} = 
\frac{1}{3} \ \delta B^2 \ l^2 \
\left( \sum_{k=1}^{{N'}_{\rm \!\!near}} (n_{\rm e}^2)_{ik}
+ N_{\rm near}^2 \sum_{k={N'}_{\rm \!\!near}+1}^{N_{{\rm layer},i}}
\frac{(n_{\rm e}^2)_{ik}}{k^2} 
\right) \ ,
\end{equation}
where the first term represents the contribution from the ${N'}_{\rm \!\!near}$ 
nearest layers, which contain a single cell ($N_{{\rm cell},ik} = 1$),
and the second term represents the contribution from 
the $N_{{\rm layer},i} - {N'}_{\rm \!\!near}$ more distant layers,
which contain more than one cell ($N_{{\rm cell},ik} = (k / N_{\rm near})^2$).
It emerges from Eq.~(\ref{eq_sigma_turb_2terms}) that $\sigma_{{\rm turb},i}$ 
varies linearly with $\delta B$.
This is because $\sigma_{{\rm turb},i}$ depends on $\delta B$ 
only through the FD of individual turbulent cells, $\sigma_{\rm cell}$, 
which is a linear function of $\delta B$ (see Eq.~(\ref{eq_sigma_cell})).
Variations with $l$ are a little more subtle: 
in addition to an explicit linear variation similar to that found for $\delta B$,
there is a more complex implicit variation through 
$(n_{\rm e}^2)_{ik} = (n_{\rm e}^2)_i (s\!=\!k\,l)$
and $N_{{\rm layer},i} = L_i / l$.
The former arises through $\sigma_{\rm cell}$, which is a linear function of $l$ 
(see Eq.~(\ref{eq_sigma_cell})),
while the latter results from the way the individual $\sigma_{\rm cell}$
combine together to produce a net $\sigma_{{\rm turb},i}$.
As we now show, the implicit variation with $l$ is actually quite weak.
Indeed, our chosen parameter values pertain to the limit 
$L_i \gg L_{\rm tr}$ or, equivalently, $N_{{\rm layer},i} \gg N_{\rm near}$,
which has two important consequences.
First, the exact value of $N_{{\rm layer},i}$, which enters 
Eq.~(\ref{eq_sigma_turb_2terms}) only as the upper bound of a rapidly 
converging series, has hardly any impact. 
Physically, only the ${N'}_{\rm \!\!near}$ nearest layers, with a single cell,
and the next few layers, with a small number of cells,
contribute significantly to $\sigma_{{\rm turb},i}$;
more distant layers have their contributions increasingly reduced 
by averaging over an increasing number of cells
(prefactor $1/N_{{\rm cell},ik}$ in Eq.~(\ref{eq_sigma_layer}),
leading to the factor $1/k^2$ in the second term of Eq.~(\ref{eq_sigma_turb_2terms})).
Second, the value of $(n_{\rm e}^2)_{ik}$ in the first few,
significantly-contributing layers is not very different from 
the local value of $n_{\rm e}^2$, and hence not very sensitive to $l$.
Altogether, $\sigma_{{\rm turb},i}$ varies a little less than linearly with $l$.

Plotted in Fig.~\ref{figure_uncertainty} is the total uncertainty 
in $\overline{\rm FD}_{{\rm obs},i}$, $\sigma_i$ (thick magenta line),
for the 356 bins with latitude $|b| \ge 10^\circ$,
together with the contributions from the reconstruction 
of \cite{oppermann&jge_15}, $\sigma_{{\rm rec},i}$ (red line),
and from turbulent fluctuations, $\sigma_{{\rm turb},i}$ (brown line).
Also shown are the contributions to $\sigma_{{\rm turb},i}$
from the ${N'}_{\rm \!\!near}$ nearest layers, which contain a single turbulent cell 
(thin green line), and from the $N_{{\rm layer},i} - {N'}_{\rm \!\!near}$ 
more distant layers, which contain more than one cell (thin blue line).
It appears that, for our choice of parameter values, $\sigma_{{\rm turb},i}$ 
is typically one order of magnitude larger than $\sigma_{{\rm rec},i}$
and $\sigma_{{\rm turb},i}$ generally has comparable contributions
from nearby and distant layers.
There is a general tendency for $\sigma_{{\rm rec},i}$ to decrease
with increasing $|b|$ (toward the ends of the $x$-axis)
and a weaker tendency for $\sigma_{{\rm turb},i}$ to do so above the midplane 
(in the right half of the figure).
In addition, both $\sigma_{{\rm rec},i}$ and $\sigma_{{\rm turb},i}$ 
undergo modulations with longitude, $\ell$.
The modulation of $\sigma_{{\rm rec},i}$, mostly apparent below the midplane 
(in the left half of the figure),
arises from a lack of data points with declination $\delta < -40^\circ$.
In contrast, the periodic fluctuations of $\sigma_{{\rm turb},i}$,
visible at all latitudes, are linked to the spatial (mainly longitudinal) 
variations of the free-electron density in the local ISM.
Finally, the sharp peaks around $i = 150$ and $i = 258$ find their origin 
in strong, localized enhancements in the free-electron density 
associated with nearby interstellar structures 
(mainly the Gum nebula and the Vela supernova remnant; 
see \cite{purcell&gsc_15} and references therein). 
These peaks hardly affect the results of our analysis: they only reduce the weight 
of the corresponding bins in the expression of $\chi^2$ (Eq.~\ref{eq_chi}).

\section{\label{appendix_param}Impact of the model parameters 
on the $\overline{\rm FD}_{\rm mod}$ map}

For compactness, the present discussion focuses on the halo-field models, 
A, B, C, and D (presented in Sect.~\ref{models_halo}), 
but all our conclusions also hold for the corresponding disk-field models, 
Ad, Bd, and Dd (presented in Sect.~\ref{models_disk})
-- remember that there is no model Cd.

All the field models are expressed in terms of a small number of free parameters,
which are related to either the shape of field lines
or the field strength distribution.
The parameters governing the spiral shape of field lines 
($p_0$, $H_p$, $L_p$; see Eq.~(\ref{eq_shiftedwindingfc}))
are common to all the models;
the parameters governing the shape of poloidal field lines 
($r_1$, $|z_1|$, $a$, $n$; see Eqs.~(\ref{eq_mfl_A}), (\ref{eq_mfl_B}),
(\ref{eq_mfl_C}), (\ref{eq_mfl_D})) apply each to one pair of models 
(A and B, C and D, A and C, B and D, respectively);
and the parameters governing the field strength distribution
($B_1$, $H$, $L$, $m$, $\varphi_{\star}$; see Eqs.~(\ref{eq_B1_AB}) 
and (\ref{eq_B1_CD})) enter either all the models or only one pair of models 
(all, A and B, C and D, all, all, respectively).
The azimuthal wavenumber, $m$, stands apart for its discrete values, 
which we further restricted in Sect.~\ref{results_remarks} 
to $m = 0$ (axisymmetric) and $m = 1$ (bisymmetric).
In this appendix, we successively consider these two values, 
and for each, we discuss the impact of all the other free parameters 
on the modeled $\overline{\rm FD}_{\rm mod}$ map,
noting that $\varphi_{\star}$ is relevant only when $m \not= 0$.
Because the discussion of the $m = 0$ case will serve as a basis 
for the $m = 1$ case, we prefer to keep model~A in the former, 
even though the inherent singularity of model~A can only be removed 
in non-axisymmetric configurations (see Sect.~\ref{models_modelA}).

\subsection{\label{appendix_param_axisym}Axisymmetric ($m = 0$) case}

We first consider the parameters governing the field strength distribution:
the normalization field strength, $B_1$ 
(in Eqs.~(\ref{eq_B1_AB}) and (\ref{eq_B1_CD})), 
the exponential scale height, $H$ (in Eq.~(\ref{eq_B1_AB}) for models~A and B),
and the exponential scale length, $L$ (in Eq.~(\ref{eq_B1_CD}) for models~C and D).
The effect of $B_1$ is straightforward.
The two poloidal components of the magnetic field 
(given in Sect.~\ref{models_poloidal}), as well as its azimuthal component
(given by Eq.~(\ref{eq_Bphi_p})), vary linearly with $B_1$.
The same must hold true for the line-of-sight field component, 
and hence for the Galactic FD (Eq.~\ref{eq_FD}).
In consequence, $B_1$ sets the overall amplitude 
of the $\overline{\rm FD}_{\rm mod}$ map, 
with no impact on its detailed structure.
$H$ and $L$, for their parts, control the vertical and radial profiles
of the field strength on the reference surface 
and, by implication, throughout the Galaxy.
Small values of $H$ or $L$ entail a rapid fall-off of the field strength 
with $|z|$ or $r$, which, in turn, tends to confine the large FD values 
to low $|b|$ or small $\ell$, respectively.
Accordingly, the $\overline{\rm FD}_{\rm mod}$ map tends to be dominated 
by a band along the Galactic plane (with longitudinal modulation; 
see next paragraph) 
or along the prime ($\ell = 0^\circ$) meridian, respectively.
Larger values of $H$ or $L$ tend to maintain large FD values 
up to higher $|b|$ or out to larger $\ell$, 
so that the $\overline{\rm FD}_{\rm mod}$ map tends to show a distribution 
that is more extended in latitude or in longitude, respectively.
It should be emphasized, though, that the effect of $H$ and $L$ 
is weighted by the free-electron density
and, therefore, becomes increasingly weak as $H$ or $L$ grows above 
the free-electron scale height or scale length, respectively.

We now turn to the parameters governing the spiral shape of field lines,
i.e., the parameters involved in the shifted winding function 
(Eq.~\ref{eq_shiftedwindingfc}): the pitch angle at the origin, $p_0$, 
the vertical scale height, $H_p$, and the radial scale length, $L_p$.
As indicated by Eq.~(\ref{eq_Bphi_p}), the ratio of azimuthal-to-poloidal field 
varies with $r$ and $z$ and goes to zero for $r \to \infty$ or $|z| \to \infty$.
In the limit $p_0 \to 0^\circ$, 
the field is purely azimuthal throughout the Galaxy, so that the Galactic FD 
(Eq.~\ref{eq_FD}) reverses sign at $\ell = 0^\circ, 180^\circ$
and peaks at intermediate $\ell$ in the inner Galactic quadrants.
The $\overline{\rm FD}_{\rm mod}$ map is then approximately\footnote{
not perfectly, because the free-electron density distribution 
is not perfectly axisymmetric.
\label{foot_sym}
}
antisymmetric with respect to the prime meridian
and dominated (in each hemisphere) by two patches of opposite signs 
on either side of the prime meridian.
Conversely, in one of the three limits $|p_0| \to 90^\circ$, 
$H_p \to 0$, or $L_p \to 0$, the field is purely poloidal everywhere
(except at $z = 0$, when only $H_p \to 0$), 
so that the Galactic FD generally peaks around $\ell = 0^\circ$,
with a secondary peak around $\ell = 180^\circ$,
and reverses sign at intermediate $\ell$. 
The $\overline{\rm FD}_{\rm mod}$ map is then 
approximately$^{\mbox{\scriptsize \ref{foot_sym}}}$
symmetric with respect to the prime meridian
and dominated (in each hemisphere) by one patch straddling it.
Between these two extremes, an increase in $|p_0|$ or a decrease 
in $H_p$ or $L_p$ tends to make the field globally more poloidal, 
thereby shifting the longitudes of FD reversals 
from $\ell = 0^\circ, 180^\circ$ toward intermediate $\ell$
and the longitudes of peak FD from intermediate $\ell$ 
toward $\ell = 0^\circ$ (stronger peak) and $\ell = 180^\circ$ (weaker peak).
While the value of $|p_0|$ affects the whole sky,
the values of $H_p$ and $L_p$ have more limited effects,
which are mainly felt at high $|b|$ and large $|\ell|$, respectively.

We finish with the parameters governing the shape of poloidal field lines:
the reference radius, $r_1$ (in models~A and B),
the positive reference height, $|z_1|$ (in models~C and D),
the opening parameter of parabolic field lines, $a$ 
(in Eq.~(\ref{eq_mfl_A}) for model~A and Eq.~(\ref{eq_mfl_C}) for model~C),
and the power-law index, $n$
(in Eq.~(\ref{eq_mfl_B}) for model~B and Eq.~(\ref{eq_mfl_D}) for model~D).
In the present study, the reference radius is set to $r_1 = 3~{\rm kpc}$
(a plausible value for the inner radius of the Galactic disk),
and the positive reference height is set to $|z_1| = 0$ in model~C 
and $|z_1| = 1.5~{\rm kpc}$ (a plausible value for the transition height 
between the disk and the halo) in model~D.
The power-law index turns out to have very little impact 
on the $\overline{\rm FD}_{\rm mod}$ map up to at least $n = 5$.
This allows us, for the purpose of speeding up the fitting process,
to fix $n$ at its smallest acceptable value, i.e.,
$n = 1$ in model~B, $n = 0.5$ in model~D, 
$n = 2$ in model~Bd, and $n = 0.5$ in model~Dd. 
The opening parameter is more critical, as we now discuss along general lines.
To simplify the discussion, we focus on one hemisphere, keeping in mind 
that the situation in the other hemisphere is either the same or opposite 
according to whether the magnetic field is symmetric or antisymmetric.

\noindent
$\bullet$ In model~A, small values of $a$ imply that the poloidal field 
is nearly radial and the total field nearly horizontal.
The Galactic FD then has a sinusoidal-like variation with $\ell$,
weighted toward the inner Galaxy and with zero point depending on pitch angle.
Accordingly, the $\overline{\rm FD}_{\rm mod}$ map is dominated by two patches
of opposite signs, with the innermost (lower $|\ell|$) patch being predominant.
For larger values of $a$, poloidal field lines are curved toward the rotation axis, 
and their spacing in azimuthal planes increases with increasing $r$, 
hence a faster decline in poloidal field strength 
and (since $B_\varphi$ is linearly related to $B_r$ and $B_z$; 
see Eq.~(\ref{eq_Bphi_p})) in total field strength.
In the $\overline{\rm FD}_{\rm mod}$ map, the two dominant patches 
become weaker -- especially the outermost patch, which rapidly fades away --
and the line of sign reversal between them (where the field is on average
perpendicular to the line of sight) moves away from the pole, 
toward the innermost patch.

\noindent
$\bullet$ In model~C, 
small values of $a$ imply that the poloidal field is nearly vertical.
The associated Galactic FD keeps the same sign throughout the considered hemisphere, 
but a generally larger FD, which reverses sign across the prime meridian
arises from the azimuthal field.
Accordingly, the $\overline{\rm FD}_{\rm mod}$ map has a single-sign background, 
onto which are superimposed two patches of opposite signs 
on either side of the prime meridian.
For larger values of $a$, poloidal field lines are curved toward the Galactic plane, 
and their spacing in azimuthal planes increases with increasing $|z|$, 
hence a faster decline in total field strength. 
In the $\overline{\rm FD}_{\rm mod}$ map, the two patches can be either enhanced 
or weakened, and the line of sign reversal between them moves toward the pole.

\noindent
In both models~A and C, the regions where field lines are most affected 
by a change in $a$ happen to be the regions where the free-electron density is lowest.
This automatically places a limit to the influence of $a$.

\subsection{\label{appendix_param_bisym}Bisymmetric ($m = 1$) case}

The bisymmetric case is generally more complex.
The parameters governing the field strength poloidal distribution
($B_1$, $H$, $L$) have basically the same impact
on the $\overline{\rm FD}_{\rm mod}$ map as in the axisymmetric case.
The other parameters have the same effect on the spiral and poloidal shapes 
of field lines, but their exact impact on the $\overline{\rm FD}_{\rm mod}$ map
is more difficult to determine, because it now depends in a strong and fine way
on a combination between the shifted winding function, $g_\varphi$ 
(Eq.~\ref{eq_shiftedwindingfc}) and the orientation angle of the azimuthal pattern, 
$\varphi_{\star}$ (in Eqs.~(\ref{eq_B1_AB}) and (\ref{eq_B1_CD})).

As it turns out, clear predictions can be made only in the limit 
$|p_0| \to 90^{\circ}$ (or, almost equivalently, $H_p \to 0$ or $L_p \to 0$),
where $g_\varphi \to 0$ and, therefore, the magnetic field is purely poloidal 
(see Eq.~(\ref{eq_Bphi_g}) or (\ref{eq_Bphi_p})).
In this limit, if $\varphi_{\star} = 0^\circ$ or $180^\circ$, 
the field azimuthal modulation 
(given by the cosine factor in Eqs.~(\ref{eq_B1_AB}) and (\ref{eq_B1_CD})) 
reaches its maximum amplitude in the azimuthal plane through the Sun 
($\varphi = 0^\circ$), 
i.e., in the plane $\ell = 0^\circ, 180^\circ$, where the projection factor 
of the poloidal field onto the line of sight is also maximum.
It then follows that the $\overline{\rm FD}_{\rm mod}$ map 
resembles that obtained in the axisymmetric case: 
it is approximately symmetric with respect to the prime meridian,
with (in each hemisphere) a dominant patch straddling it
and a line of sign reversal on either side of this patch.
In addition, FD has nearly the same peak value, 
but falls off faster than in the axisymmetric case.
If $\varphi_{\star} = \pm 90^\circ$, the field azimuthal modulation reaches 
its maximum amplitude in the azimuthal plane parallel to the plane of the sky
($\varphi = \pm 90^\circ$) and goes through zero in the azimuthal plane 
through the Sun ($\varphi = 0^\circ$),
so that FD vanishes at $\ell = 0^\circ, 180^\circ$.
FD also vanishes at two intermediate $\ell$ of opposite signs 
where the field is on average perpendicular to the line of sight.
The $\overline{\rm FD}_{\rm mod}$ map is then approximately antisymmetric 
with respect to the prime meridian
and composed (in each hemisphere) of four longitudinal sectors of alternating signs.
Moreover, the peak FD values are smaller than in the axisymmetric case,
because the poloidal field in the plane of maximum amplitude 
loses a fraction of its strength upon projection onto the line of sight.
Intermediate values of $\varphi_{\star}$ lead to asymmetric configurations
with two or four lines of sign reversal 
and with peak FD values smaller than in the axisymmetric case.

When $|p_0| \not= 90^\circ$, the magnetic field has a non-vanishing azimuthal
component, which gives field lines a spiral shape.
The azimuthal modulation, imposed on the reference surface, 
is carried along the spiraling field lines, 
such that the field generally reverses direction one or several times 
along any given line of sight.
These field reversals lead to an overall reduction of the FD values 
and to a more structured $\overline{\rm FD}_{\rm mod}$ map, 
whose details depend sensitively on the exact 
$g_\varphi - \varphi_{\star}$ combination.

\section{\label{appendix_results}Confidence intervals and correlations}

\subsection{\label{appendix_results_intervals}Confidence intervals}

The different best-fit parameters discussed in Sect.~\ref{results_bestfit} 
are obtained with very different degrees of accuracy 
(see Table~\ref{table_bestfit}).
The most-accurately determined parameter is the pitch angle at the origin, 
$p_0$, which is obtained to better than $\approx \pm 3^\circ$
at the $1\sigma$ confidence level when the halo field is axisymmetric (model~C0)
and better than $\approx \pm 1.3^\circ$ when the halo field is bisymmetric
(model~C1).
At the other extreme, the orientation angle of the azimuthal pattern,
$\varphi_\star$, for a bisymmetric field is always very poorly constrained:
when the halo field is axisymmetric, the $1\sigma$ confidence interval 
of $(\varphi_\star)_{\rm disk}$ covers the entire $\pm 180^\circ$ range,
and when the halo field is bisymmetric, the $1\sigma$ confidence intervals
of $(\varphi_\star)_{\rm disk}$ and $(\varphi_\star)_{\rm halo}$ 
are nearly the same and both $\approx 100^\circ - 160^\circ$.

The parameters $B_1$, $H$, $L$, and $\!\sqrt{a}$ are obtained
to better than a factor $\approx 2$, with two exceptions.
First, in the combination C0-Dd1, $L_{\rm disk}$ approaches 
the free-electron scale length, $L_{\rm e} \simeq 11~{\rm kpc}$ 
(see Sect.~\ref{method_simul}), such that its $1\sigma$ upper confidence limit 
can only be inferred to be $\gg L_{\rm e}$
(for that reason, it is set to $\infty$ in Table~\ref{table_bestfit})
and its $1\sigma$ lower confidence limit is also quite uncertain.
Second, in C1-Ad1, the $1\sigma$ lower confidence limit of $\!\sqrt{a_{\rm disk}}$ 
approaches zero, corresponding to purely horizontal field lines. 

The scale height of the winding function, $H_p$, is well constrained
as long as $H_p \not\gg H_{\rm e}$,
with an accuracy better than $\approx \pm 15\%$ when the halo field is bisymmetric
and $\approx \pm 25\%$ in C0-Dd1.
However, in C0-Ad1 and C0-Bd1, where $H_p \gg H_{\rm e}$, 
neither the best-fit value nor the $1\sigma$ upper limit of $H_p$ 
can be derived with any accuracy;
only an approximate $1\sigma$ lower limit can be drawn.
Similarly for the scale length of the winding function, $L_p$:
in all the total-field models, $L_p \gg L_{\rm e}$ 
and only an approximate $1\sigma$ lower limit can be drawn.
This is because, when $H_p \gg H_{\rm e}$ or $L_p \gg L_{\rm e}$,
the magnetic field inside the free-electron region (where the Galactic FD arises)
depends only weakly on the exact value of $H_p$ or $L_p$, respectively,
which, therefore, can hardly be constrained by the $\overline{\rm FD}_{\rm obs}$ map.
As it turns out, this weak dependence on $H_p$ or $L_p$ extends to 
the whole Galactic region with non-negligible magnetic field,
which means that the large uncertainties in $H_p$ or $L_p$ 
are luckily not critical for our magnetic field models.
Thus, we may set $H_p$ or $L_p$ to a somewhat arbitrary value 
that is large compared to the size of the Galactic magnetic region
and yet small compared to the distance to the external intergalactic medium 
where field lines are supposed to be anchored 
(see Sect.~\ref{models_azimuthal}).
For instance, we may let $L_p = 50~{\rm kpc}$ in all the total-field models
and $H_p = 20~{\rm kpc}$ in C0-Ad1 and C0-Bd1.
In the latter models, Eq.~(\ref{eq_Bphi_p}) then implies that the azimuthal 
field component is simply given by $B_\varphi \simeq \cot p_0 \ B_r$, 
i.e., the pitch angle is nearly constant, throughout the magnetic region.

Another issue that should be addressed here concerns the restrictions 
imposed on the large-scale magnetic field at the Sun, 
namely, $B_\odot \in [1,2]~\mu{\rm G}$ and $p_\odot \in [-15^\circ, -4^\circ]$
(see Sect.~\ref{method_simul}).
The imposed range of $p_\odot$ turns out to have no impact on the final results,
as the $1\sigma$ confidence intervals of $p_\odot$ in all six total-field models
fall well inside the imposed range of $[-15^\circ, -4^\circ]$.
Note that $p_\odot$ is always close to $p_0$:
the best-fit values differ by only $\simeq 0.7^\circ - 0.8^\circ$ 
and the histograms look very similar.
This is obviously a direct consequence of the large values of $L_p$.
On the other hand, the imposed range of $B_\odot$ directly affects our results:
the $1\sigma$ confidence intervals of $B_\odot$ are clearly cut off 
either at $2~\mu{\rm G}$ (in models C0-Ad1 and C0-Bd1)
or at both $1~\mu{\rm G}$ and $2~\mu{\rm G}$ (in the three models with C1).

\subsection{\label{appendix_results_correlations}Correlations}

The confidence intervals shown in Table~\ref{table_bestfit} 
ignore the possible correlations and degeneracies between parameters.
To uncover the important correlations, we consider all the parameters two by two, 
plot the associated 2D marginalized point densities 
from the second halves of the relevant Markov chains
(see Sect.~\ref{method_fitting}),
visually examine the 2D density plots,
and compute the Pearson and Spearman correlation coefficients
as indicators of linear and nonlinear relations, respectively.

In all models, we find strong anti-correlations between $B_1$ 
and either $H$ (in models~Ad1 and Bd1) or $L$ (in models~C0, C1, and Dd1),
as well as somewhat weaker anti-correlations between $p_0$ and both $H_p$ and $L_p$.
These conform to the expected anti-correlation between the normalization value 
and the scale height/length of a governing quantity.
We also find strong anti-correlations between $(B_1)_{\rm disk}$ and $p_0$,
plus, in some models, weaker anti-correlations between $(B_1)_{\rm halo}$ and $p_0$.
Physically, an increase in $p_0$, corresponding to a decrease in $|p_0|$, 
entails an increase in $|B_\varphi|$ (especially at low $|z|$
(see Eq.~(\ref{eq_Bphi_p})), i.e., in the disk), 
and hence a global increase in $|\overline{\rm FD}_{\rm mod}|$ 
(especially at low $|b|$ (see Eq.~(\ref{eq_FD}), with $B_\parallel$ having 
a contribution $\propto \cos b$ from $B_\varphi$), i.e., mostly through the disk). 
To recover a good fit to the observational $\overline{\rm FD}_{\rm obs}$ map,
this global increase in $|\overline{\rm FD}_{\rm mod}|$ must be counterbalanced
by a decrease in $(B_1)_{\rm disk}$, 
sometimes accompanied by a decrease in $(B_1)_{\rm halo}$.

The correlations between the disk and halo field parameters are generally weak
and not necessarily negative (as might naively be expected).
This is because the contributions from the disk and halo fields 
to $\overline{\rm FD}_{\rm mod}$ generally add up in some regions 
and cancel out in other regions.
There is, however, one notable exception: when the halo field is bisymmetric, 
the orientation angles $(\varphi_\star)_{\rm disk}$ and $(\varphi_\star)_{\rm halo}$
are nearly perfectly correlated.
This means that the azimuthal patterns of the disk and halo fields
have a narrowly-constrained relative orientation,
so that they remain locked to each other even as their combined pattern rotates 
about its best-fit orientation and as the other parameters 
vary about their best-fit values.

The orientation angle $(\varphi_\star)_{\rm disk}$ is also strongly correlated 
with the parameters of the winding function, $p_0$, $H_p$, and $L_p$.
Remember that $(\varphi_\star)_{\rm disk}$ gives the orientation 
of the bisymmetric azimuthal pattern of the disk field at infinity
(see Sect.~\ref{models_azimuthal}, below Eq.~(\ref{eq_mfl_winding_surf})),
whereas the $\overline{\rm FD}_{\rm mod}$ map is most sensitive to 
its orientation in the nearby outer Galaxy
(see north-south symmetry argument in Sect.~\ref{FD_data_trends}).
Both orientations are linked through the winding function.
Therefore, a change in $(\varphi_\star)_{\rm disk}$ can roughly preserve 
the azimuthal pattern of the disk field in the nearby outer Galaxy 
if it is offset by appropriate changes in $p_0$, $H_p$, $L_p$.
In practice, due to the tight winding of field lines, a small change in $p_0$
can be sufficient to offset any change in $(\varphi_\star)_{\rm disk}$.
This explains the narrow ($\lesssim \pm 3^\circ$) and total ($\pm 180^\circ$)
$1\sigma$ confidence intervals of $p_0$ and $(\varphi_\star)_{\rm disk}$,
respectively, when the halo field is axisymmetric.
When the halo field is bisymmetric, the $\overline{\rm FD}_{\rm mod}$ map 
is also very sensitive to the orientation of its azimuthal pattern 
in the inner Galaxy, 
which is linked, through the winding function, to $(\varphi_\star)_{\rm halo}$,
which, in turn, is nearly perfectly correlated to $(\varphi_\star)_{\rm disk}$.
To maintain a good fit to the $\overline{\rm FD}_{\rm obs}$ map, 
a change in $(\varphi_\star)_{\rm disk}$ must now roughly preserve
the azimuthal patterns of both the disk field in the nearby outer Galaxy
and the halo field in the inner Galaxy, 
which is again possible with appropriate changes in $p_0$, $H_p$, $L_p$, 
but only over a limited ($\approx 100^\circ - 160^\circ$) interval 
of $(\varphi_\star)_{\rm disk}$. 
Accordingly, the $1\sigma$ confidence interval of $p_0$ is narrower
($\lesssim \pm 1.3^\circ$) than when the halo field is axisymmetric,
and so are the confidence intervals of $H_p$ and $L_p$.
Another consequence of the locking between the azimuthal patterns of the disk 
and halo fields is that all correlations tend to be tighter when the halo field 
is bisymmetric.

\begin{acknowledgement}{}
We express our deep gratitude to our colleagues, Tess Jaffe and Pierre Jean, 
who provided valuable help with the MCMC simulations,
and to Maik Wolleben, who kindly sent us the FD data 
for the \cite{wolleben&flc_10} bubble.
\end{acknowledgement}

\bibliographystyle{aa}
\bibliography{BibTex}

\end{document}